%% file: main_final.tex
\documentclass[journal ]{new-aiaa}
\usepackage[utf8]{inputenc}
\usepackage{textcomp}
\usepackage{gensymb}
\usepackage{graphicx}
\usepackage{amsmath}
\usepackage[version=4]{mhchem}
\usepackage{siunitx}
\usepackage{longtable,tabularx}
\usepackage{booktabs}
\usepackage{placeins}
\usepackage{caption}
\usepackage[hypcap=true]{subcaption}
\usepackage{multirow}
\usepackage{bm}
\usepackage[nameinlink,capitalise]{cleveref} %Automatic references
\usepackage{apptools}
\crefname{section}{section}{sections}
\crefname{appendix}{\IfAppendix{section}{appendix}}{\IfAppendix{sections}{appendices}s}

\newcommand{\mb}[1]{\mathbf{#1}}
\newcommand{\mr}[1]{\mathrm{#1}}
\newcommand{\mi}[1]{\mathit{#1}}
\newcommand{\mref}[1]{\mbox{\cref{#1}}}
\newcommand{\Mref}[1]{\mbox{\Cref{#1}}}

\input{review.tex}

\title{Data-Driven Modeling of Hypersonic Reentry Flow with Heat and Mass Transfer}

\author{Leonidas Gkimisis\footnote{Research Master Student, Aeronautics and Aerospace Department. Currently: Graduate Student, Max Planck Institute for Dynamics of Complex Technical Systems, Germany.}, Bruno Dias\footnote{Postdoctoral Researcher, Aeronautics and Aerospace Department. Currently: Postdoctoral Research Fellow, NASA Ames Research Center, Moffett Field, CA, USA. AIAA Member.}, James B. Scoggins\footnote{Postdoctoral Researcher, Aeronautics and Aerospace Department.  Currently: Aerospace Research Engineer, Aerothermodynamics Branch, NASA Langley Research Center, Hampton, VA, USA. AIAA Member.}, Thierry Magin \footnote{Professor, Aeronautics and Aerospace Department. AIAA Senior Member.}, Miguel A. Mendez\footnote{Assistant Professor, Environmental and Applied Fluid Dynamics Department.}, Alessandro Turchi\footnote{Research Expert, Aeronautics and Aerospace Department. Currently: Aerospace Engineer, Italian Space Agency, Science and Research Directorate, Rome, Italy. Corresponding author: alessandro.turchi@asi.it. AIAA Senior Member.}}

\affil{von Karman Institute for Fluid Dynamics, Rhode-Saint-Gen\`{e}se, Belgium}

\begin{document}

\maketitle

\begin{abstract}
The entry phase constitutes a design driver for aerospace systems that include such a critical step. This phase is characterized by hypersonic flows encompassing multiscale phenomena that require advanced modeling capabilities. However, since high fidelity simulations are often computationally prohibitive, simplified models are needed in multidisciplinary analyses requiring fast predictions.
This work proposes data-driven surrogate models to predict the flow, and mixture properties along the stagnation streamline of hypersonic flows past spherical objects. Surrogate models are designed to predict velocity, pressure, temperature, density and air composition as a function of the object's radius, velocity, reentry altitude and surface temperature. These models are trained with data produced by numerical simulation of the quasi-one-dimensional Navier-Stokes formulation and a selected Earth atmospheric model. Physics-constrained parametric functions are constructed for each flow variable of interest, and artificial neural networks are used to map the model parameters to the model's inputs. Surrogate models were also developed to predict surface quantities of interest for the case of nonreacting or ablative carbon-based surfaces, providing alternatives to semiempirical correlations. A validation study is presented for all the developed models, and their predictive capabilities are showcased along selected reentry trajectories of space debris from low-Earth orbits.

\end{abstract}

\section*{Nomenclature}

{\renewcommand\arraystretch{1.0}
\noindent\begin{longtable*}{@{}l @{\quad=\quad} ll@{}}
\multicolumn{3}{@{}l}{\textit{Roman symbols}}\\
$\mi{Bo}$ & Bodenstein number & - \\
$C_p$ & specific heat & \si{J\per (kg.K)} \\
$\mi{Da}$ & Damk\"{o}hler number & - \\
$\mi{Ec}$ & Eckert number & - \\
$f$ & flow field quantity CFD solution & - \\
$f'$ & flow field quantity prediction & - \\
$h$ & altitude & $\si{km}$ \\
$h_K$ & Nondimensional altitude & $-$ \\
$H$ & mixture enthalpy & $\si{J/kg}$ \\
$H_K$ & K\'{a}rm\'{a}n line altitude & $\si{km}$ \\
$j$ & diffusive fluxes & $\si{kg/(m^2s)}$ \\
$J$ & objective function & - \\
$\mi{M}$ & Mach number & - \\
$\dot{m}$ & surface mass blowing flux& \si{kg \per (m^2.s)} \\
$n$ & Input vector dimension & - \\
$m$ & Regression output vector dimension & - \\
$N$ & Number of CFD grid points & - \\
$p$ & pressure & $\si{Pa}$ \\
$\mi{Pe}$ & Pecklet number & - \\
$\mi{Pr}$ & Prandtl number & - \\
$q$ & heat flux & $\si{W/m^2}$ \\
$r$ & radial position & $\si{m}$ \\
$R$ & radius & $\si{m}$ \\
$\mi{Re}$ & Reynolds number & - \\
$S$ & Parameters regression function & - \\
$t$ & time & $\si{s}$ \\
$T$ & temperature & $\si{K}$ \\
$u$ & velocity (radial component) & $\si{m/s}$ \\
$\bm{u}$ & dimensional inputs vector & - \\
$\bm{\hat{u}}$ & dimensionless inputs vector & - \\
$v$ & velocity (angular component) & $\si{m/s}$ \\
$\bm{w}$ & predicted parameters & -  \\
$\bm{w_g}$ & curve-fitting "ground truth" parameters & -  \\

\multicolumn{3}{@{}l}{\textit{Greek symbols}}\\
$\alpha$ & speed of sound & $\si{m/s}$ \\
$\theta$ & dimensionless temperature & -\\
$\rho$ & density & $\si{kg/m^3}$ \\
$\tau$ & stress tensor components & $\si{Pa}$ \\
$\phi$ & angular direction & -\\
$\Phi$ & regression functions & -\\
$\dot{\omega}$ & chemical source term & $\si{kg/(m^3.s)}$ \\
\end{longtable*}}

\section{Introduction}
\label{S:1}

Hypersonic flows (roughly identified as having a Mach number greater than 5) are one of the critical topics in planetary entry studies. The simulation of these flows past blunt bodies finds application to several crucial aerospace research fields. Examples include the design of entry vehicle thermal protection systems \cite{Uyanna2020}, the disintegration analysis \cite{Patera1998, Riley2017} or the trajectory prediction of decommissioned satellites and space debris upon their atmospheric reentry \cite{Ailor2007}. 

Several challenges arise in the numerical simulation of hypersonic flows past reentry objects. These are linked to the sharp gradients across the detached normal shock \cite{Moretti1966, Prakash2011, AndersonJr.2006}, the modelling of chemical reactions among the species under high temperature and pressure in the post-shock conditions \cite{Najm1998,Schwer2003} and the treatment of gas-surface interaction at the wall \cite{Dias2020a, Dias2020, Turchi2015}. These difficulties motivate the development of simplified prediction tools for the hypersonic flow around blunt bodies. Semianalytical relations have been proposed over the past decades to predict crucial pointwise properties \cite{FAY1958, Billig1967}, with particular attention to the flow conditions at the stagnation point, receiving the highest thermal loads \cite{LEES1956}. Because of the high computational cost of high fidelity CFD simulations \cite{Padilla2005}, semianalytical correlations \cite{Lips2005} remain the main tools to account for the aerothermochemistry in multidisciplinary investigations \cite{Murray} incorporating thermal structural and break up analyses \cite{Wu2011, Bose2009, Martin2004}. 

Nevertheless, the range of validity of these correlations is often limited \cite{Park2021}. This motivates the interest in developing data-driven models, potentially leveraging on available data and modern techniques for regression and dimensionality reduction from machine learning.  

Data-driven methods offer promising solutions for bridging the gap between low cost and high accuracy for high-dimensional problems \cite{Yu2019, Kutz2017, Brunton2016, Peherstorfer2016, BenGHP20,brunton2021data}. 
These tools are becoming increasingly popular in fluid dynamics and, more recently, in hypersonic and reactive flows \cite{Huang2018}. Examples are reduced models for predicting surface properties for reentering aerospace vehicles \cite{Crowell2010,Chen2015} and space debris \cite{Drouet2021}, and reduced models based on operator inference, developed in Ref. \cite{McQuarrie2021} for a 2D reacting and subsonic flow. Focusing on hypersonic flow predictions, recent work \cite{Mao} presented operator-predicting Neural Networks (DeepONets) to retrieve post-shock chemical composition along the stagnation streamline, based on velocity and temperature measurements for Mach number values ranging from 8 to 10.

When the focus is on the analysis of hypersonic flows for atmospheric entry, the reconstruction of wall quantities such as the incoming convective heat flux or the stagnation pressure, often main quantities of interest, is of obvious significance.
To this end, the application of data-driven techniques to build efficient regressors for these wall quantities is a promising approach. 
In addition, the ability to obtain quick yet accurate reconstructions of flow variables along the stagnation streamline can be appealing for a variety of applications.
A few examples of them, which could benefit from the ability to obtain insight on the flow field along an entire reentry trajectory in a relatively quick manner, could be: the one-way coupled evaluation of the incoming radiative heat flux \cite{Soucasse2016,Dias2020}; the implementation of gas-surface interaction models, which could use the reconstructed boundary-layer edge composition as input; and the acquisition of detailed information on the degree of thermochemical nonequilibrium in the shock layer, e.g., useful for the verification of the existence of the basic requirements for the application of the Local Heat transfer Simulation methodology \cite{kolesnikov:2000,Turchi2015} when defining relevant ground-testing conditions for inductively-coupled plasma torches \cite{kolesnikov_aiaa}.

This work presents the development of surrogate models to predict the main flow features along the stagnation streamline of hypersonic flows past spherical bodies. These models include regressors for the flow velocity, pressure, temperature, density, and species partial densities, as well as the heat and mass fluxes at the stagnation point.
The article is structured as follows.  \Cref{S:2} presents the numerical model used to compute stagnation line flow fields, including the governing equations of the numerical model, their nondimensionalization and solution procedure, and the strategy used to generate a machine learning dataset.  In \Cref{S:3} and \Cref{S:4}, we present the development and validation, respectively, of the surrogate models themselves. 
\Cref{S:5} presents an application of the developed models to analyze the reentry of space debris from low-Earth orbits.  Finally, \Cref{S:6} draws the conclusions and discusses possible further developments.

\section{Theoretical model and numerical approach for data collection}
\label{S:2}

\subsection{Stagnation streamline formulation}
\label{sub:eqs}

\begin{figure}
\begin{center}$
\begin{array}{rr}
\includegraphics[width=75mm]{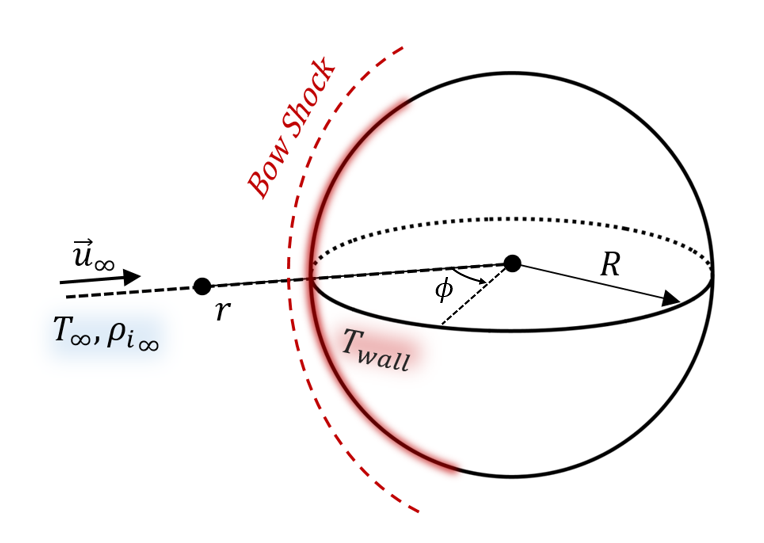}&
\includegraphics[width=45mm]{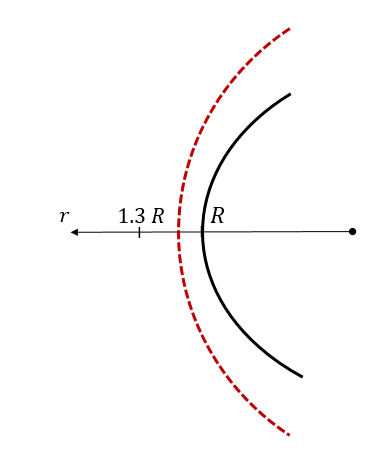}
\end{array}$
\end{center}
\caption{Schematic with the main parameters to study the hypersonic reentry of a spherical object. The figure on the left recalls the main variables in 3D; the figure on the right shows the 1D domain for the stagnation streamline formulation used in this work.} \label{fig:stag}
\end{figure}

We consider the three-dimensional flow past a spherical object of radius $R$, reentering at a velocity $u_\infty$ (see \cref{fig:stag}). The object is at an altitude $h$, where the air in the free stream is a composition of $i=1,\dots n_I$ species, each with density $\rho_{i_\infty}$, at a temperature $T_\infty$. Assuming spherical symmetry of the problem, we adopt the quasi-one-dimensional formulation of the Navier-Stokes equations along the stagnation streamline as derived by Klomfass and M\"uller \cite{Klomfass}. This formulation is derived via variable separation for the Navier-Stokes equations in spherical coordinates. 
Specifically, writing every variable $a$ as $a(r,\phi)=\overline{a}(r) a'(\phi)$ and under the assumption of rotationally-symmetric flow, it is possible to derive the specific governing equations for the flow along the stagnation streamline, i.e., $\overline{a}(r)$ at $\phi=0$.

These equations read as follows:

\begin{equation}
\label{syst}
\begin{array}{c}
\frac{\partial}{\partial t}\left(\begin{array}{c}
\rho_{i} \\
\rho u \\
\rho v \\
\rho E
\end{array}\right)+\frac{\partial}{\partial r}\left(\begin{array}{c}
\rho_{i} u+j_{i} \\
\rho u^{2}+p+\tau_{r r} \\
\rho v u+\tau_{r \phi} \\
\rho u H+q+\tau_{r r} u
\end{array}\right)+ 
\frac{1}{r}\left(\begin{array}{c}
2 \rho_{i}(u+v)+2 j_{i} \\
2 \rho u(u+v)+2\left(\tau_{r r}-\tau_{\phi \phi}+\tau_{r \phi}\right) \\
3 \rho v(u+v)-2 (p-p_\infty)-\tau_{\phi \phi}+3 \tau_{r \phi} \\
2 \rho H(u+v)+2\left(q+u \tau_{r r}+v \tau_{\phi \phi}+u \tau_{r \phi}\right)
\end{array}\right)=\left(\begin{array}{c}
\dot{\omega}_{i} \\
0 \\
0 \\
0
\end{array}\right)\,,
\end{array}
\end{equation} 
where we have dropped the overline notation given that all the variables are now solely a function of the radial coordinate $r$.
In \cref{syst}, $\rho_i$ is the density of the $i^\text{th}$ species; $u$ and $v$ are linked to the linear-velocity components in the radial (i.e., along the stagnation line) and angular (i.e., normal to the stagnation line) directions\footnote{The radial component is equivalent to the actual physical velocity in that direction, whereas the angular one is a result of the variable decomposition in the stagnation-line formulation (see \cite{Klomfass}) and does not correspond to a physical velocity normal to the stagnation line, which is obviously zero.};
$j_i$ are the diffusive fluxes of each species; $\tau_{r r}$,$\tau_{r \phi}$,$\tau_{\phi \phi}$ are the entries of the 2D viscous stress tensor; $q$ is the heat flux; and $H$ is the specific mixture enthalpy. 
We omit writing explicitly these terms and refer the reader to \cite{Klomfass} and \cite{Munafo2014} for more details on the quasi-1D stagnation streamline formulation. 
It is worth remarking that \cref{syst} includes $n_I$ mass continuity equations and, following the assumption of thermal equilibrium, a single energy equation.
%The dimensionless formulation of this equation, as used in this work, is reported in appendix \ref{appex1}. The reader should note that the resulting system of equations (\cref{syst}) includes as many mass conservation equations as the number of species in the considered air mixture. Moreover, only a single energy equation is considered since ionization is neglected.

%whereas the radial one is a result of the variable decomposition and does not correspond to a physical velocity, obviously zero in the direction normal to the stagnation line for the stagnation streamline.

The integration of \cref{syst} is carried out in a 1D domain from $r_\mr{in}=R$ to $r_\mr{out}>R$, 
$R$ being the radius of the spherical body under analysis.
The boundary condition on $r_{\mr{out}}$ corresponds to a supersonic free-stream with imposed velocity ($u_\infty$), temperature ($T_\infty$) and air composition ($\rho_{i_\infty}$). 
The free-stream thermodynamic state can be defined using an Earth atmospheric model \cite{NGDC1992} and assuming chemical equilibrium. Therefore, the free-stream temperature and composition are entirely defined by the altitude $h$. On the inner boundary, $r_{\mr{in}}$, the boundary condition is that of a solid wall. For the present analysis, this corresponds to a prescribed wall temperature $T_{\mr{wall}}$ and null or nonzero normal velocity for a nonreacting/catalytic or ablative wall respectively. The wall velocity in case of ablative walls is set using the model discussed in Ref. \cite{Dias2015}. 
Based on the above, four variables uniquely define a solution of \cref{syst}: the object altitude $h$, the free-stream velocity $u_{\infty}$, the object radius $R$, and wall temperature $T_{\mr{wall}}$.
These quantities define the dimensional input vector $\bm{u}=[h,u_{\infty},R,T_{\mr{wall}}]^T$.

\subsubsection{Nondimensional analysis}
\label{subs:nda}

The governing equations in \cref{syst} were scaled to identify the dimensionless numbers governing the scaling of the problem. Moreover, this allowed for formulating the surrogate models (see \cref{S:3}) in dimensionless form, hence ensuring dimensional consistency of the prediction and enabling physical insights and verifications. 

The derivation of the nondimensional system is detailed in Appendix \ref{appex1}.
The scaling of the continuity equation(s) relies on the Mach number $M={u_\infty}/{c}$ ($c$ being the speed of sound), while the chemistry and mass diffusion are scaled through the Damköhler ($\mi{Da_i}={k_i R}/{\rho_\infty u_\infty}$ where $k_i$ is a reference reaction rate for species $i$) and Bodenstein numbers ($\mi{Bo}_i={D_i}/{\rho_\infty u_\infty R}$ where $D_i$ is the diffusion coefficient of species $i$). The scaling of the momentum equation introduces the Reynolds number ($\mi{Re}={u_\infty R}/{\nu}$) while the scaling of the energy equation involves the Pecklet ($Pe={u_\infty R}/{a}$ with $a$ being the thermal diffusivity) and the Eckert ($\mi{Ec}={{u_\infty}^{2}}/{(C_p (T_{wall}- T_\infty))}$) numbers as well as the heat capacity ratio (${C_p}_i/C_p$) for each chemical species.

Having linked the fluid properties to an atmospheric model, the quantities $\mi{Pe}$, $\mi{Da}_i$, $\mi{Bo}_i$ and ${C_p}_i/C_p$ depend only on the reentry free-stream conditions. Therefore, the dimensionless set associated to the reentry parameters $\bm{u}$, herein denoted as $\bm{\hat{u}}$, requires three parameters for the scaling in absence of chemistry and one parameter defining the air composition, namely a dimensionless altitude. We consider $\mi{M, Re, Ec}$ for the first three and $h_K=h/H_K$, with $H_K$ the von Karman altitude, to obtain $\bm{\hat{u}}=[\mi{M,Re,Ec,h_K}]$. It can be shown that the above selection allows for a bijective representation from $\bm{u}$ to $\bm{\hat{u}}$. Therefore, we are entitled to use the nondimensional inputs $\bm{\hat{u}}(\bm{u})$ in place of the dimensional inputs $\bm{u}$ for our model.

\subsection{Numerical solver and data collection}
\label{CFD_na}
The stagnation-line solver employed to solve \cref{syst} was originally developed by Munaf\`o and Magin \cite{Munafo2014} and previously employed for the reentry simulation of meteors \cite{Dias2015} and aerospace vehicles \cite{Turchi2015}. 
The code is coupled to the VKI \textsc{Mutation++} library \cite{Scoggins2020}, which is used for the computation of the thermodynamic and transport properties.
The library also provides evaluation of the gas-phase chemical source terms and the mass diffusion fluxes $j_{i}$.

Using the \textsc{Mutation++} library, which implements a generalized gas-surface interaction (GSI) model \cite{bellas2017development}, the stagnation-line solver features a variety of surface boundary conditions, ranging from nonreacting to ablative. 
Taking advantage of this flexibility, our study analyzes two scenarios.
The first is the reentry of spherical bodies with a nonreactive surface.
The second is the reentry of spherical carbon-based bodies with an ablative surface.

\subsubsection{Case of nonreactive bodies}
\label{subs:nom}
Our interest lies in mild reentry conditions of man-made objects, e.g., space debris.
Therefore, keeping the reentry velocity below $10$ km/s \cite{Bose2009} justifies neglecting the species ionization and using a classic five-species air mixture (\ce{N,O,NO,N_2,O_2}) with gas-chemistry data for all simulations. 
Moreover, as for \cref{syst}, the flow was assumed to be in thermal equilibrium. This simplifying assumption was made to ease the CFD simulations and could potentially have a minor quantitative effect on the CFD results. Generally speaking, the shape of the internal temperature profile in the case of a multi-temperature model is not expected to differ significantly from the translational temperature one. Similarly, it is expected that the qualitative shape of the other flowfiled quantities, and thus their corresponding parametrization, will not differ noticeably between the equilibrium and the non-equilibrium case. Therefore, it is reasonable to assume that the current work could be extended to the multi-temperature case without major modifications.
A total of 2091 simulations were carried out, varying the reentry parameters within the ranges summarized in \cref{tabin}. The parameters for each simulation were randomly sampled within these ranges. 
The wall temperature in the range \SIrange{1000}{3000}{\kelvin} was selected to be in line with other reentry studies \cite{Klinkrad2013}, and the range of velocities to be compatible with the object size in the range \SIrange{0.1}{1}{m} \cite{Klinkrad2013, Opiela2009}. 
Stagnation line distributions for velocity, pressure, density, and species densities as well as the stagnation point convective heat flux were then extracted from each solution to form the model dataset for nonreactive bodies.

\begin{table}[t!]
\footnotesize
\centering
\renewcommand*{\arraystretch}{1.5}
\caption{CFD simulation dimensional input range. These dimensional quantities ($\bm{u}$) were scaled into their nondimensional counterparts ($\bm{\hat{u}}$).}
\label{tabin}
\begin{tabular}{lccccc}
\toprule
\toprule
& $h$ (km) &  $u$ (km/s)  &  $T_{\mr{wall}}$ (K) &  $R$ (m) \\
\midrule
& $[30, 70]$ &$[3, 10]$ &$[1000, 3000]$& $[0.1, 1]$\\
\bottomrule 
\bottomrule 
\end{tabular}
\end{table}

\subsubsection{Case of carbon-based ablative bodies}
\label{subs:extend}

When ablation occurs one should consider that i) mass is injected into the boundary layer through a convective flux outgoing from the surface -- hence the velocity normal to the wall is no longer zero -- and ii) additional heat-flux contributions appear in the energy balance over the surface \cite{Turchi2015}. The flux balance must thus include gas conduction ($q_{\mr{con}}$), diffusion ($q_{\mr{dif}}$) and blowing ($q_{\mr{bl}}$):

\begin{equation}
\underbrace{k_{\mr{w}}\, \left. \frac{\partial T}{\partial r}\right|_{\mr{w}}}_{\mbox{gas conduction}}+ 
\underbrace{{\sum_i^{N} \left(h_{i} \,\rho_i \,v_{i}^{\mr{d}}\right)_{\mr{w}} }}_{\mbox{diffusion}}-
 \underbrace{{\dot m}\,(h_{{\mr{w}}}-h_{s})}_{\mbox{blowing}}
   = 
\underbrace{k_{s}  \left.\frac{\partial T}{\partial r}\right|_{s}}_{\mbox{solid conduction}},
\label{eq:bound1}
\end{equation} where the subscripts $w$ and $s$ denote the fluid and solid sides at the wall, $k$ is the thermal conductivity, $\dot{m}$ is the blowing mass flux, $h_i$ and $v^d_i$ are the specific enthalpy and the drift velocity of the $i$-th species and $N$ is the number of species involved in the ablation process. 

Six additional carbon-based species, products or byproducts of the surface ablation, were introduced in the mixture: \ce{C,C_2,C_3,CN,CO,CO_2}. 
Additional gas-phase reactions, involving the new species, were taken from Ref. \cite{olynick1999aerothermodynamics} and considered in the simulations. 
As for the surface process, the following reactions were considered: i) carbon oxidation due to atomic oxygen, and sublimation of solid carbon into \ce{C3} with data from Refs. \mbox{\cite{park1976effects,park1989nonequilibrium}}; ii) carbon nitridation due to atomic nitrogen with data from Ref. \mbox{\cite{helber2017determination}}; and iii) nitrogen recombination (considered with an efficiency of \num{1d-3}). Note that the selection of this phenomenological ablation model was considered appropriate for the present study, without loosing generality. However, we would remark that data-driven models of the type presented in this work could also be obtained from CFD simulations that implements more complex gas-surface interaction models \cite{prata2022air}.

The surrogate model developed for the flow-field mixture quantities (i.e., velocity, pressure, and density) in the case of nonreactive surfaces could also be extended to the ablative case if the species models are developed accordingly. However, this extension is left to future work. In the case of ablative surfaces,  we here focus on the development of surrogate models to predict the three contributions to the heat flux balance on the left-hand side of \eqref{eq:bound1} from the input parameters $\hat{\bm{u}}$. A total of 1738 cases were run considering carbon-based ablative bodies spanning the same range of parameters analyzed for the nonreactive surfaces in Table \ref{tabin}.

\subsubsection{Grid convergence study}
\begin{figure}[!htp]
\center
\includegraphics[width=0.5\columnwidth]{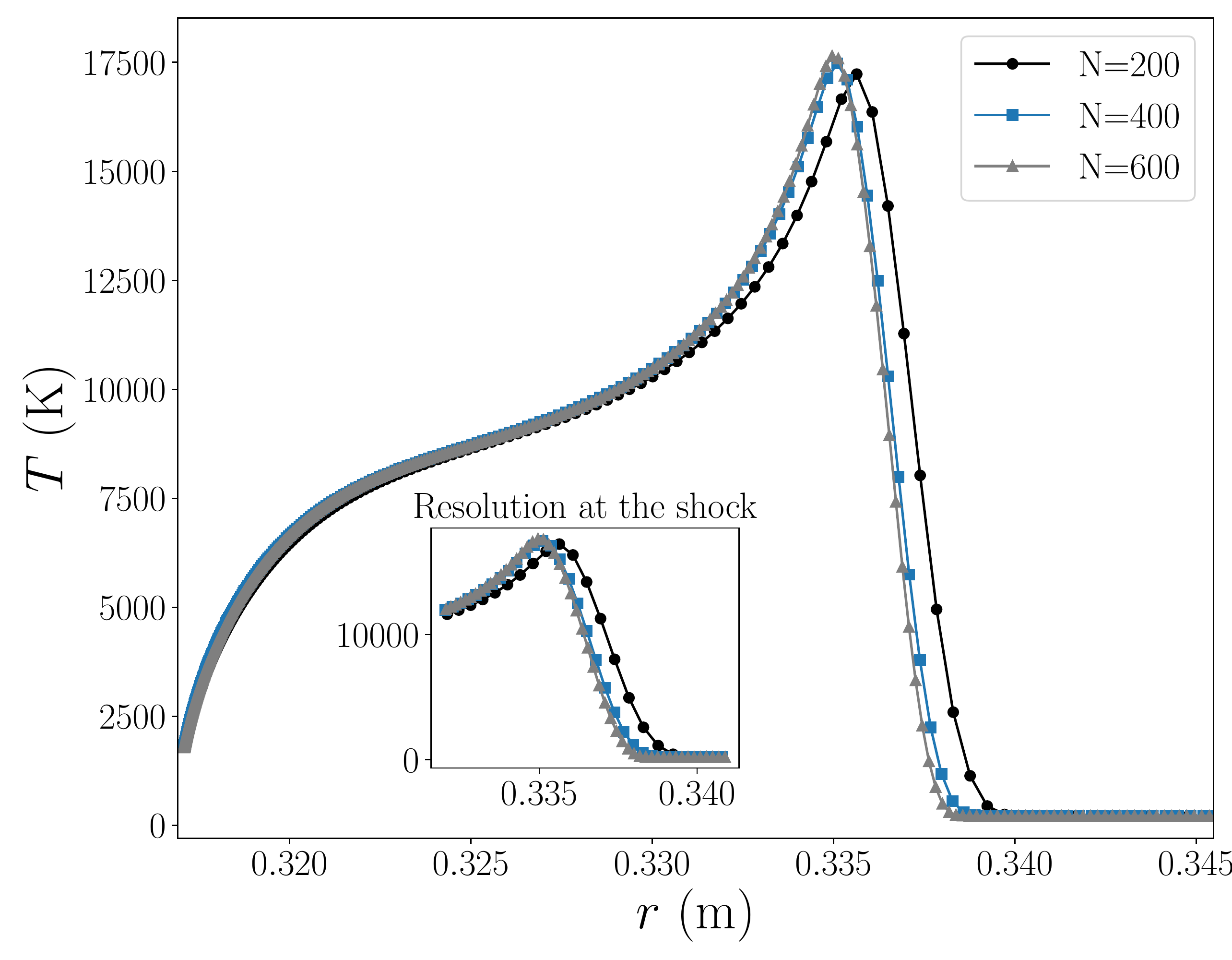}
\caption{Temperature distribution computed from the 1D stagnation line formulation in \cref{syst} for various grids. The profiles corresponds to a case with $\bm{u} =[71.64  \; \rm{km}, 8.42  \; \rm{km/s}, 0.317  \; \rm{m}, 1785  \; \rm{K}]$. This shows that a grid with $\bm{N=400}$ nodes is sufficient to obtain a mesh-converged solution.} 
\label{fig:refcomp}
\end{figure}

The selected domain for all the simulations ranged from $r=R$ to $r=1.3R$, adequate for capturing the normal shock position under the conditions of \cref{tabin}.
The 1D computational mesh featured a mesh refinement in the wall region to ensure sufficient resolution of the near-wall phenomena.
A first simulation was performed with $N=200$ points, then the mesh was progressively refined using the same stretching law but considering an increasing number of mesh points, i.e., $N=400$ and $N=600$.
 At each level of the refinement, the solution from the coarser mesh was used as a starting condition.
\Cref{fig:refcomp} depicts the temperature distribution for the three meshes, showcasing the mesh convergence and the good resolution both near the shock and the wall. 
An example of the evolution of the computed conductive heat flux for a nonreactive surface is given in \cref{mesh_ref}. 

\begin{table}[!htb]
\footnotesize
\centering
\renewcommand*{\arraystretch}{1.5}
\caption{CFD results for wall conductive heat flux, for three different grids ($\bm{N=200, 400, 600}$).}
\label{mesh_ref}
\begin{tabular}{lccccc}
\toprule
\toprule
%row1
& &  $N=200$  &  $N=400$  &  $N=600$ \\
\midrule
%next row
& $q_{\mr{con}}$ (MW/m$^2$) & $0.556$& $0.626$ & $0.625$\\
% row 3
\bottomrule 
\bottomrule 
\end{tabular}
\end{table}

\section{Surrogate model development}
\label{S:3}

\subsection{Overview of the data-driven model development strategy}
\label{sub:overview}

\begin{figure}[!htb]
\center
\includegraphics[width=1\columnwidth]{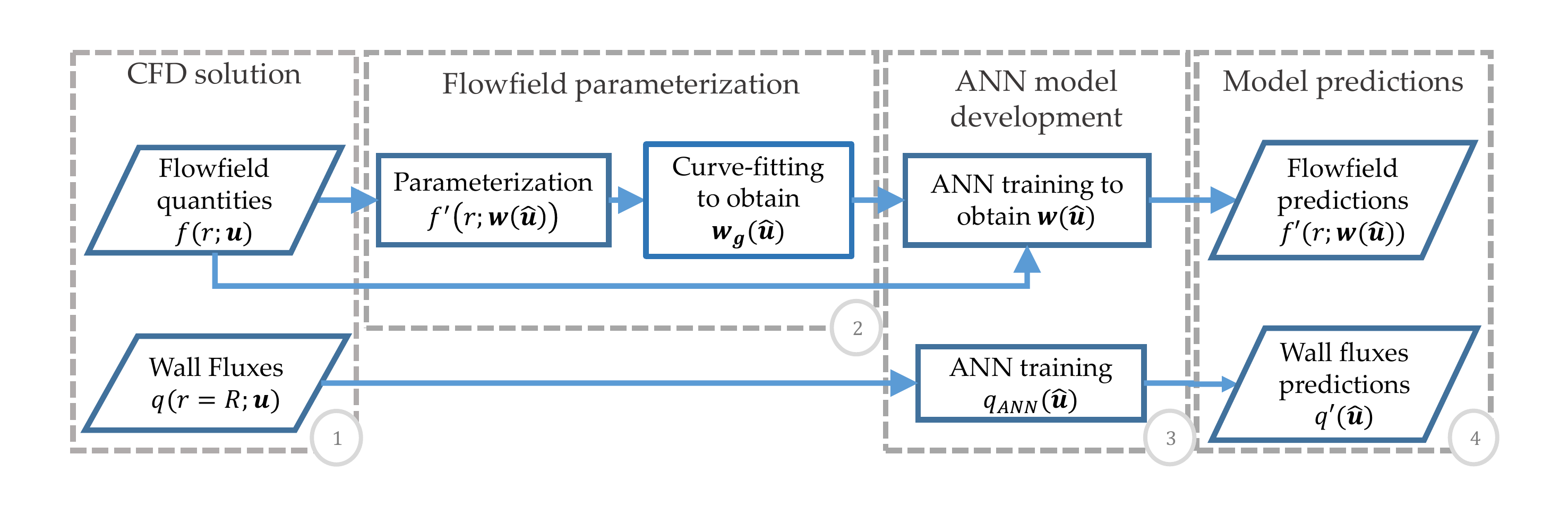}
\caption{Data-driven model flowchart: Different approaches used for flow field quantities along the stagnation streamline and wall heat and mass fluxes.} 
\label{flowchart}
\end{figure}

Figure \ref{flowchart} provides an overview of the model's architecture and the methodology for its derivation. 
As common in the development and application of data-driven models, the flowchart's building blocks include i) the production of a numerical data set through a deterministic numerical tool (a CFD code in the present case); ii) the development of a data-driven model, which in the present case includes first a flow field parametrization and then a more classical training step; and iii) the verification and the application of the model.

Starting from the CFD solutions presented in \Cref{S:2} (block 1 in Fig. \ref{flowchart}), we denote a general flow field variable as a function $f(r;\mathbf{u})$, where we recall $r$ is the distance from the stagnation point along the stagnation line and $\mathbf{u}$ is the set of reentry parameters for each CFD simulation. The wall heat and mass fluxes are denoted as $q(r=R,\mathbf{u})$, with $R$ the object's radius (see \cref{fig:stag}). 

The prediction of the distributions $f(r;\mathbf{u})$, namely the mixture temperature, pressure, density, velocity components and chemical species partial densities, was carried out by first introducing ad hoc parametric functions $f'(r; \bm{w}(\bm{\hat{u}}))$ (block 2 in \cref{flowchart}), where $\bm{w}$ are the parameters controlling the functions. These parameters are linked to the dimensionless formulation of reentry parameters $\mathbf{u}$, here denoted as $\hat{\bm{u}}$. The derivation of these parametric functions and their link to physical features is presented in the next subsection. The training of the model consists in finding the link between the function parameters and the reentry parameters, i.e., $\bm{w}(\hat{\bm{u}})$. This regression was carried out using Artificial Neural Networks (ANN, block 3 in Figure \ref{flowchart}) which are known to be universal function approximators \cite{Goodfellow2016} and are becoming increasingly popular in fluid mechanics (see \cite{Brunton2020,Mendez2022}) for solving regression (e.g., Refs. \cite{Dominique2022,Fiore2022}) or control problems (e.g., Ref. \cite{Pino2022}). The network architecture and the methodology for its training are described in \cref{ANN_subs}. The accuracy of model predictions $f'(r; \bm{w}(\bm{\hat{u}}))$ and $q'(\bm{\hat{u}})$ is then evaluated in \cref{S:4} (block 4 of \cref{flowchart}). The best and worst ANN predictions are investigated, as well as the effect of the dataset size to the results. Moreover, by training an ensemble of ANNs, we also estimate the predictions uncertainty. 

The derivation of wall flux predictions from a combined ANN prediction of temperature, pressure and air composition is feasible, however could potentially lead to error accumulation in the surrogate models. For example, the prediction of the conductive heat flux is highly sensitive to the error made on the temperature gradient at the wall. Thus, those wall fluxes are regressed independently of the flow field variables (lower part of \cref{flowchart}). This approach is selected to reduce the involved error, given also the significance of wall fluxes as design-driving factors in reentry vehicles and satellites.

\subsection{Parametric representation of flow field quantities}
\label{subsPS}

Linear decomposition methods are known to be rather inefficient for transport-dominated problems, providing poor convergence in such conditions (see Refs. \cite{Chauvat, Benn}). The current application corresponds to such a transport-dominated problem, since the detached shock represents a discontinuity of varying smoothness, moving along the input space (depending on the Mach number). However, the hypersonic stagnation streamline flow presents some dominant, physically interpretable features, discussed later in detail. Influenced by these features, we developed an ad-hoc, nonlinear function parametrization, denoted as $f'(r;\bm{w})$ in \cref{flowchart}, to describe the flow quantities of interest.

The parametrization was designed to account for the main features visible in the distributions of each of these quantities: a normal shock, a post-shock peak, a transition region, and a boundary layer.
Denoting as $\bm{w}\in\mathbb{R}^{n_w}$ the set of parameters controlling the prediction of the flow variables of interest, the regression problem becomes that of predicting $\bm{w}$ as a function of $\hat{\bm{u}}$. 
The construction of the parametrization for mixture and species-specific quantities is detailed in the following. 
For both sets, we considered a dimensionless spatial coordinate $\hat{x}=r/R-1$.

\subsubsection{Flow mixture quantities}\label{hfm}

First, the functions $\rho(r)$, $u(r)$, $v(r)$, $p(r)$ and $T(r)$ are scaled with respect to the reference quantities in \cref{refq}. Hence, their dimensionless counterparts are $\hat{\rho}=\rho/\rho_{\infty}$,  $\hat{u}=u/u_{\infty}$, $\hat{v}=v/u_{\infty}$, $\hat{p}=p/(\rho_{\infty }u^2_{\infty})$ and $\hat{\theta}=(T-T_{wall})/(T_{wall}-T_{\infty})$. Then, the regression of these quantities is performed by combining four functions, each describing a certain feature of the flow. The boundary layer near the wall is modelled by an exponential law of the form:
\begin{equation}
\label{bl1}
\Phi_1(\hat{x};\bm{w}_1)=b e^{c \hat{x}}-1 \quad i.e.\quad  \bm{w}_1=[b, c]\,,
\end{equation} 
with the parameter $c$ controlling the boundary layer thickness. The sharp variation due to the shock, located at $\hat{x}=s$, is modelled via a sigmoid function of the form 
\begin{equation}
\label{bl4}
\Phi_2(\hat{x};\mathbf{w}_2)=\frac{1-b}{1+e^{k (\hat{x}-s)}}\quad i.e.\quad  \bm{w}_2=[b, k, s]\,, 
\end{equation} 
while the transition between the two regions is modelled with a second-order polynomial:
\begin{equation}
\label{bl3}
\Phi_3(\hat{x};\mathbf{w}_3)=1+u \hat{x}+p \hat{x}^2 \quad i.e.\quad  \bm{w}_3=[u, p]\,.
\end{equation} 
Finally, the post shock peak is modelled via a Gaussian centered at the shock location $s$, hence
\begin{equation}
\label{bl2}
\Phi_4(\hat{x};\mathbf{w}_3)=1+f \,e^{-g (\hat{x}-s )^2 }\quad i.e.\quad  \bm{w}_3=[f, g, s]\,.
\end{equation}
Note that this sigmoid function can represent the finite but narrow width of the shock (see \cite{Yee1990}). 

\begin{figure}[!tbp]
\center
\begin{center}$
\begin{array}{rr}
\includegraphics[width=72mm]{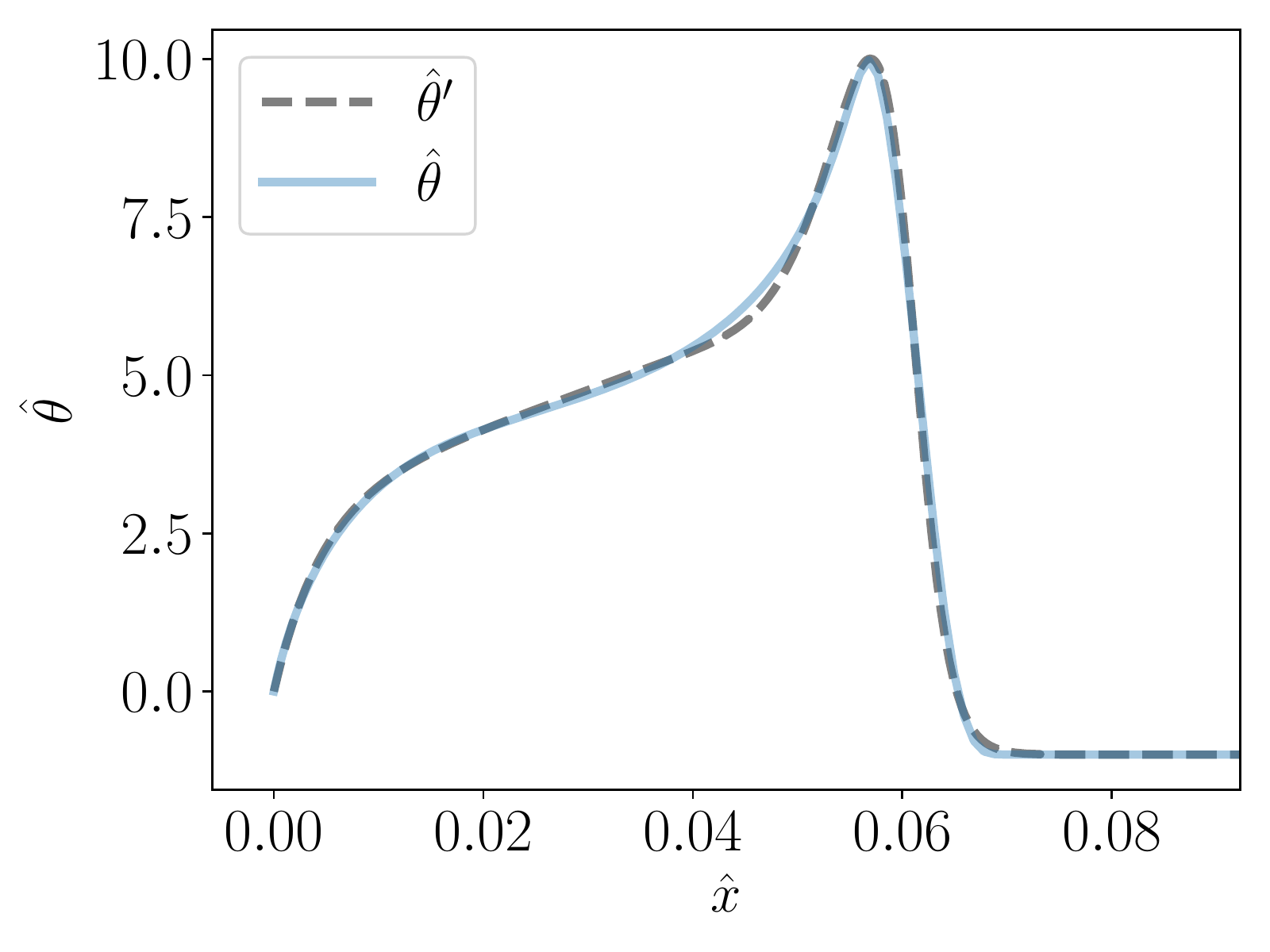}&
\includegraphics[width=72mm]{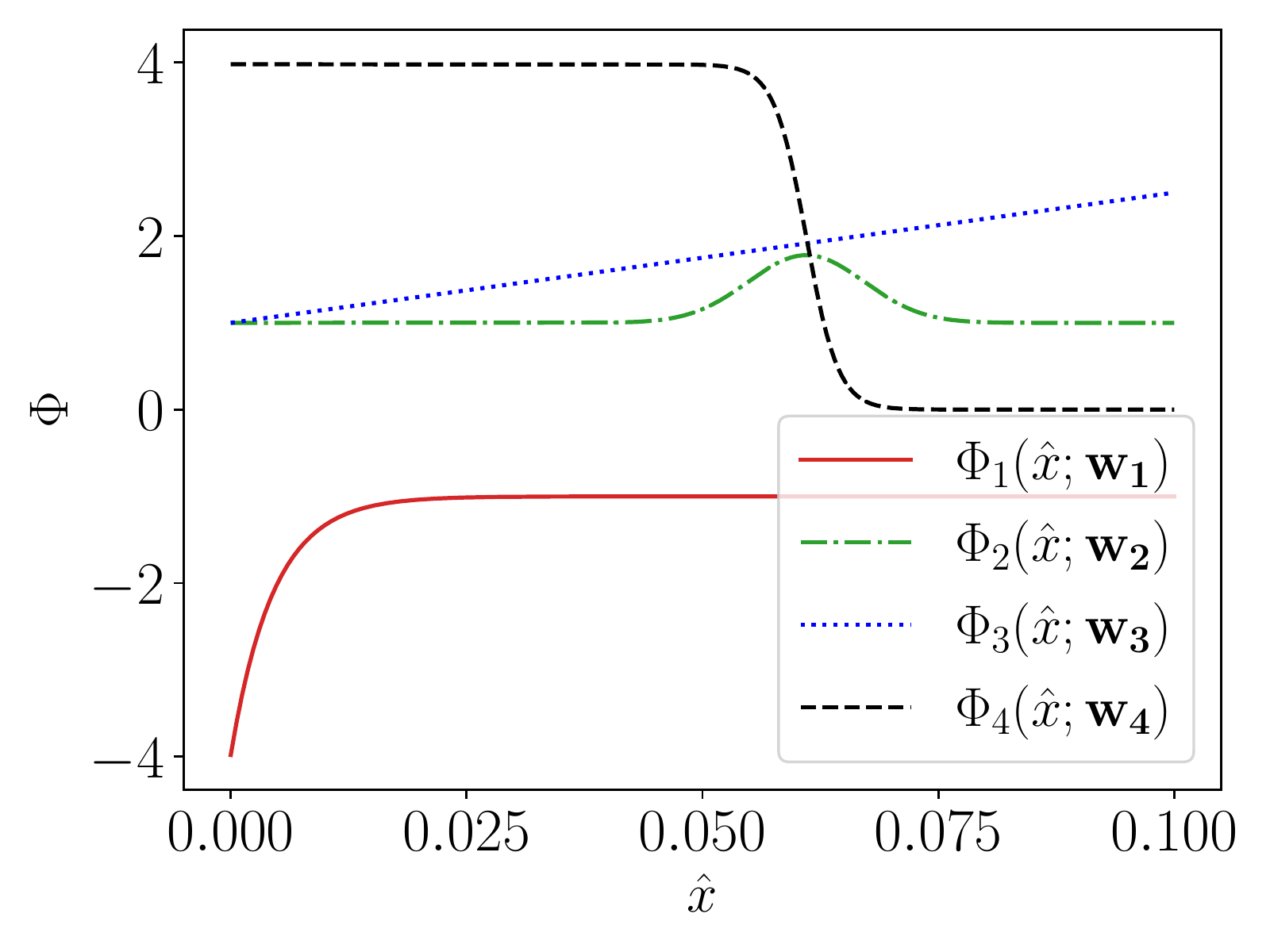}
\end{array}$
\end{center}
\caption{Regression of the dimensionless temperature profile from the conditions in \cref{fig:refcomp} using the parametric function from \cref{phi}. The figure on the left compares the numerical result $\bm{\hat{\theta}}$ with the regression $\bm{\hat{\theta}'}$. The figure on the right shows the four functions $\Phi_1$, $\Phi_2$, $\Phi_3$ and $\Phi_4$ included in the regression from \cref{phi}.}
\label{typical}
\end{figure}

The above four functions (\crefrange{bl1}{bl2}) are combined in various fashions to represent the stagnation-line behavior of the different flow variables. For the dimensionless temperature, for example, we propose 
\begin{equation}
\label{phi}
\hat{\theta}'(\hat{x};\bm{w})=\Phi_1(\hat{x};\bm{w}_1)+\Phi_2(\hat{x};\bm{w}_2)\, \Phi_3(\hat{x};\bm{w}_3)\, \Phi_4(\hat{x};\bm{w}_4) \quad \mbox{with} \quad \bm{w}=[\mathbf{w}_1 \cup \mathbf{w}_2 \cup \mathbf{w}_3 \cup \mathbf{w}_4]\,,
\end{equation} where the prime hereinafter denotes the parametric approximation. 
\Cref{typical} illustrates the regression of a temperature profile using \cref{phi} for the conditions illustrated in \cref{fig:refcomp}. 
Note that the combinations of the the $\Phi$ functions are designed to asymptotically satisfy the prescribed boundary conditions of each parametrized quantity.
For example, in the case of the temperature distribution in \cref{phi} (see \cref{typical}) one has $\hat{\theta}'\rightarrow -1$ for $\hat{x}\rightarrow +\infty$ and $\hat{\theta}'\rightarrow 0$ for $\hat{x}\rightarrow 0$ as long as $k\,s\gg1$ and $g\, s^2 \gg 1$. 
Furthermore, besides enabling an accurate and differentiable representation of the temperature distribution over the full range of reentry parameters, the proposed parametric functions have the main advantage of retaining a phenomenological link to the underlying physics of the problem and thus provide insights into different flow features as a function of the input parameters. These are briefly discussed in Appendix \ref{appex2}.

Similar procedures are followed for the regression of the other variables. Without going into the details of these derivations, the complete set of parametric equations for the quantities of reads:

\begin{equation}
\label{rhohat}
\hat{\rho}'(\hat{x};\mathbf{w}_\rho)=b_{\rho} e^{c_{\rho} \hat{x}}+\frac{a_{\rho} \hat{x}^{2}+d_{\rho}}{1+e^{k_{\rho}\left(\hat{x}-s\right)}}+1,
\end{equation}
with $\mathbf{w}_{\rho}=[a_{\rho},b_{\rho},c_{\rho},d_{\rho}, k_{\rho},s]$;

\begin{equation}
\label{uhat}
\hat{u}'(\hat{x};\mathbf{w}_u)=\frac{a_{u} \hat{x}^{2}+b_{u} \hat{x}-1}{1+e^{k_{u}\left(\hat{x}-s\right)}}+1,
\end{equation}
with $\mathbf{w}_{u}=[a_{u},b_{u}, k_{u},s]$;

\begin{equation}
\label{vhat}
\hat{v}'(\hat{x};\mathbf{w}_v)=\frac{a_{v}\left(1-e^{-b_{v} \hat{x}}\right)+c_{v} \hat{x}+d_{v}}{1+e^{k_{v}(\hat{x}-s)}}+1,
\end{equation}
with $d_v=-(1+b_u/2)$,  $c_v=(d_v-a_v)/{s}$, and $\mathbf{w}_{v}=[a_{v},b_{v}, k_{v},s]$; 

\begin{equation}
\label{phat}
\hat{p}'(\hat{x};\mathbf{w}_p)=\frac{a_{p} \hat{x}^{2}+c_{p}}{1+e^{k_{p}\left(\hat{x}-s\right)}}+\hat{p}_{\infty},
\end{equation}
with $\mathbf{w}_{p}=[a_{p},c_{p}, k_{p},s]$; and

\begin{equation}
\label{temhat}
\hat{\theta}'(\hat{x};\mathbf{w}_\theta)=b_{T} e^{c_{T} \hat{x}}-1+\left(1-b_{T}\right) \frac{1+f_{T} e^{-g_{T}\left(\hat{x}-s\right)^{2}}}{1+e^{k_{T}\left(\hat{x}-s\right)}}\left(1+u_{T} \hat{x}+p_{T} \hat{x}^{2}\right),
\end{equation}
with $\mathbf{w}_{T}=[b_{T}, c_T, f_T, g_T, u_T,p_T, k_{T},s]$. The coefficients are labeled with a subscript since they differ for the different functions (e.g., $b_{\rho} \neq b_T$) expect for the parameter $s$,primarily linked to the shock position and equal in all functions. 

The parametric functions in \crefrange{rhohat}{temhat} respect by construction the physical boundary conditions at the limits of the sigmoid parameters $k\,s\gg1$. 
For the radial velocity the parametric functions give $\hat{u}'\rightarrow 0$ for $\hat{x}\rightarrow 0$ if $k_{u} s\gg 1$, while for the angular component, in line with the continuity equation (see also Appendix \ref{appex1}), one has
\begin{equation}
\label{uv}
\frac{d \hat{u}}{d \hat{x}}\biggr|_{\hat{x}=0}=-2 \hat{v}\biggr|_{\hat{x}=0}\,\,
\end{equation}
if $k_v s\gg 1$. 
Note that, although not physically relevant as velocity, the angular component has been retained in the reduced model as it can be used to compute derived quantities such as the stagnation-line velocity gradient, which is linked to a combination of $u$ and $v$ in the stagnation-line formulation.
Similarly, the pressure function gives $d \hat{p}/d\hat{r}=0$ at $\hat{x}=0$, in line with the momentum equation with the viscous terms neglected, if $k_p s\gg 1$. 
On the free-stream side, that is $\hat{x}\rightarrow +\infty$, the free-stream values of the quantities of interest are also matched by construction, as one has 
$\hat{\rho}'\rightarrow 1$, $\hat{u}'\rightarrow 1$, $\hat{v}'\rightarrow 1$, $\hat{p}'\rightarrow \hat{p}_{\infty}$ and $\hat{\theta}'\rightarrow -1$.

The parametric functions in \crefrange{rhohat}{temhat} depend on a total of 22 parameters $\mathbf{w}=\bm{w}_{\rho} \cup \bm{w}_{u} \cup \bm{w}_{v} \cup \bm{w}_{p} \cup \bm{w}_{T}$. 
From these parameters, these functions allow reconstructing the flow variables along the 1D domain for an arbitrary grid. 
For each of the 2091 nonreactive surface simulations, the best coefficients $\bm{w}$ were identified solving a nonlinear least square regression using the \textsc{cuve\_fit} library from \textsc{scipy} \cite{2020SciPy-NMeth} in \textsc{Python}. To evaluate the quality of the regression, \cref{acctab} shows the average $R^2$ coefficient obtained across the 2091 test cases. The regression is satisfactory for all quantities.

\begin{table}[!htb]
\footnotesize
\centering
\caption{Average $\mb{R^2}$ values for the regression of mixture quantities using the parametric functions in \crefrange{rhohat}{temhat}.}
\label{acctab}
\begin{center}
\begin{tabular}{lccccc}
\toprule
\toprule
%row1
& $\hat{\theta}$ 
&  $\hat{u}$ &  $\hat{v}$  &  $\hat{p}$ &  $\hat{\rho}$  \\
\midrule
%next row
&$99.41$ & $99.40$ & $99.62$ & $97.36$ & $97.13$\\
\bottomrule
\bottomrule
\end{tabular}
\end{center}
\end{table}

\subsubsection{Species partial densities}\label{hypscecies}

For the nonreactive surfaces we also construct parametric functions to predict the partial densities of \ce{N,O,NO,N_2,O_2} along the stagnation streamline. We recall that these five species compose the air mixture under the considered mild velocities, that allow neglecting ionization. 

Within the investigated reentry conditions, the species partial densities span a range of approximately 8 orders of magnitude (see also \cite{Mao}), as reported in \cref{tabscale}.
This challenges the regression and the definition of accuracy in the prediction. 
A first investigated workaround was a transformation of these variables to their common logarithm, on which predictions are made. However, since these variables span a large range within the stagnation streamline, for given reentry conditions, this approach results in a nonlinear mapping that amplifies errors differently in different regions of the domain. 
Therefore, we resort to a heuristic nonlinear scaling which depends on the reentry parameters and which is thus linear within a given simulation. The scaling takes the form 

\begin{equation}
\label{rhoi_nd}
\hat{\rho}_{i}=\frac{{\rho}_{i}-{\rho}_{i_\infty}}{\rho_{\infty}(\hat{u}_j^{a_i}+b_i)}\,,
\end{equation} where $\hat{u}_j$ denotes a selected suitable nondimensional number of those within $\bm{\hat{u}}=[\mi{M,Re,Ec,h_K}]$, and the coefficients $a_i,b_i$ differ for each species and are identified through the regression analysis. \Cref{tabscale} shows the original range for the partial densities of the five species, the chosen $\hat{u}_j$, and obtained variable range following the scaling in \cref{rhoi_nd}. 
Remarkably, the employed transformation leads to a range compression of approximately 3 orders of magnitude.

\begin{table}[tb]
\footnotesize
\centering
\renewcommand*{\arraystretch}{1.5}
\caption{Normalization of the partial densities of the five species in the air mixture. The table shows the range for each variable (in kg/m$^3$), first row, the parameters used for the scaling according to \eqref{rhoi_nd} (second row) and the range for the dimensionless variables after the scaling; this is reduced from 8 to 3 orders of magnitude.}
\label{tabscale}
\begin{center}
\begin{tabular}{lccccc}
\toprule
\toprule
%row1
& $\ce{N}$ &  $\ce{O}$ &  $\ce{NO}$  &  $\ce{N_2}$ &  $\ce{O_2}$  \\
\midrule
%next row
 $\rho_i$ range  & \multirow{1}{*}{$[6.48\times10^{-10},0.16]$} & \multirow{1}{*}{$[3.74\times10^{-6},0.17]$} & \multirow{1}{*}{$[9.03\times10^{-8},0.33]$} & \multirow{1}{*}{$[3.09\times10^{-4},2.62]$} & \multirow{1}{*}{$[3.82\times10^{-5},1.03]$}\\
%next
 $(\hat{u}_j, a_i, b_i)$   & \multirow{1}{*}{$(\mi{M},3,10^{-5})$} & \multirow{1}{*}{$(\mi{M},1,0.1)$} & \multirow{1}{*}{$(\mi{Ec},2,0.01)$} & \multirow{1}{*}{$-$} & \multirow{1}{*}{$-$}\\

% row 3
 \cref{rhoi_nd} range  & \multirow{1}{*}{$[0.65,97.20]$} & \multirow{1}{*}{$[0.41,48.72]$} & \multirow{1}{*}{$[0.10,170]$} & \multirow{1}{*}{$[3.45,169.39]$} & \multirow{1}{*}{$[0.06,92.44]$}\\
\bottomrule
\bottomrule
\end{tabular}
\end{center}
\end{table}

Finally, the partial densities scaled according to \cref{rhoi_nd} are modeled via parametric functions similarly to the mixture quantities in Section \ref{subsPS}.
The proposed parametric forms are:

\begin{equation}
\label{rhoN}
\hat{\rho}'_{\ce{N}}(\hat{x};\mathbf{w}_{\ce{N}})=\left(b_{\ce{N}} e^{c_{\ce{N}} \hat{x}}+ l_{\ce{N}} e^{-m_{\ce{N}}\left(\hat{x}-n_{\ce{N}}\right)^{2}}+d_{\ce{N}}\right) \frac{1+u_{\ce{N}} \hat{x}+p_{\ce{N}} \hat{x}^{2}}{1+e^{k_{\ce{N}}\left(\hat{x}-s\right)}},
\end{equation}
with  $\bm{w}_{\ce{N}}=[b_{\ce{N}},c_{\ce{N}},l_{\ce{N}},m_{\ce{N}},n_{\ce{N}},d_{\ce{N}},u_{\ce{N}},p_{\ce{N}},k_{\ce{N}},s]$;

\begin{equation}
\label{rhoO}
\hat{\rho}'_{\ce{O}}(\hat{x};\mathbf{w}_{\ce{O}})=\left(b_{\ce{O}} e^{c_{\ce{O}} \hat{x}}+ l_{\ce{O}} e^{-m_{\ce{O}}\left(\hat{x}-n_{\ce{O}}\right)^{2}}+d_{\ce{O}}\right) \frac{1+p_{\ce{O}} \hat{x}^{2}}{1+e^{k_{\ce{O}}\left(\hat{x}-s\right)}},
\end{equation} 
with $\bm{w}_{\ce{O}}=[b_{\ce{O}},c_{\ce{O}},l_{\ce{O}},m_{\ce{O}},n_{\ce{O}},d_{\ce{O}},p_{\ce{O}},k_{\ce{O}},s]$;

\begin{equation}
\label{rhoNO}
\hat{\rho}'_{\ce{NO}}(\hat{x};\mathbf{w}_{\ce{NO}})=\left(b_{\ce{NO}} e^{c_{\ce{NO}} \hat{x}}+ l_{\ce{NO}} e^{-m_{\ce{NO}}\left(\hat{x}-n_{\ce{NO}}\right)^{2}}+d_{\ce{NO}}+f_{\ce{NO}} e^{-g_{\ce{NO}}\left(\hat{x}-s\right)^{2}}\right) \frac{1+p_{\ce{NO}} \hat{x}^{2}}{1+e^{k_{\ce{NO}}\left(\hat{x}-s\right)}},
\end{equation} 
with $\bm{w}_{\ce{NO}}=[b_{\ce{NO}},c_{\ce{NO}},l_{\ce{NO}},m_{\ce{NO}},n_{\ce{NO}},d_{\ce{NO}},f_{\ce{NO}},g_{\ce{NO}},p_{\ce{NO}},k_{\ce{NO}},s]$;

\begin{equation}
\label{rhoN2}
\hat{\rho}'_{\ce{N_2}}(\hat{x};\mathbf{w}_{\ce{N_2}})=\left(b_{\ce{N_2}} e^{c_{\ce{N_2}} \hat{x}}+ f_{\ce{N_2}} e^{-g_{\ce{N_2}}\left(\hat{x}-s\right)^{2}}+d_{\ce{N_2}}\right) \frac{1+p_{\ce{N_2}} \hat{x}^{2}}{1+e^{k_{\ce{N_2}}\left(\hat{x}-s\right)}},
\end{equation}
with $\bm{w}_{\ce{N_2}}=[b_{\ce{N_2}},c_{\ce{N_2}},f_{\ce{N_2}},g_{\ce{N_2}},d_{\ce{N_2}},p_{\ce{N_2}},k_{\ce{N_2}},s]$; and

\begin{equation}
\label{rhoO2}
\hat{\rho}'_{\ce{O_2}}(\hat{x};\mathbf{w}_{\ce{O_2}})=\left(b_{\ce{O_2}} e^{c_{\ce{O_2}} \hat{x}}+f_{\ce{O_2}} e^{-g_{\ce{O_2}}\left(\hat{x}-s\right)^{2}}+d_{\ce{O_2}}\right) \frac{1+u_{\ce{O_2}} \hat{x}}{1+e^{k_{\ce{O_2}}\left(\hat{x}-s\right)}},
\end{equation}
with $\bm{w}_{\ce{O_2}}=[b_{\ce{O_2}},c_{\ce{O_2}},f_{\ce{O_2}},g_{\ce{O_2}},d_{\ce{O_2}},u_{\ce{O_2}},k_{\ce{O_2}},s]$.

All functions lead to $\hat{\rho}'_i=0$ for $\hat{x}\rightarrow \infty$, hence recovering the imposed mixture density in the free stream.
The parametrization of the partial densities is more challenging than for the mixture quantities requiring a total of 42 parameters in $\mathbf{w}_{\rho_i}=\bm{w}_{\ce{N}}\cup\bm{w}_{\ce{O}}\cup\bm{w}_{\ce{NO}}\cup\bm{w}_{\ce{N}_2}\cup \bm{w}_{\ce{O_2}}$. Nevertheless, the accuracy of the regression, carried out using the same numerical tool as in the previous section, yields a satisfactory accuracy for all the species (see \cref{tabRho}).

\begin{table}[!htb]
\footnotesize
\centering
\caption{Same as Table \ref{acctab}, but considering the average $R^2$ values for the regression of the species partial densities according to the parametric functions in \eqref{rhoN}-\eqref{rhoO2} for the 2091 investigated test cases.}
\label{tabRho}
\begin{center}
\begin{tabular}{lccccc}
\toprule
\toprule
& $\hat{\rho}_{\ce{N}}$ &  $\hat{\rho}_{\ce{O}}$ &  $\hat{\rho}_{\ce{NO}}$  &  $\hat{\rho}_{\ce{N_2}}$ &  $\hat{\rho}_{\ce{O_2}}$  \\
\midrule
%next row
& $99.34$ & $99.31$ & $99.29$ & $96.78$ & $98.02$\\
% row 3
\bottomrule
\bottomrule
\end{tabular}
\end{center}
\end{table}

\subsection{Artificial Neural Networks for flow field predictions}
\label{ANN_subs}

Two artificial neural networks (ANNs) were designed and trained to predict the parameters of the flow mixture quantities (network N1 for \crefrange{rhohat}{temhat}) and those of the species partial densities (network N2 for \crefrange{rhoN}{rhoO2}). 
These predictions are based on the dimensionless input $\hat{\mathbf{u}}$ and the learned functions are thus of the kind $f\!\!:\hat{\mathbf{u}}\in \mathbb{R}^{4}\rightarrow \mathbf{w}\in\mathbb{R}^{n_w}$; with $n_w=22$ for the mixture quantities
and $n_w=42$ for the species partial densities.
This is a classic supervised learning problem, with the labeled data on the weights taken from the least square regression of the quantities profiles $f'(\hat{x};\bm{w})$ described earlier.
Both networks are fully connected feed-forward nets constructed in the \textsc{Python}'s API \textsc{tensorflow} \cite{tensorflow2015}. 

\begin{figure}[!htb]
\center
\includegraphics[width=0.5\columnwidth]{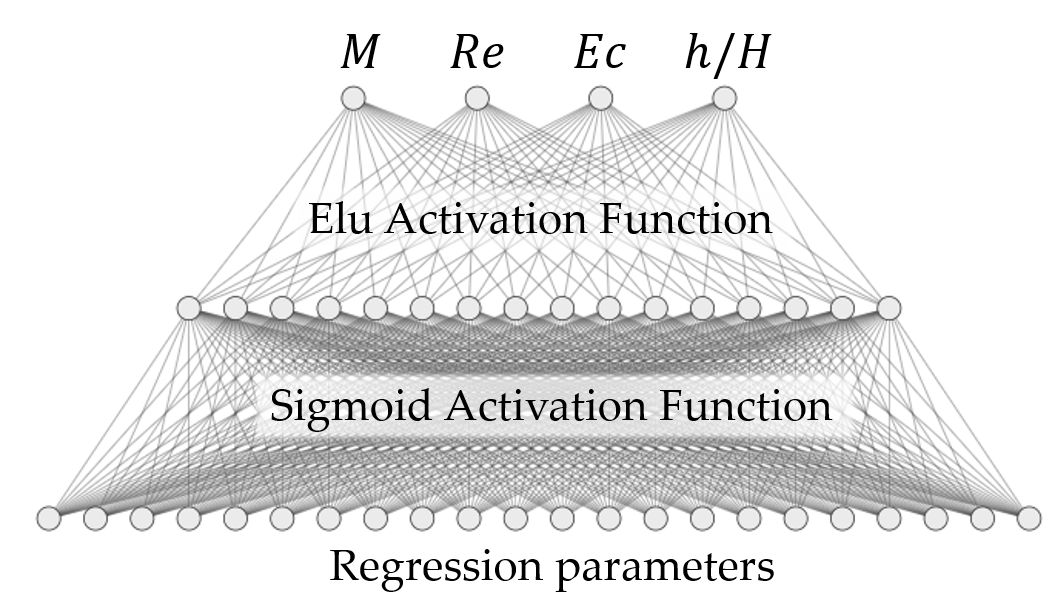}
\caption{Schematic representation of ANN structure: Functions $\bm{w}(\bm{\hat{u}})$ are obtained by training the ANNs.}
\label{fig:anns}
\end{figure}

The architecture for the network N1 is illustrated in \cref{fig:anns}. 
The reader is referred \cite{Mendez2022} and \cite{Dominique2022} for a concise review of the ANN terminology and to \cite{Goodfellow2016} for a more detailed introduction.
Activation functions are exponential linear unit (elu) in the first layer and sigmoid in the second. 
This allows for smooth and bounded predictions while facilitating the backpropagation of the gradient during the training. 
Similar architectures have been investigated in \cite{Dominique2022} and \cite{Fiore2022}. 
The second layer (or hidden layer, in the machine learning terminology) consists of 17 neurons. Both inputs and outputs of the network are normalized in the range [0,1] using a min-max scaling and the training is performed using a classic ADAM \cite{Kingma2014} optimizer with mini-batches. 
The main parameters for the two networks architecture are listed in \cref{anncoef}, along with the batch size used during training. Having parameterized the flow quantity profiles as shown in the previous subsections, two "shallow" neural network architectures proved sufficient for the prediction task.
The learning rate was set to 0.001, with a reduction factor of 0.7 in case the loss function does not reduce more than 0.01\% within 200 epochs.

\begin{table}[bth]
\renewcommand*{\arraystretch}{1.5}
\centering
\footnotesize
\caption{Flow field quantities ANNs architecture and hyper-parameters selection.}
\label{anncoef}
\begin{center}
\begin{tabular}{lcccc}
\toprule \toprule
%row1
&  Nodes (input/middle/output) 
&  Activation functions  &  Batch size &  No. of epochs  \\
\midrule
%next row
 Mixture quantities (N1)  & $4/17/22$ & linear / elu / sigmoid & $100$ & $5000$\\
% row 3
\midrule
% row5
 Species densities (N2)  & $4/28/44/42$&linear / relu / elu / sigmoid & $60$ & $20000$\\
% row 3
\bottomrule
\bottomrule
\end{tabular}
\end{center}
\end{table}

To introduce the loss function used in the training of these two networks, let us denote as $(\hat{\bm{u}}_k,\mathbf{w}_{k})$ the set of $k=1,\dots n_p$ training points, with the weights $\mathbf{w}_{k}$ obtained during the nonlinear least square regression of the CFD simulations. Let $\mathbf{w}'(\hat{\bm{u}}_k;\bm{\theta})$ denote the prediction of a network with parameters $\bm{\theta}$ and $f_j'(r; \mathbf{w}'(\hat{\bm{u}}_k;\bm{\theta}))$ be the corresponding prediction of the mixture flow quantities $j=1,\dots n_j$ according to \crefrange{rhohat}{temhat} (hence $j=1,2,3,4,5$ for $\rho$, $u$, $v$, $p$, $T$) or the species partial densities according to \crefrange{rhoN}{rhoO2} (hence $j=1,2,3,4,5$ for $\hat{\rho}_{\ce{N}}$, $\hat{\rho}_{\ce{O}}$, $\hat{\rho}_{\ce{NO}}$, $\hat{\rho}_{\ce{N_2}}$, $\hat{\rho}_{\ce{O_2}}$). Finally, let $f_j(r; \hat{\bm{u}}_k)$ be the CFD prediction of the quantity $j$ for the input vector (i.e., flight condition) $\hat{\bm{u}}_k$.

The above quantities allow defining various contributions to the loss function, which reads 

\begin{equation}
J(\bm{\theta})=\sum_{k=1}^{n_p}\left( ||\bm{w}_k-\mathbf{w}'(\hat{\bm{u}}_k;\bm{\theta})||^2_2 + \sum^{n_r}_{i=1} \sum^{n_j}_{j=1} g_j\bigl (f_j(r_i;\bm{\hat{u}}_k)-f_j'(r_i; \bm{w}'(\hat{\bm{u}}_k;\bm{\theta}))\bigr )^2 + 
{g_c}\sum^{n_r}_{i=1}{\mathcal{R}_c (r_i;\mathbf{w}'(\hat{\bm{u}}_k;\bm{\theta}))}^2 \right)^2
\label{loss}
\end{equation} 
where $||\bullet||_2$ denotes the $l_2$ norm, $g_j$ and $g_c$ are weights given to the second and third terms, with the first possibly varying for the $j$-th quantity. The first term aims at promoting $\bm{w}_k \approx\mathbf{w}'(\hat{\bm{u}}_k;\bm{\theta})$, thus, the ANN predictions are close to the ones identified by the nonlinear least squares problem (see the $R^2$ in \cref{tabin} and \cref{tabRho}). The second term promotes $f_j'(r; \mathbf{w}'(\hat{\bm{u}}_k;\bm{\theta}))\approx f_j(r; \hat{\bm{u}}_k)$. This serves as a regularization, suggesting that the ANN predictions are close to the CFD predictions, while accounting for the sensitivities $\partial_{\bm{w}} f_j$ of the parametrization.

Finally, the last term can be used to promote solutions that respect as much as possible a differential constraint. This was used for the network N1 (mixture quantities) to enforce compliance to the continuity equation. Leveraging the differentiability of the parametric functions, these can be introduced in the continuity equation to derive the residual: 

\begin{equation}
\frac{d (\rho u)}{d r}+\frac{2}{r} \rho(u+v)=0 \rightarrow \frac{d }{d r} \biggl(f'_1(r_i;\bm{w}') f'_2(r_i;\bm{w}') \biggr)+\frac{2}{r_i} f'_1(r_i;\bm{w}')(f'_2(r_i;\bm{w}')+f'_3(r_i;\bm{w}'))=\mathcal{R}_c(r_i;\bm{w}'),
\end{equation} where $r_i$ with $i=1,\dots n_r$ is a grid along the $r$ coordinate.

This residual term can be evaluated analytically from \crefrange{rhohat}{temhat}, but the resulting function is cumbersome and hence not shown.
This approach is closely related to the physics-informed neural network (PINNs) context \cite{Raissi2019}, with the critical difference being that the PDE residuals are defined on the function parametrization and not on the ANN directly. This makes the ANN independent from the computational domain in which it is trained, thus allowing for better generalization. This work considers only the residual on the mixture continuity equation. Adding the residual for other conservation equations involved requires introducing the thermodynamic properties of the chemical species into the N1 structure; such an extension is not pursued in this work.

Finally, for all ANNs implemented in this work, we use an ensemble learning approach (see Refs. \cite{Abdar2021,Gawlikowski2021} to estimate the prediction uncertainties (that is those linked to uncertain determination of the ANN weights and biases). Besides providing confidence intervals, the uncertainties can be used to unveil regions of the input space lacking training data. In the ensemble formalism, an ANN is trained $n_E$ times to produce a population of $n_E$ possible network realizations. Each realization is trained using $70\%$ of the available data (randomly sampled) while the remaining $30\%$ (consisting of $n_o$ samples) is used for validation. For every input 
$\hat{\bm{u}}_k$, given $\mathbf{w}'(\hat{\bm{u}}_k;\bm{\theta}_m)$ the prediction of the m-th network, the ensemble prediction is taken as $\mu_{w}(\hat{\bm{u}}_k)=\mathbb{E}_{\sim n_E}\{\mathbf{w}'(\hat{\bm{u}}_k;\bm{\theta}_m)\}$, where $\mathbb{E}_{\sim n_E}$ denotes the expectation operator over the ensemble $m=1,\dots n_E$. Therefore, for a given input $\hat{\bm{u}}_k$ and at a given position $\hat{r}$ after the shock, the ensemble prediction is $f_j(r;\mu_{w}(\hat{\bm{u}}_k))$ and the variance \emph{within} the ensemble for a given input is:

\begin{equation}
    \label{term_1}
    \sigma_1^2(\hat{r}_i,\hat{\bm{u}}_k)=\frac{1}{ n_E}\sum^{n_E}_{m=1} \bigl(f_j(\hat{r}_i;\mu_{w}(\hat{\bm{u}}_k))-f_j(\hat{r}_i;\mathbf{w}'(\hat{\bm{u}}_k;\bm{\theta}_m)\bigr)^2\,.
\end{equation}

A second contribution to the uncertainty comes from the out-of-sample root mean square error of the ANN function for a given normalized position $\hat{r}$ after the shock. That is, the average of the bias error \emph{across} the ensemble with respect to the ground truth $f(\hat{r}_i;\hat{\bm{u}}_k)$: 

\begin{equation}
    \label{term_2}
    \sigma_2^2(\hat{r}_i)=\frac{1}{n_o n_E}\sum^{n_o}_{k=1}\sum^{n_E}_{m=1} \bigl( f(\hat{r}_i;\hat{\bm{u}}_k)-f(\hat{r}_i;\mathbf{w}'(\hat{\bm{u}}_k;\bm{\theta}_m)\bigr)^2 \,.
\end{equation} 

The uncertainty formulation is valid for positions $\hat{r}$ past the normal shock, since the introduced parameterization is by definition exact at the free stream. Assuming at a first approximation that the distribution of the ensemble population is Gaussian, the variance in the prediction is $\sigma^2(\hat{r}_i,\hat{\bm{u}}_k)=\sigma^2_1(\hat{r}_i)+\sigma^2_2(\hat{r}_i,\hat{\bm{u}}_k)$ and a $95\%$ confidence interval around the ensemble mean prediction can be computed as $f_j(\hat{r};\mu_{w}(\hat{\bm{u}}_k))\pm 1.96\sigma(\hat{r}_i,\hat{\bm{u}}_k)$.

\subsection{Point-wise predictions for wall fluxes}
\label{wallflux}

As described in Section \ref{CFD_na}, pointwise predictions of wall fluxes were considered of interest for both nonreactive and ablative surfaces. These quantities are critical for reentry analysis and the proposed surrogate models offer promising alternatives to semiempirical relations to compute wall fluxes on-the-fly during trajectory simulations.
These quantities could be derived from the flow-field solution obtained from the N1 and N2. However, this approach poses several challenges, especially for the ablative test cases, since the N2 network should be extended to include additional species in the mixture. To provide a first proof of concept for the regression based approach, two ANNs -- N3 and N4 -- were developed to predict the heat fluxes at the wall. For the case of nonreactive surfaces, the network N3 is designed to predict the conductive heat flux. For the carbon-based ablative surfaces, the network N4 is designed to predict the three heat flux components on the left hand side of \cref{eq:bound1} along with the mass blowing flux at the wall. 

The employed methodology is outlined in \cref{flowchart}. In essence, we predict an $R^4$ surface for the wall quantities of interest, using fully-connected feed-forward ANNs. 
The relevant wall fluxes are directly obtained from the CFD solution for each input $\hat{\bm{u}}$ in both the nonablative and the carbon ablation regime. 
Then, a logarithmic scaling is employed to cope with the many orders of magnitude spanned by these fluxes over the input-data range provided in \cref{tabin}. As in the case of flow-field quantities, the network inputs and outputs are normalized in the range [0,1], while the training is performed using the ADAM optimizer and a mean square error loss function. The details of the network architectures for N3 and N4 are given in \cref{anncoef2}. A learning rate reduction is employed, with a factor of 0.7 for a 0.01\% threshold loss function plateau after 200 epochs. 

Finally, the same cross-validation strategy and ensemble learning approach used for N1 and N2, as described in Section \ref{ANN_subs}, is used for the networks N3 and N4. In particular, the formulation of \cref{term_1,term_2} can be transferred to the uncertainty computation for the logarithms of pointwise fluxes (heat and mass) at the wall, simply by fixing $\hat{r}=0$, so that $\sigma(\hat{r}_i=0,\hat{\bm{u}}_k)=\sigma(\hat{\bm{u}}_k)$. The computation of the uncertainties for the actual flux values, given in the application of Section \ref{S:5}, requires a transformation of the computed uncertainties to a log-normal distribution \cite{Olsson2005}.

\begin{table}[bth]
\renewcommand*{\arraystretch}{1.5}
\centering
\footnotesize
\caption{Wall flux ANNs architecture and hyperparameters selection}
\label{anncoef2}
\begin{center}
\begin{tabular}{lcccc}
\toprule \toprule
%row1
&  Nodes (input/middle/output) 
&  Activation functions  &  Batch size &  No. of epochs  \\
\midrule
%next row
Non-reactive surfaces (N3) & $4/16/1$ & linear / elu / sigmoid & $60$ & $5000$\\
% row 3
\midrule
% row5
Ablative surfaces (N4) & $4/35/4$ &  elu / elu / sigmoid & $60$ & $8000$\\
% row 3
\bottomrule
\bottomrule
\end{tabular}
\end{center}
\end{table}

\section{Model Validation}
\label{S:4}

\subsection{Mixture quantities along the stagnation streamline (ANN N1)}

We here report on the results of the ensemble training of the N1 networks (aiming at predicting temperature, pressure, density and velocity components along the stagnation streamline) using an ensemble of $n_E=20$ realizations, trained with 
$70\%$ of the data. \Cref{res1} collects the average $R^2$ coefficient on the validation set, for all flow field quantities, for the best and the worst of the 20 \textsc{N1} realizations. The loss function value $J$ is also reported. It can be seen that the prediction performances are accurate for all quantities.

\begin{table}[!htb]
\renewcommand*{\arraystretch}{1.5}
\centering
\footnotesize

\caption{Average $\mb{R^2}$ values for mixture quantities predictions, for the best/worst N1 realization: $\mb{R^2}$ trends follow those of the loss value, but predictions remain consistently accurate.}
\label{res1}

\begin{center}
\begin{tabular}{lcccccc}
\toprule \toprule
%row1
& $J$ 
& $R^2_{\hat{\theta}}$ 
&  $R^2_{\hat{u}}$ &  $R^2_{\hat{v}}$  &  $R^2_{\hat{p}}$ &  $R^2_{\hat{\rho}}$  \\
\midrule
%next row
 $min(J)$ ANN  & $0.041$ & $95.16$& $98.15$ & $98.07$ & $95.68$ & $95.70$\\
% row 3
 $max(J)$ ANN  & $0.066$ & $93.93$ & $97.23$ & $97.26$ & $94.38$ & $95.04$\\
% row 3
\bottomrule
\bottomrule
\end{tabular}
\end{center}
\end{table}

\begin{figure}[!htb]
\begin{center}$
\begin{array}{ll}
\includegraphics[width=115mm]{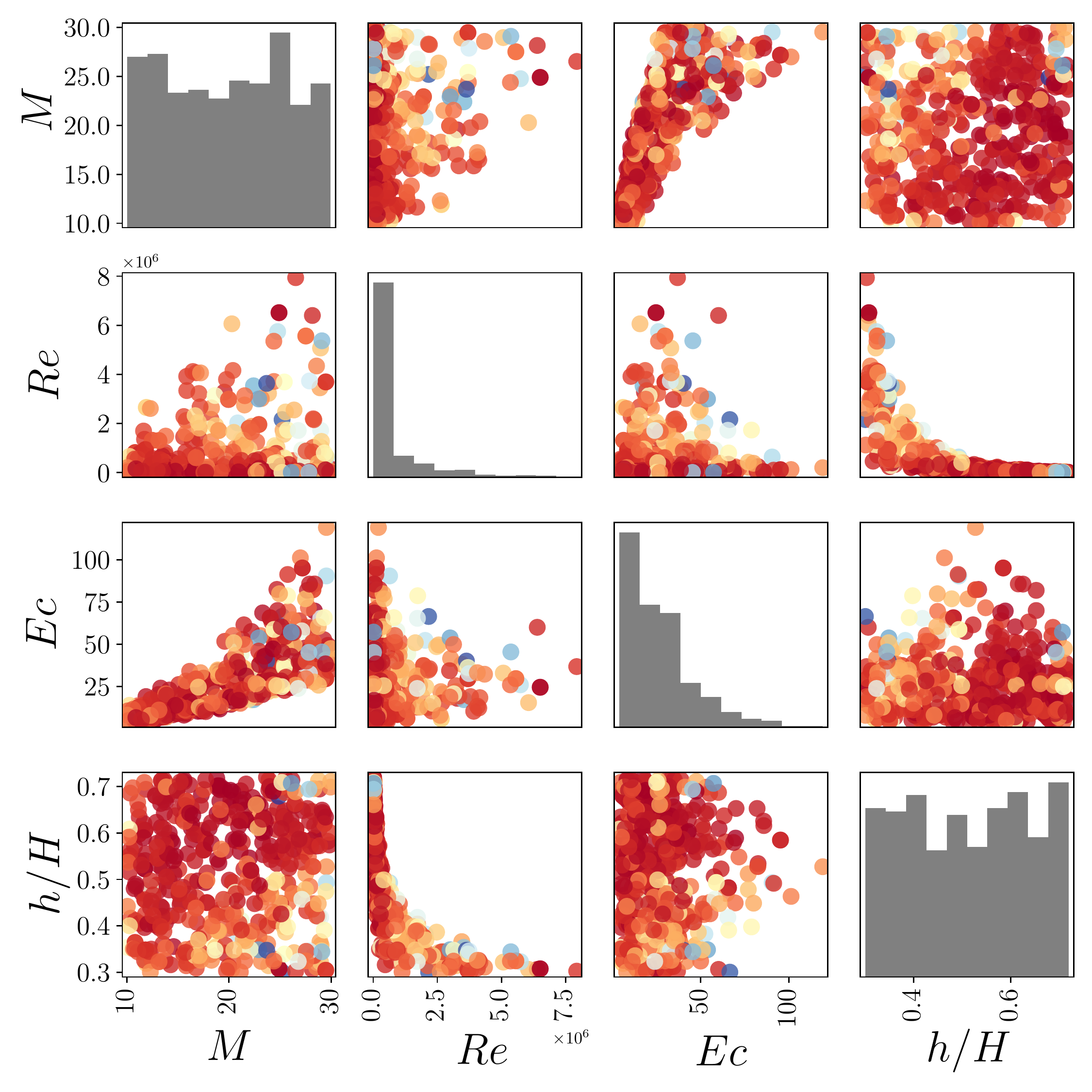}&
\includegraphics[width=19mm]{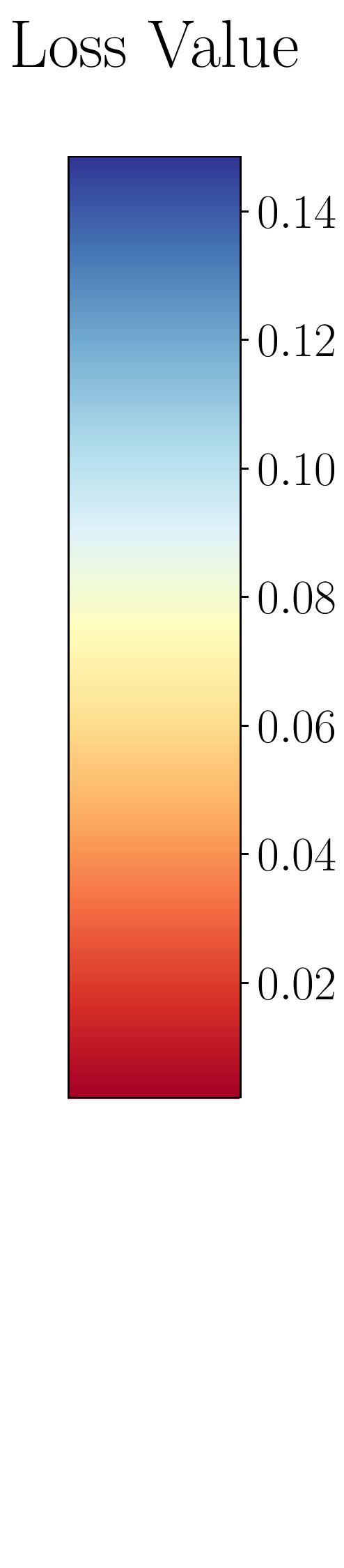}
\end{array}$
\end{center}
\caption{Scatter matrix of individual N1 validation test cases (mixture quantities). Relatively lower accuracy for high reentry velocity values.}
\label{histnab}
\end{figure}

We now focus on the best N1 realization (the one leading to the smallest loss $J$). Figure \ref{histnab} shows the scatter matrix with the loss value distribution in all the investigated conditions and as a function of all input parameters. It is evident that the training data, in the dimensionless representation, is not uniformly distributed and certain regions of the input space are entirely unexplored. However, considering the realistic range taken for the input parameters in dimensionless form (see Table \ref{tabin}), the poorly sampled area corresponds to regions not observed during reentry and in which it is not worth training the model. 

\begin{figure}[!htbp]
\begin{center}$
\begin{array}{lll}
\includegraphics[width=56mm]{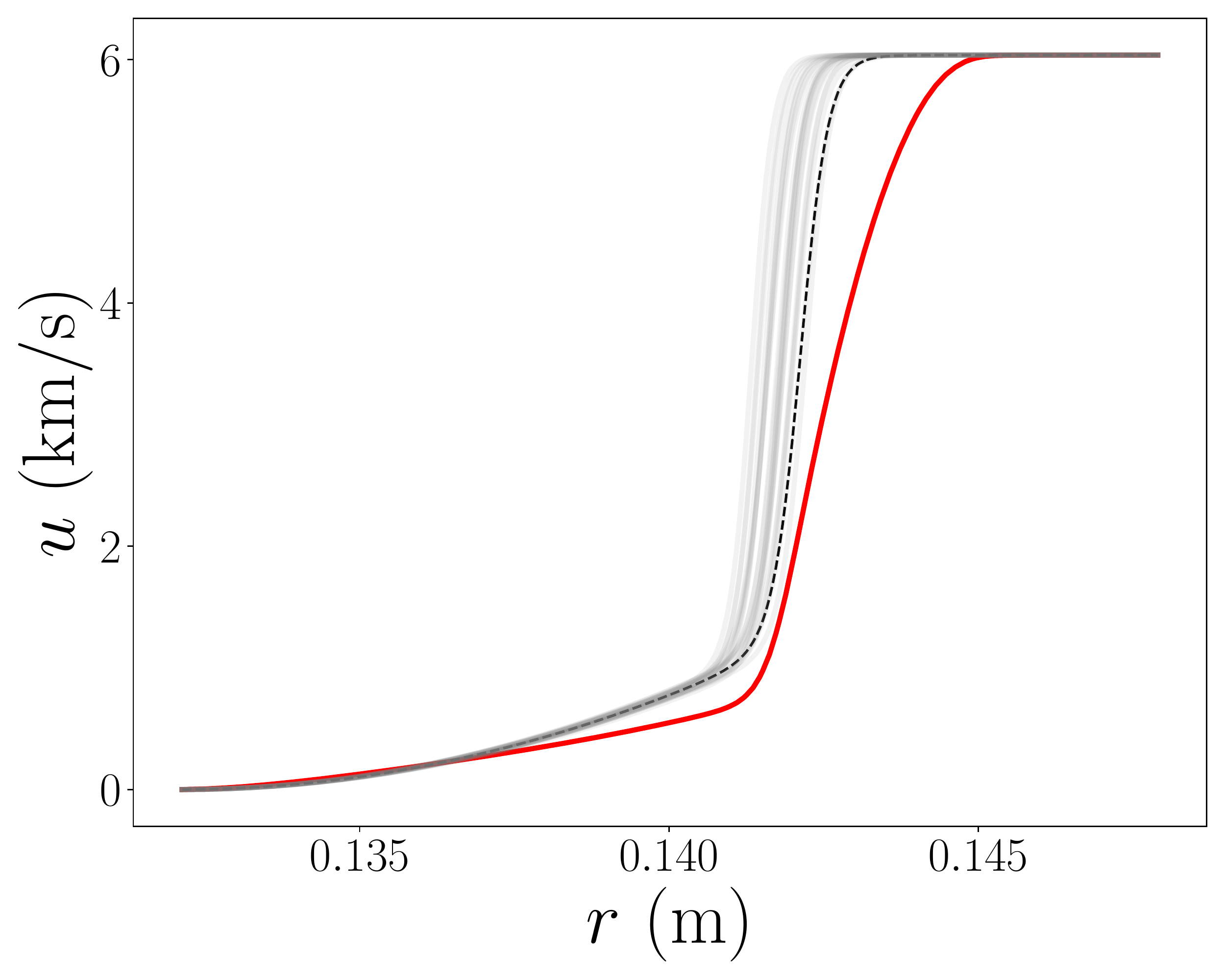}&
\includegraphics[width=56mm]{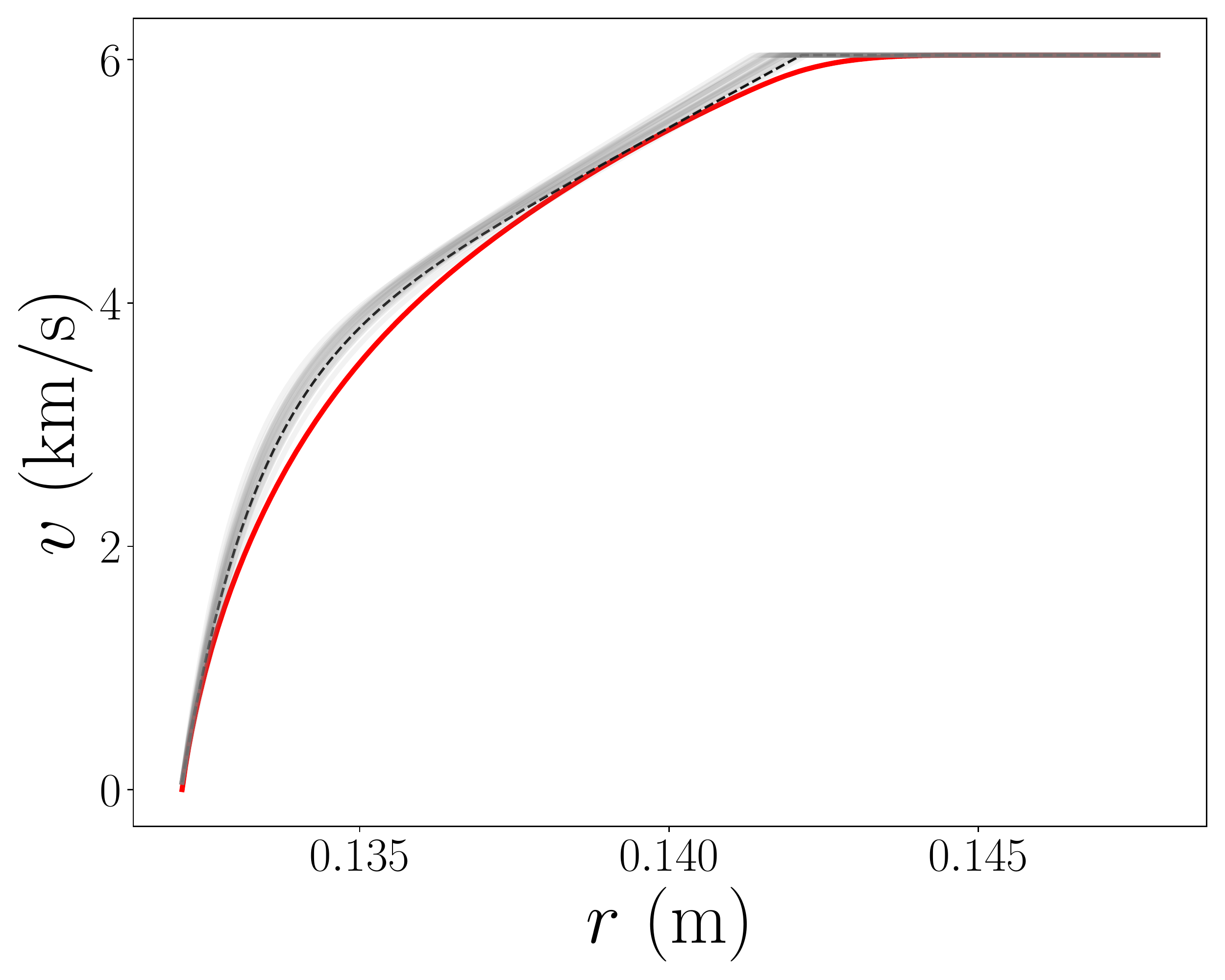}&
\includegraphics[width=56mm]{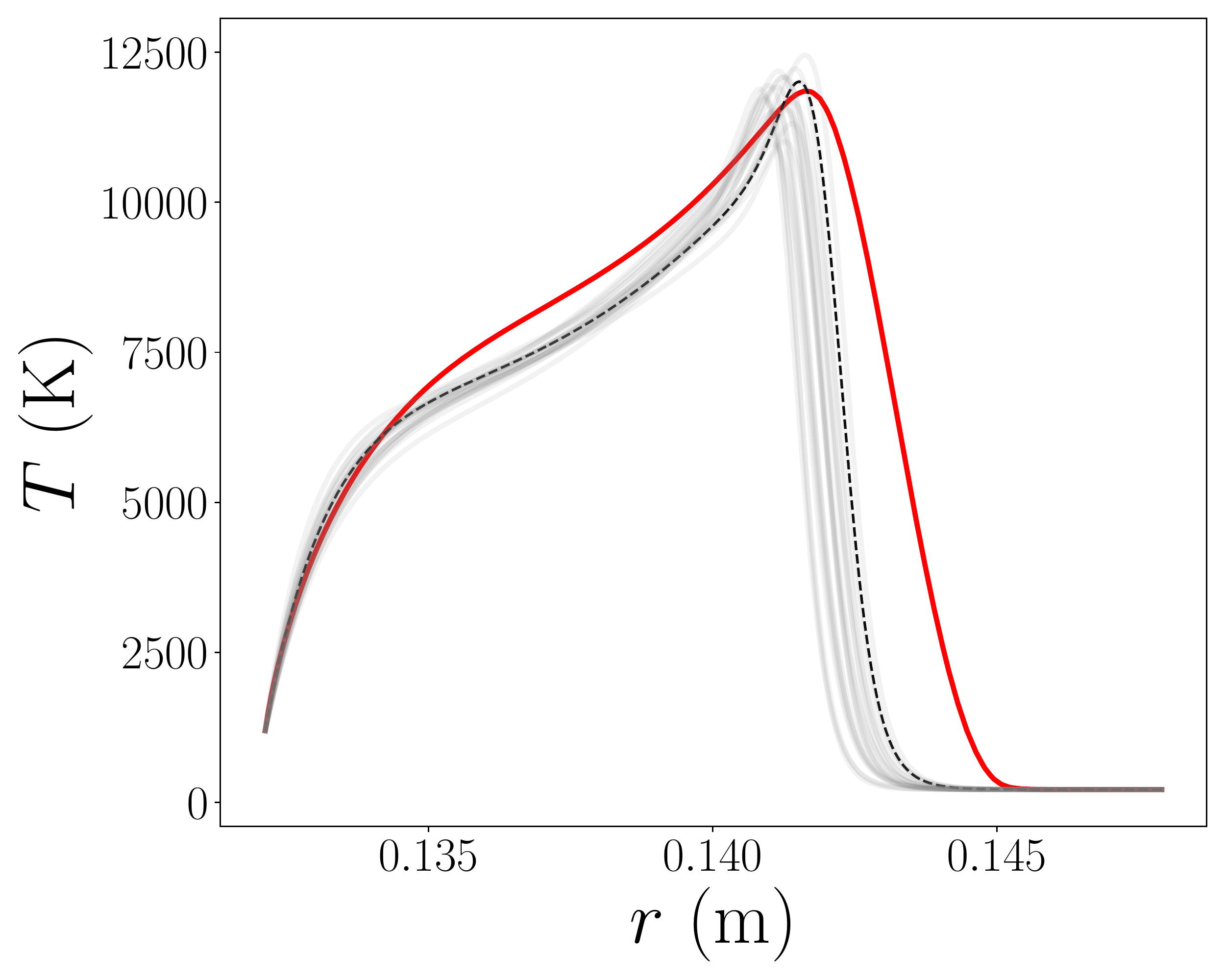}
\end{array}$
\end{center}

\begin{center}$
\begin{array}{rr}
\includegraphics[width=56mm]{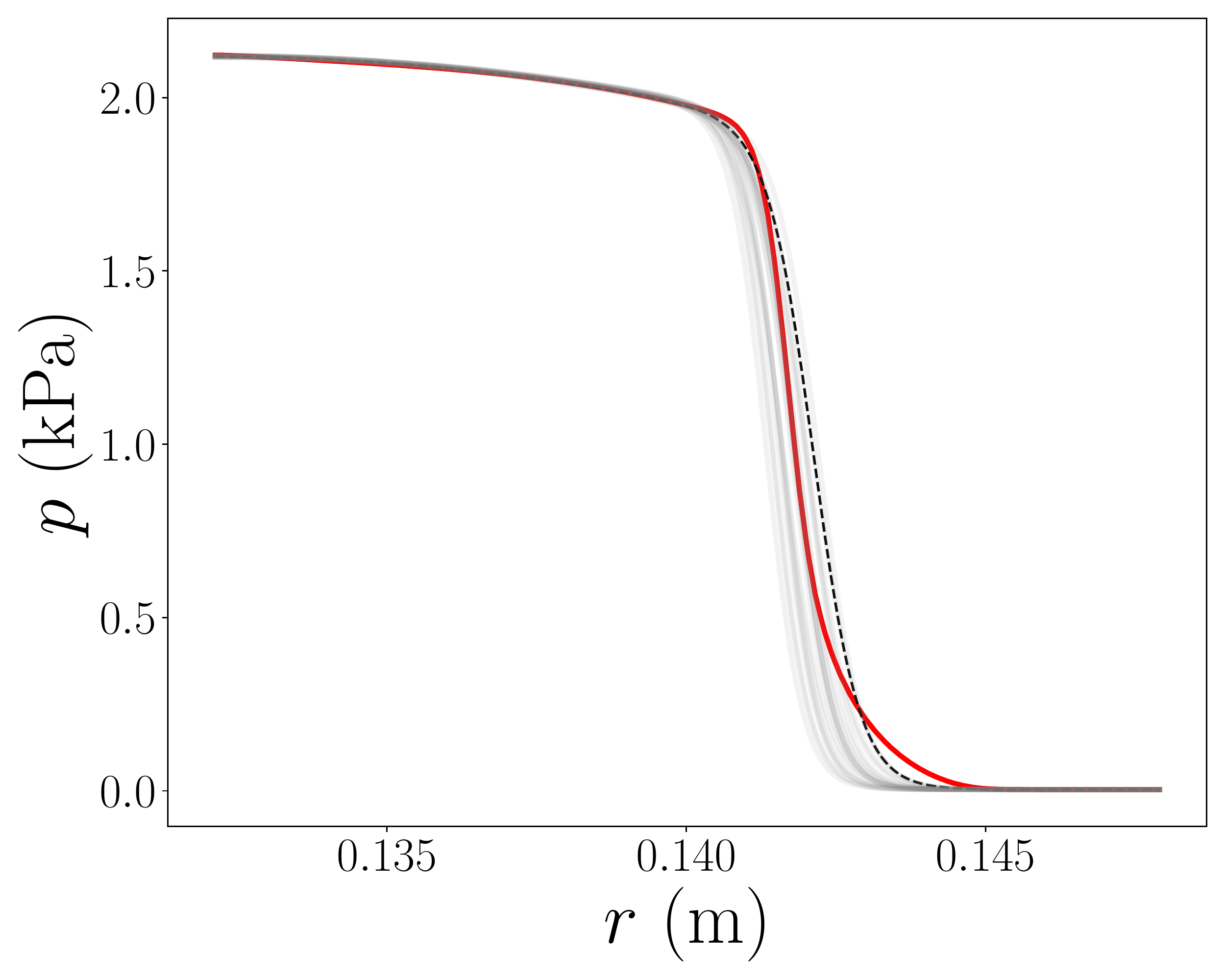}&
\includegraphics[width=56mm]{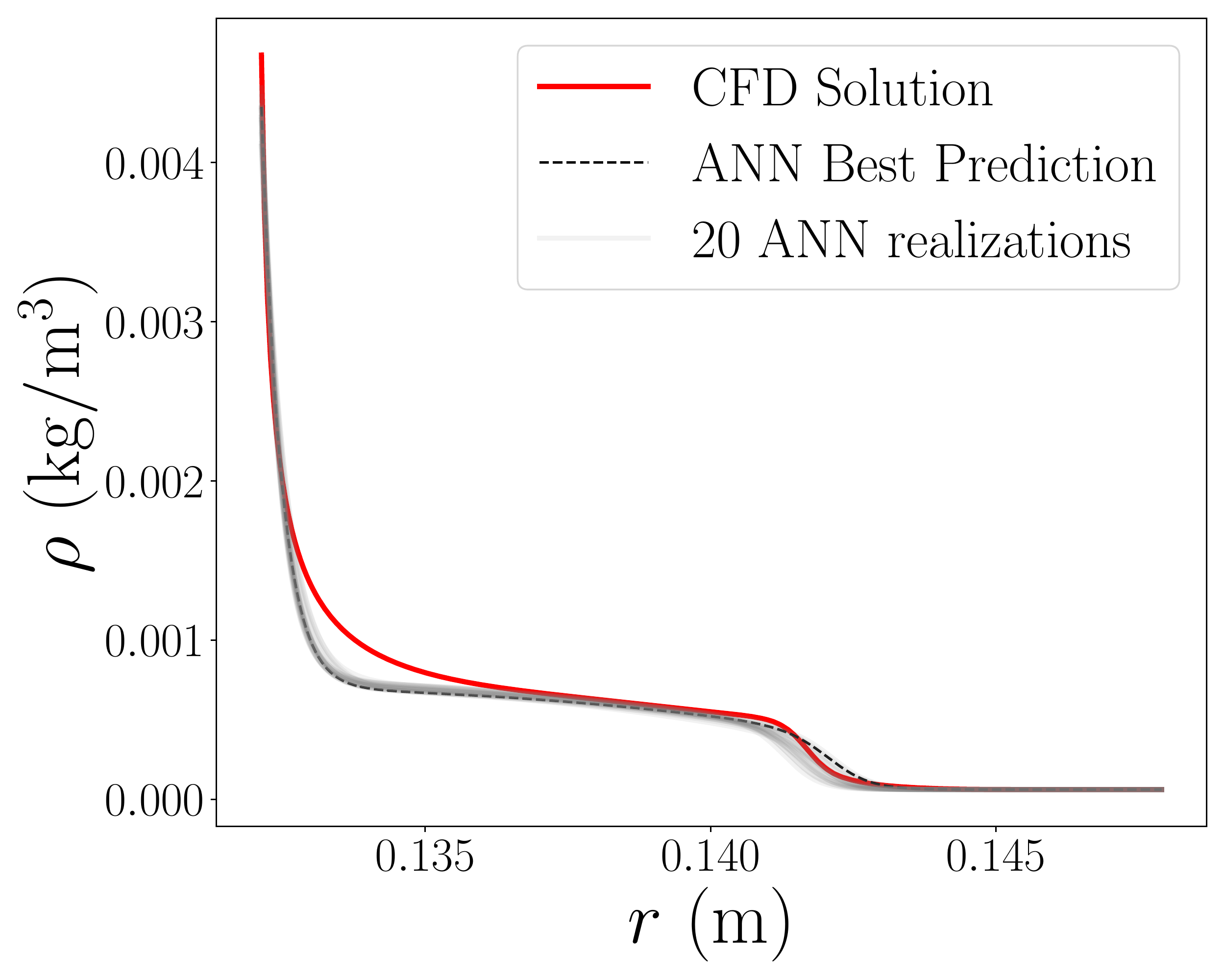}
\end{array}$
\end{center}

\caption{N1 predictions for $\bm{max(J)}$ validation test case $\mathbf{u}=[68.9  \; \rm{km}, \; 6.04  \; \rm{km/s}, \; 0.132  \; \rm{m}, \; 1194  \; \rm{K}]^T$: Mismatch in shock position prediction, high deviation among ANN realizations (testcase in the extremity of input space).}
\label{worst}
\end{figure}

\begin{figure}[!htbp]
\begin{center}$
\begin{array}{lll}
\includegraphics[width=56mm]{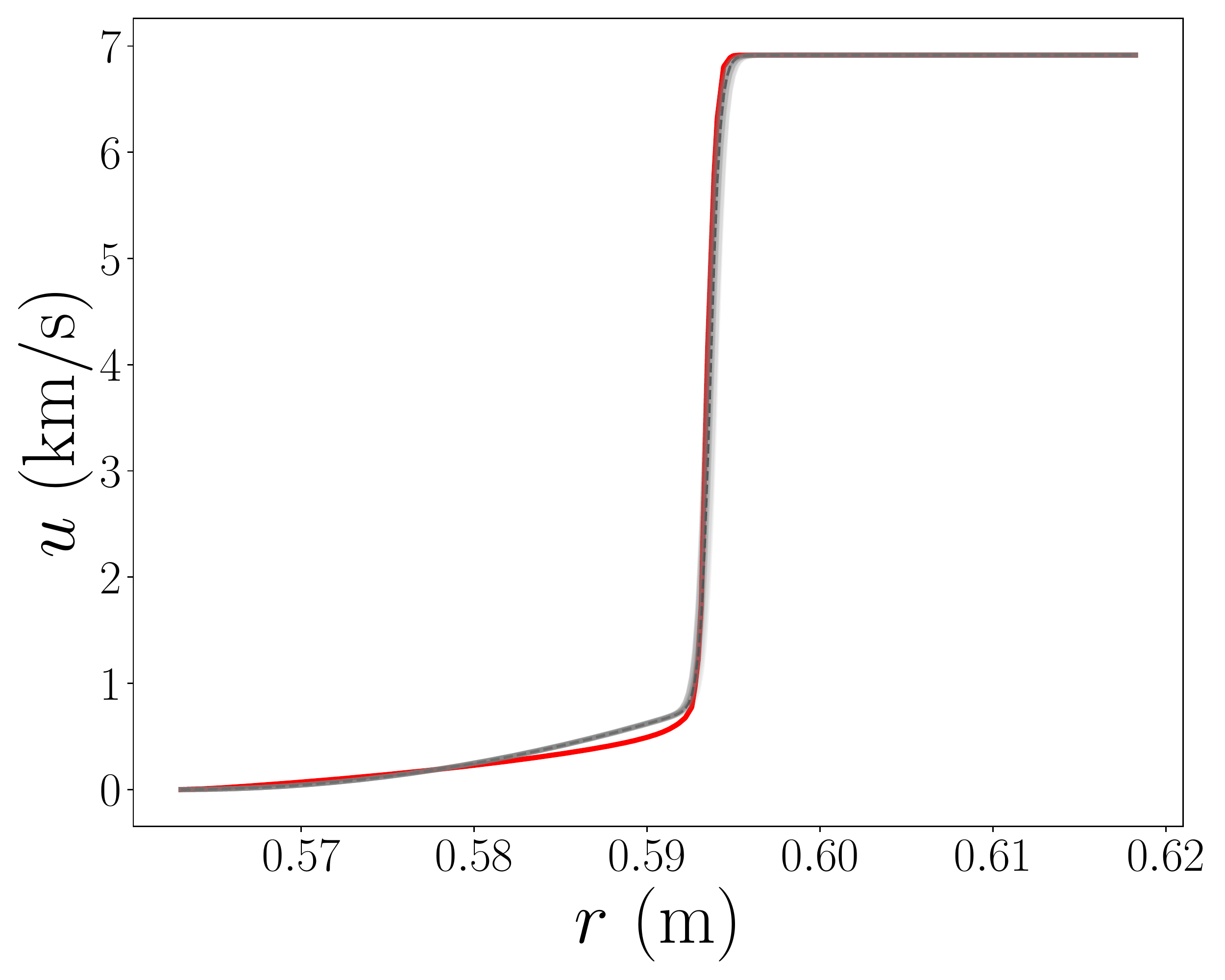}&
\includegraphics[width=56mm]{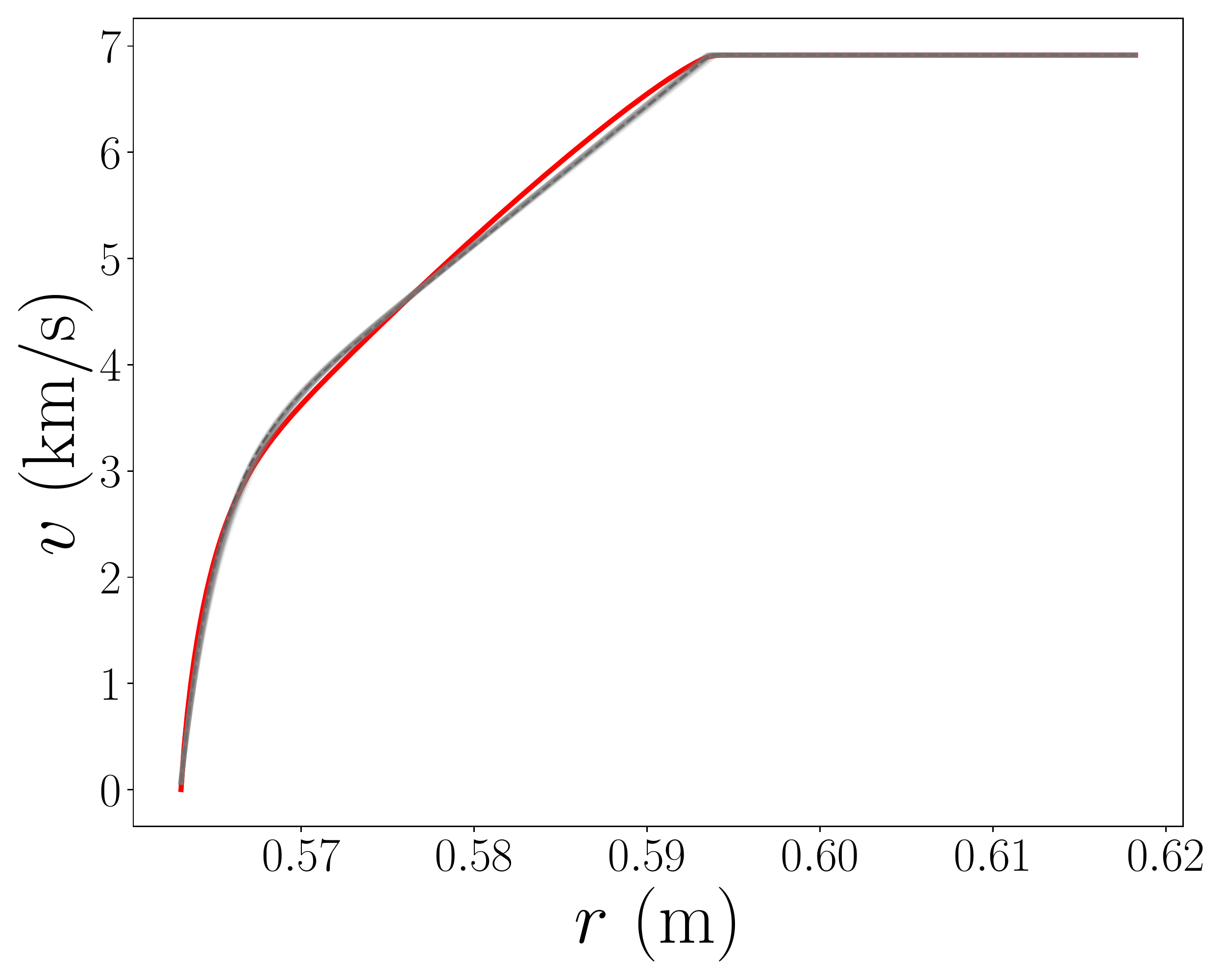}&
\includegraphics[width=56mm]{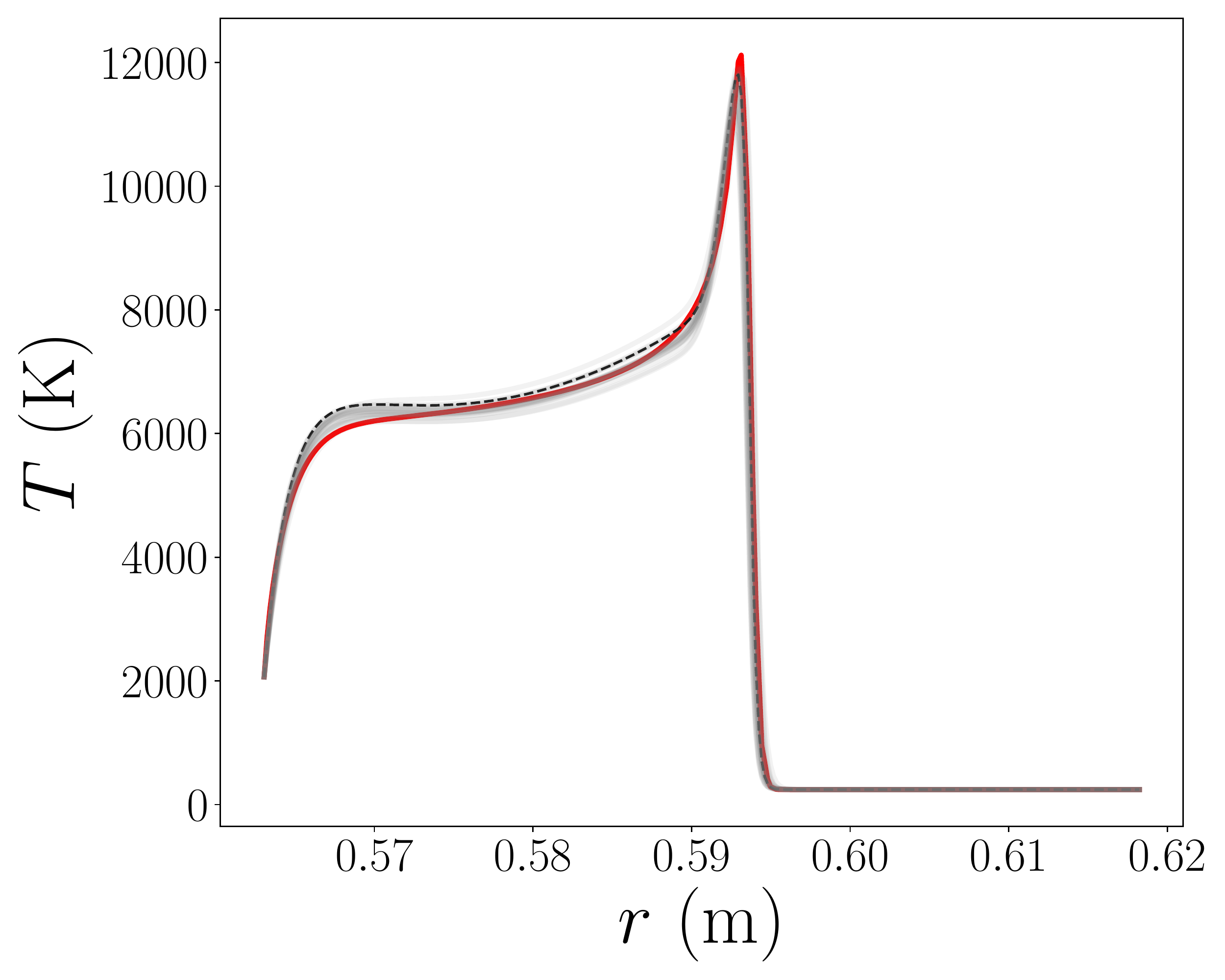}
\end{array}$
\end{center}

\begin{center}$
\begin{array}{rr}
\includegraphics[width=56mm]{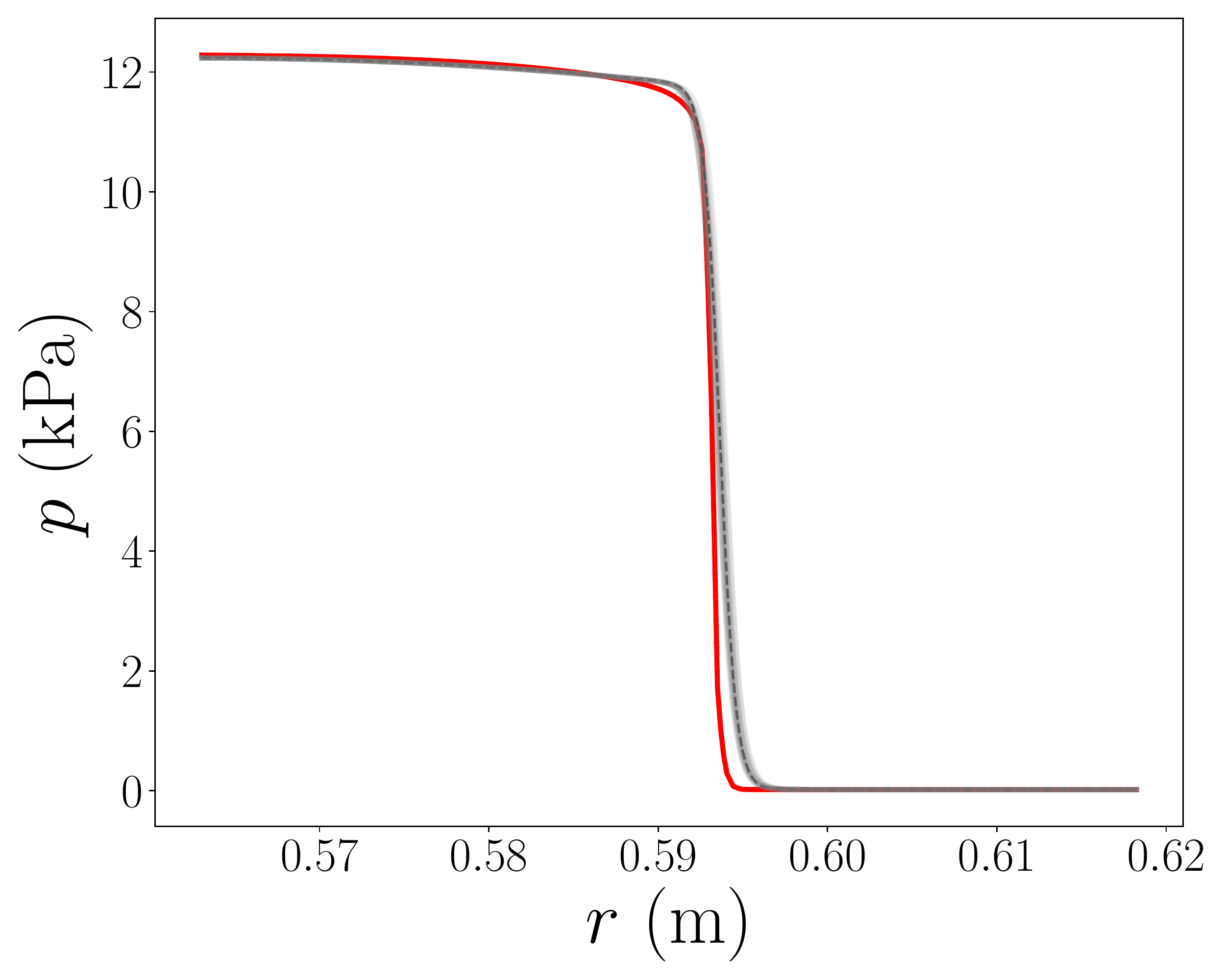}&
\includegraphics[width=56mm]{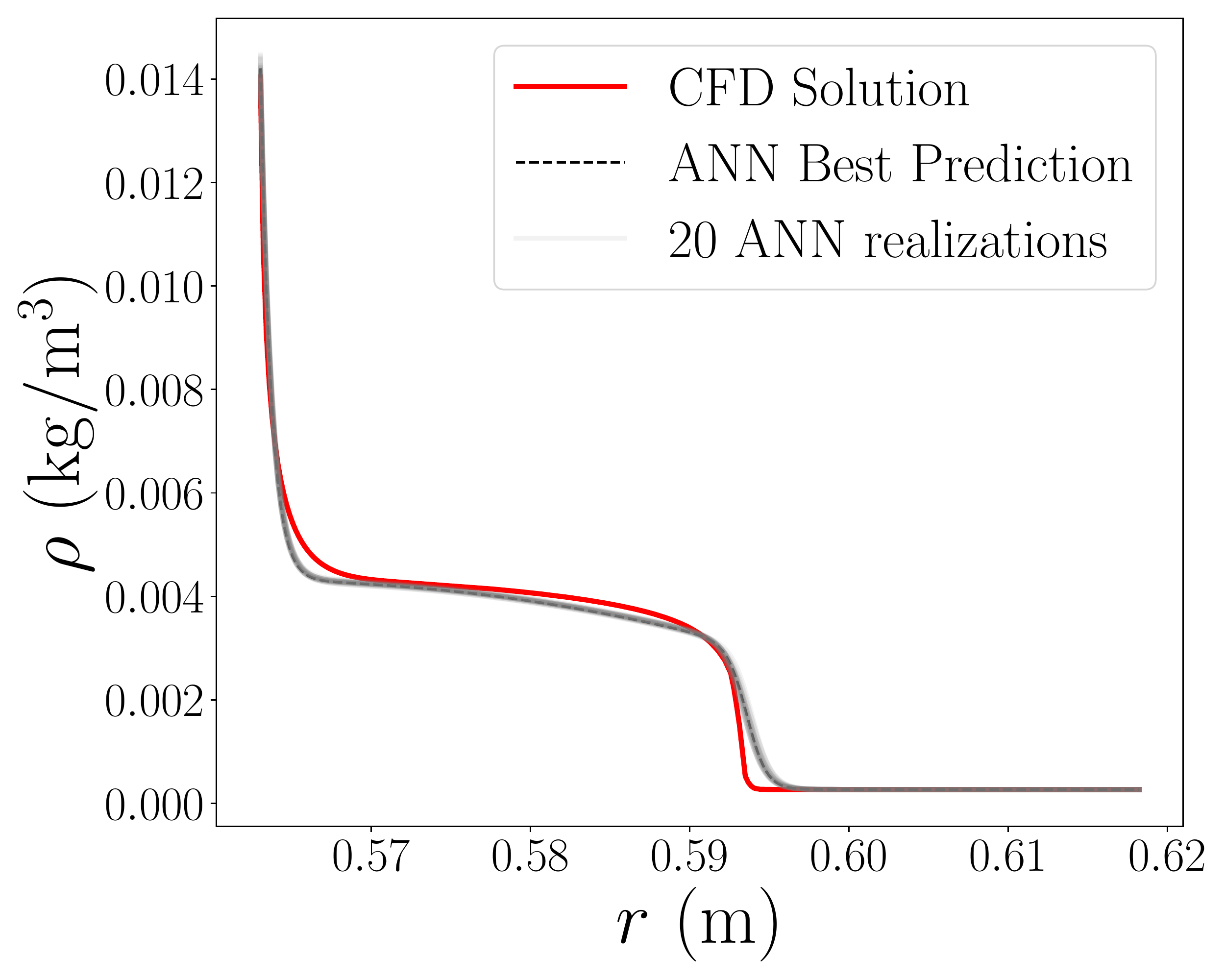}
\end{array}$
\end{center}

\caption{N1 predictions for $\bm{min(J)}$ validation test case $\mathbf{u}=[61.3  \; \rm{km},  \; 6.91  \; \rm{km/s},   \; 0.563  \; \rm{m},  \; 2063  \; \rm{K}]^T$: Highly accurate reproduction of CFD results for main flow quantities, with low deviation between ANN realizations.}
\label{best}
\end{figure}

Particularly interesting is the distribution of the loss function and its behavior near the boundaries of the domain explored by the training data. It is evident that the model is less successful at high $M$ and low $h$ but no clear trend is observed for other inputs. This suggests that the model is robust over the entire training range. To further illustrate this, we show in Figs. \ref{worst} and \ref{best} the model predictions for the worst (higher loss, $J=0.142$) and the best (lower loss,  $J=0.015$) predictions of the network for all quantities. The associated input parameters are recalled in the figure caption. In all figures, the predictions from the 20 realizations of N1 are shown with shaded grade lines, while the one coming from the 'best' N1 realization is shown as a dashed line. This gives an estimate on the main regions of uncertainty and identifies the shock position as the main source error in the performances of the N1 network. 

\begin{figure}[!htbp]
\center
\includegraphics[width=0.65\columnwidth]{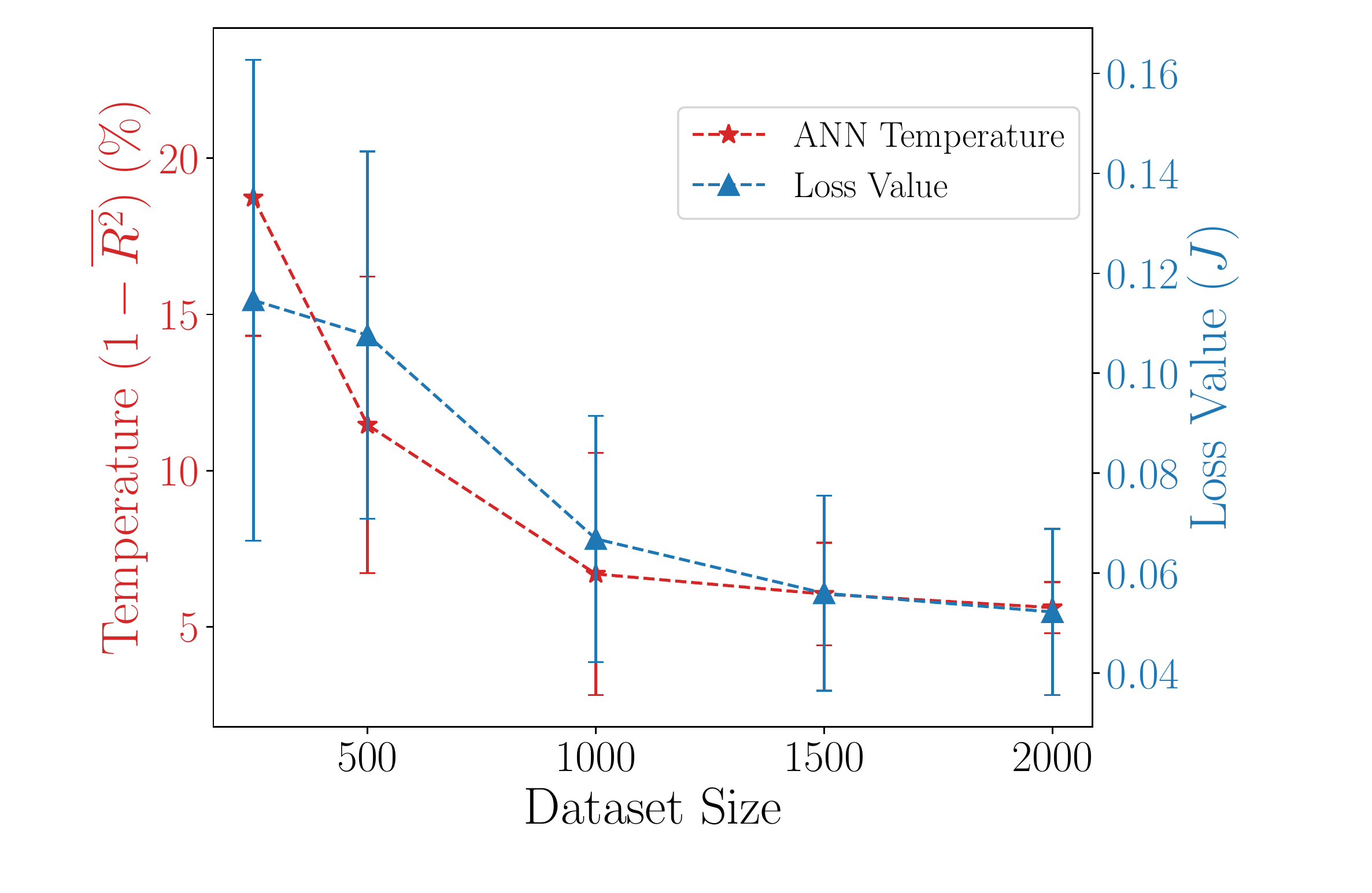}
\caption{Performance of the network N1 in the temperature prediction as a function of the dataset size used in the training. Reducing the dataset by a factor 2 leads to a minor loss of accuracy, proving that the size of the training data is appropriate.}
\label{dataset}
\end{figure}

Finally, to analyze the model performances in relation to the available training data, we train five $n_E=10$ ensembles using $250$, $500$, $1000$, $1500$ and $2000$ randomly selected test cases. All hyperparameters of the ANN were kept constant ($5000$ epochs, batch size of 30, and $70\%$ training cases) to examine the effect of the CFD dataset size. \Cref{dataset} shows the average metrics values and their uncertainty over the $n_E=10$ ANN folds ($\pm 1.96 \sigma_E$) for the temperature prediction. While it is evident that the performances deteriorate rapidly if less than about $1000$ conditions are used, the gentle improvement observed from $1000$ to $2000$ shows that the complexity of the proposed model is well tuned to the size of the dataset available, and no significant improvements are to be expected if more CFD simulations are used in training.

\subsection{Species densities of 5-species air mixture (ANN N2)}

\begin{figure}[!htb]
\begin{center}$
\begin{array}{ll}
\includegraphics[width=115mm]{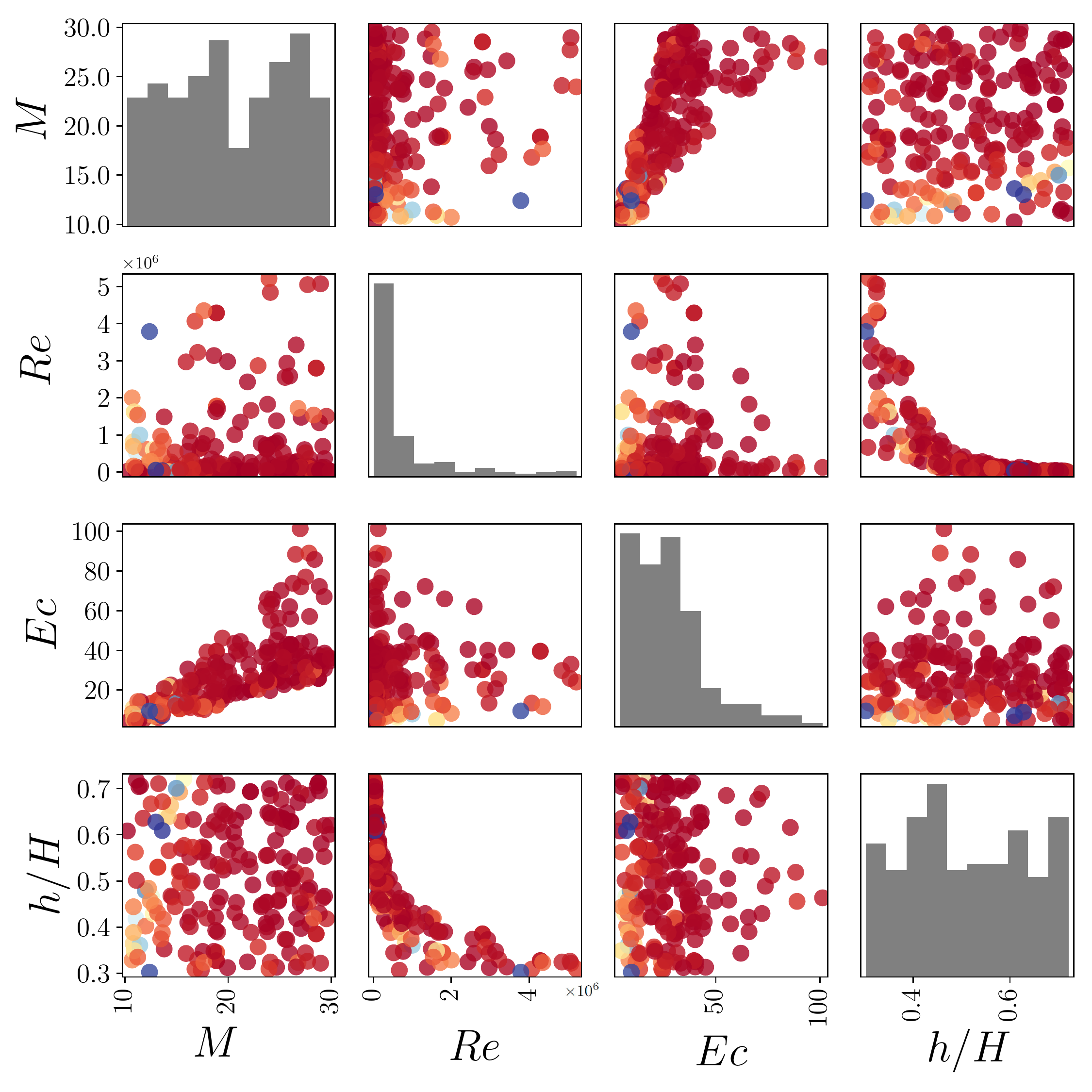}&
\includegraphics[width=19mm]{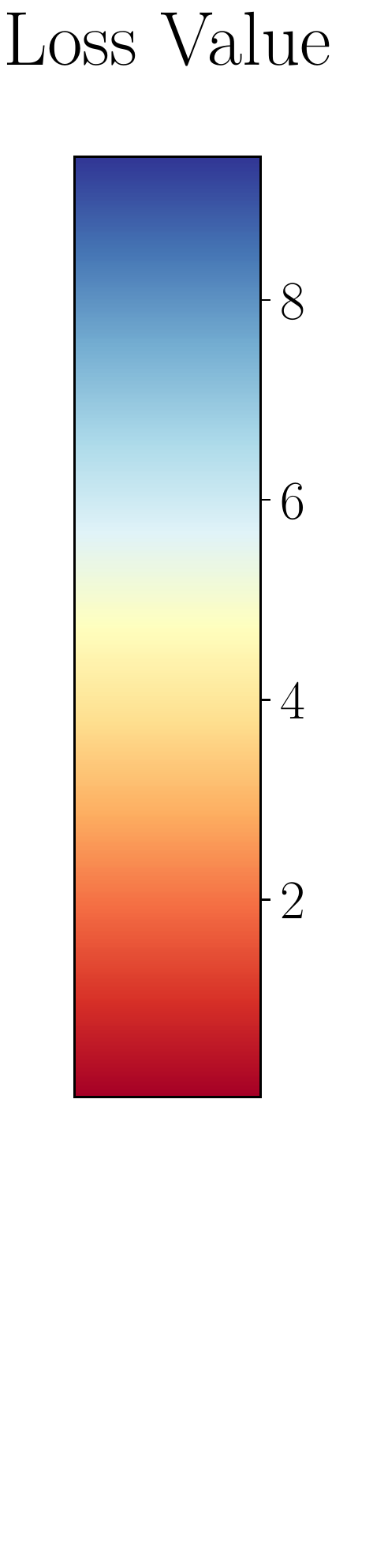}
\end{array}$
\end{center}
\caption{Scatter matrix of individual N2 validation test cases (species partial densities): Relatively lower accuracy for low $\mi{M}, \mi{Ec}$ values, due to very low \ce{N,O,NO} densities.}
\label{histrhoi}
\end{figure}

\begin{figure}[!htbp]
\begin{center}$
\begin{array}{lll}
\includegraphics[width=58mm]{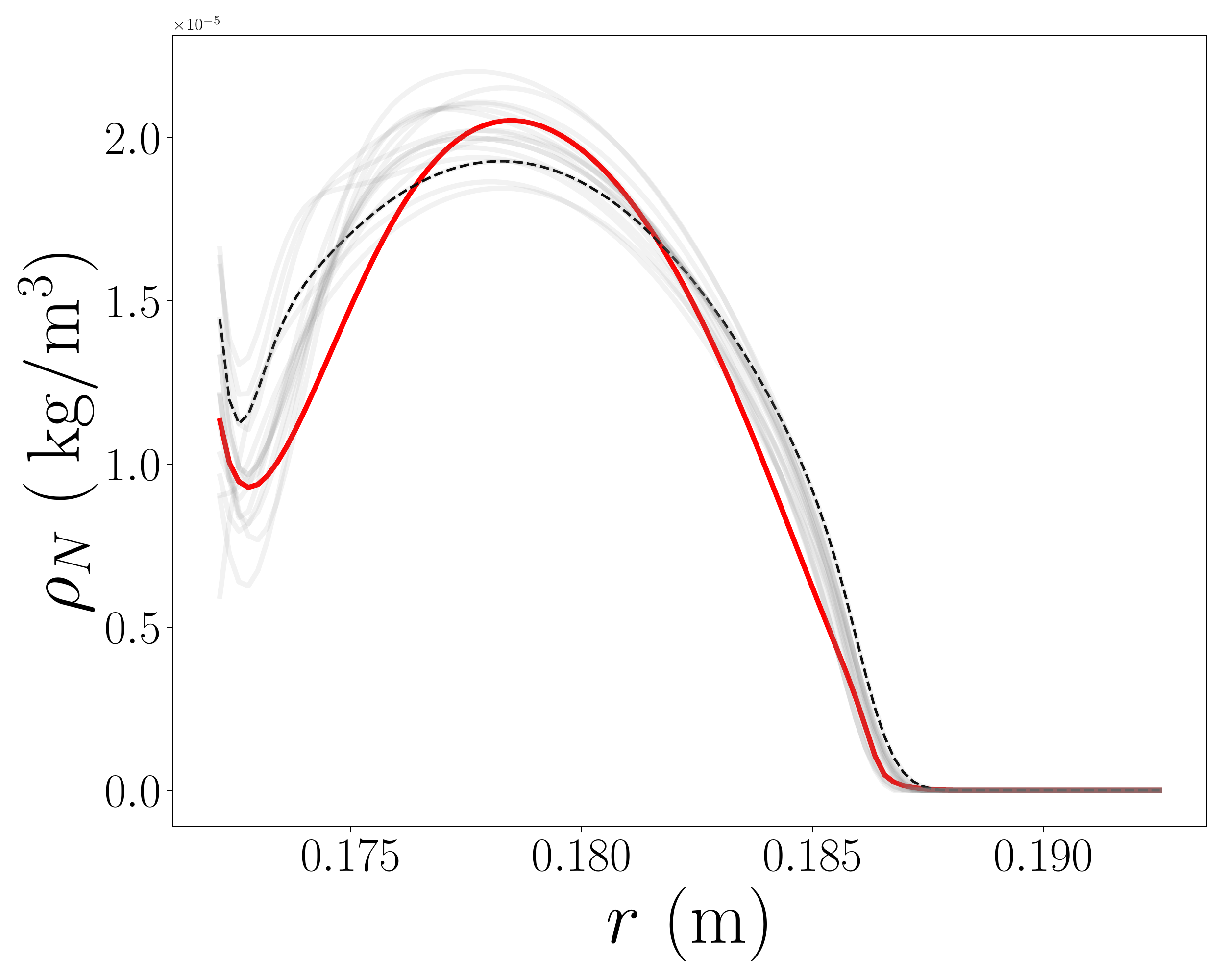}&
\includegraphics[width=58mm]{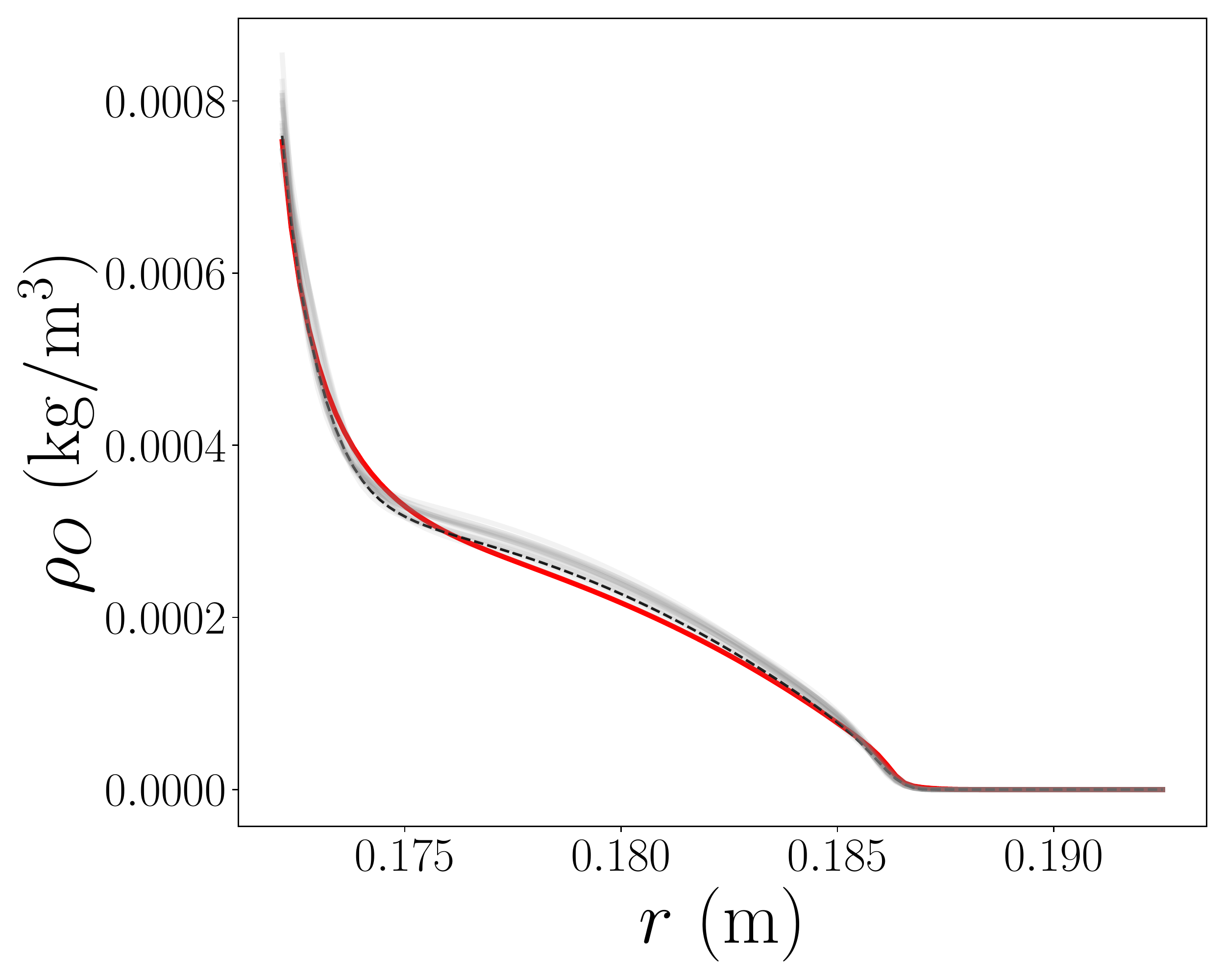}&
\includegraphics[width=58mm]{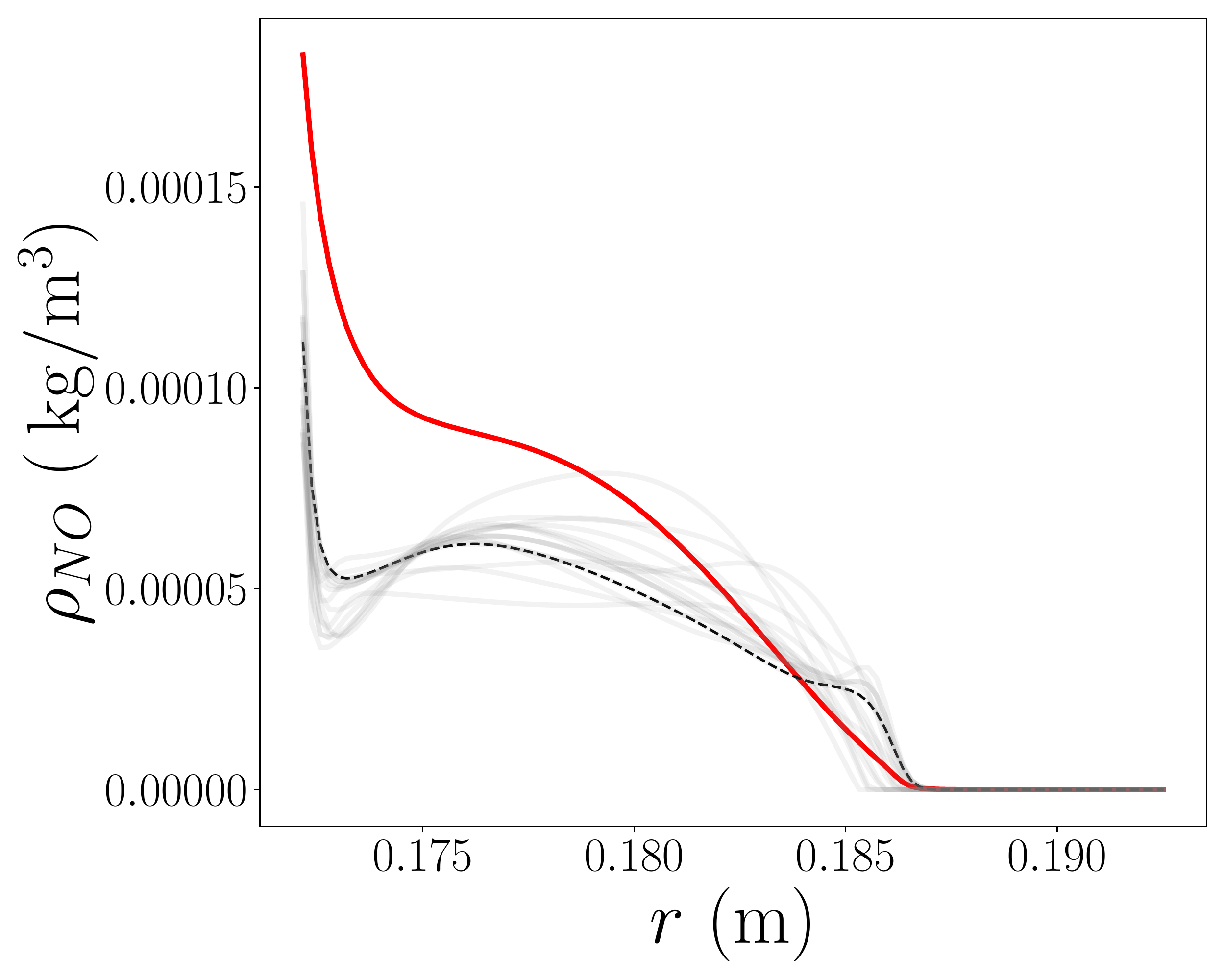}
\end{array}$
\end{center}

\begin{center}$
\begin{array}{rr}
\includegraphics[width=58mm]{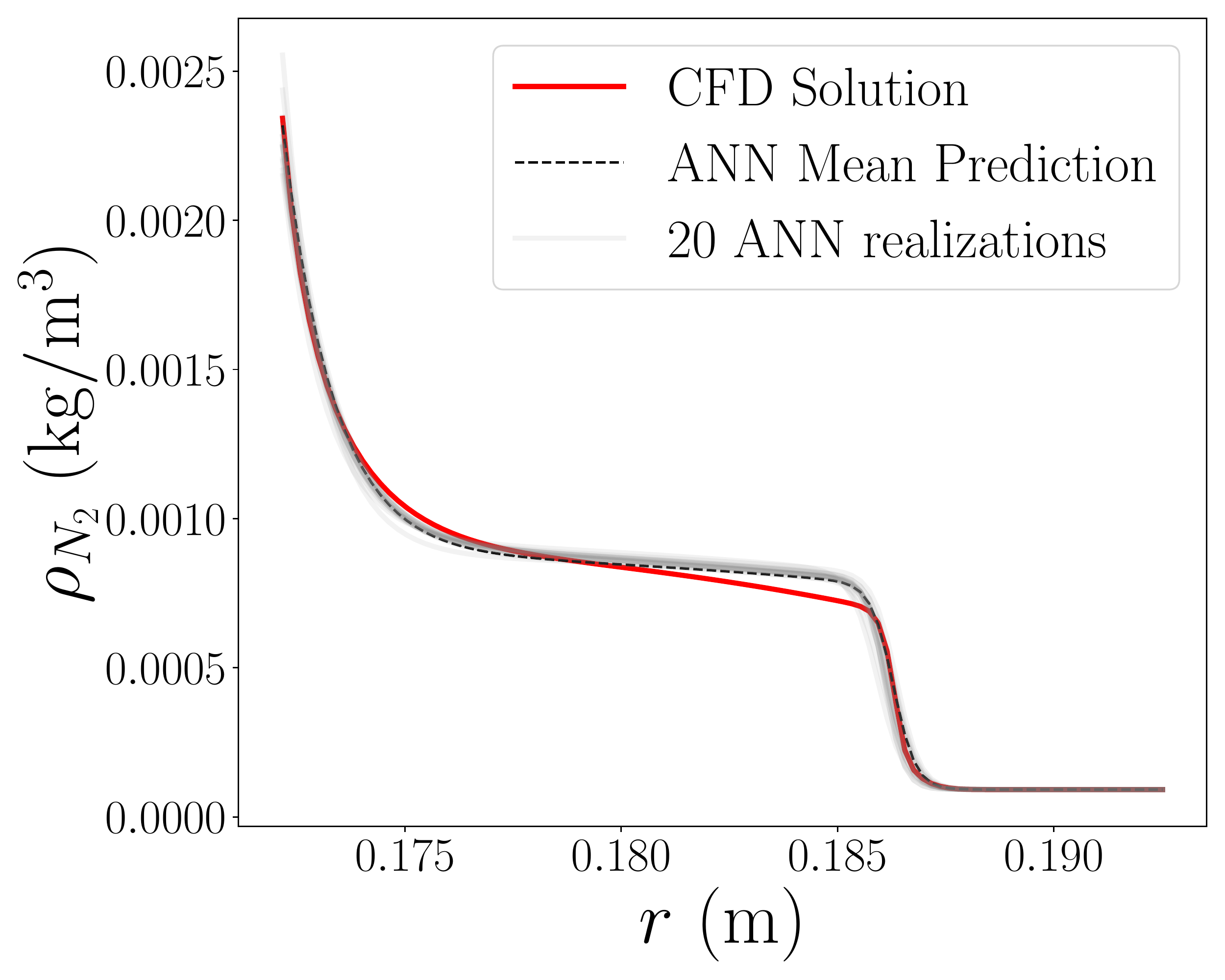}&
\includegraphics[width=58mm]{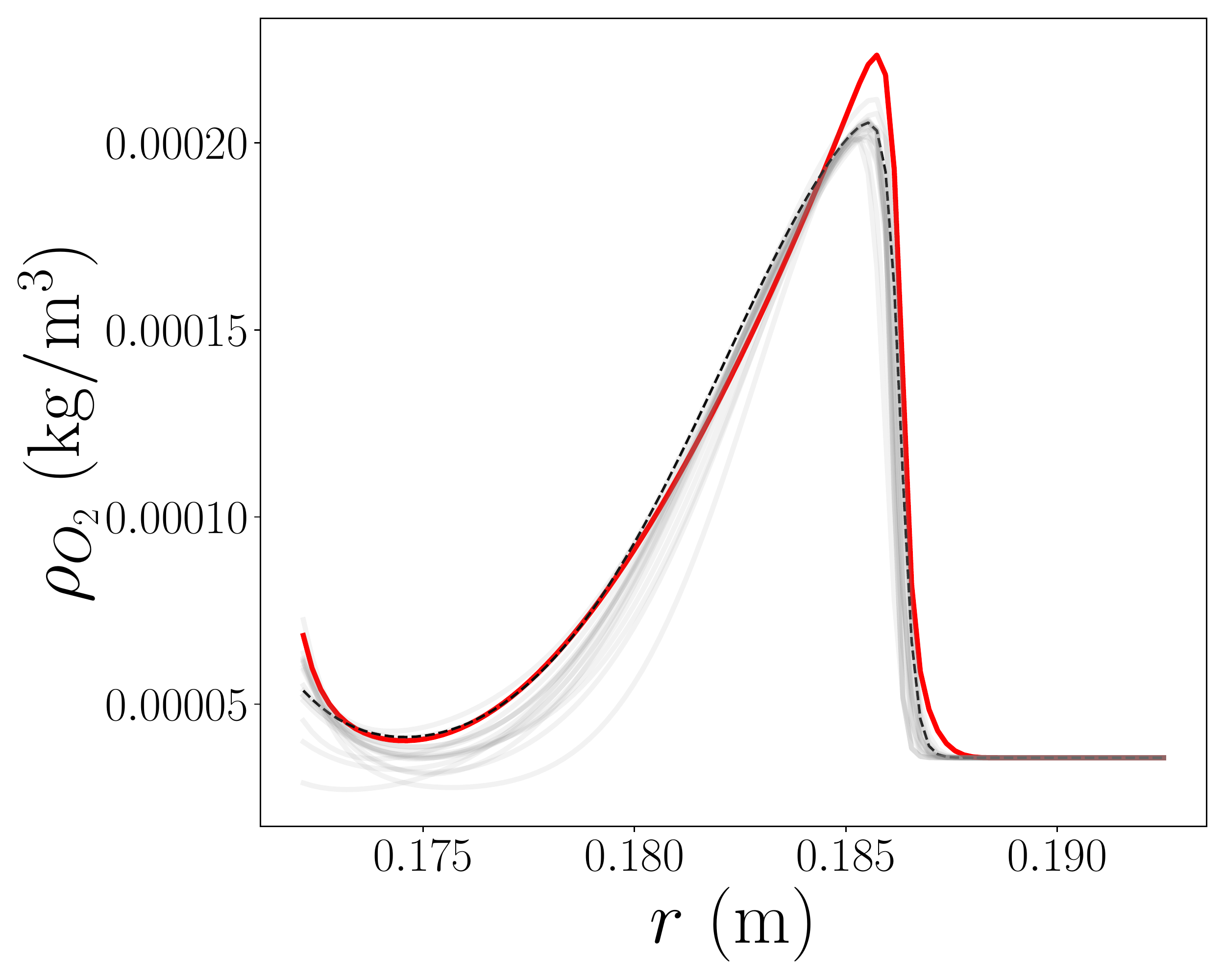}
\end{array}$
\end{center}
\vspace{-5mm}
\caption{N2 network predictions for $\bm{max(J)}$ validation test case $\mathbf{u}=[61.9  \; \rm{km}, \; 4.07  \; \rm{km/s}, \; 0.172  \; \rm{m}, \; 2383  \; \rm{K}]^T$: Inaccuracy in $N,NO$ predictions, high variance across the 20 ANN realizations.}
\label{worstrhoi}

\begin{center}$
\begin{array}{lll}
\includegraphics[width=58mm]{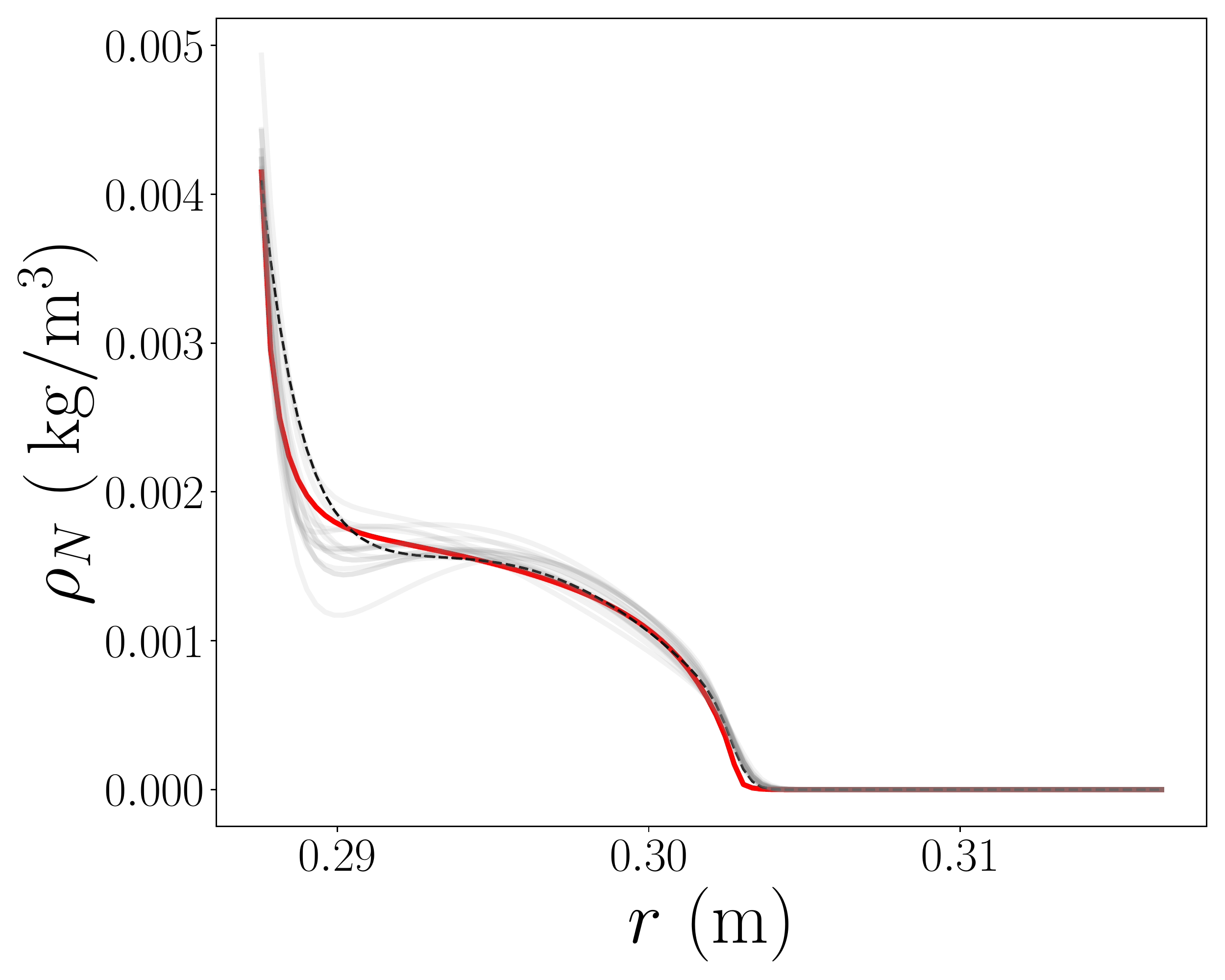}&
\includegraphics[width=58mm]{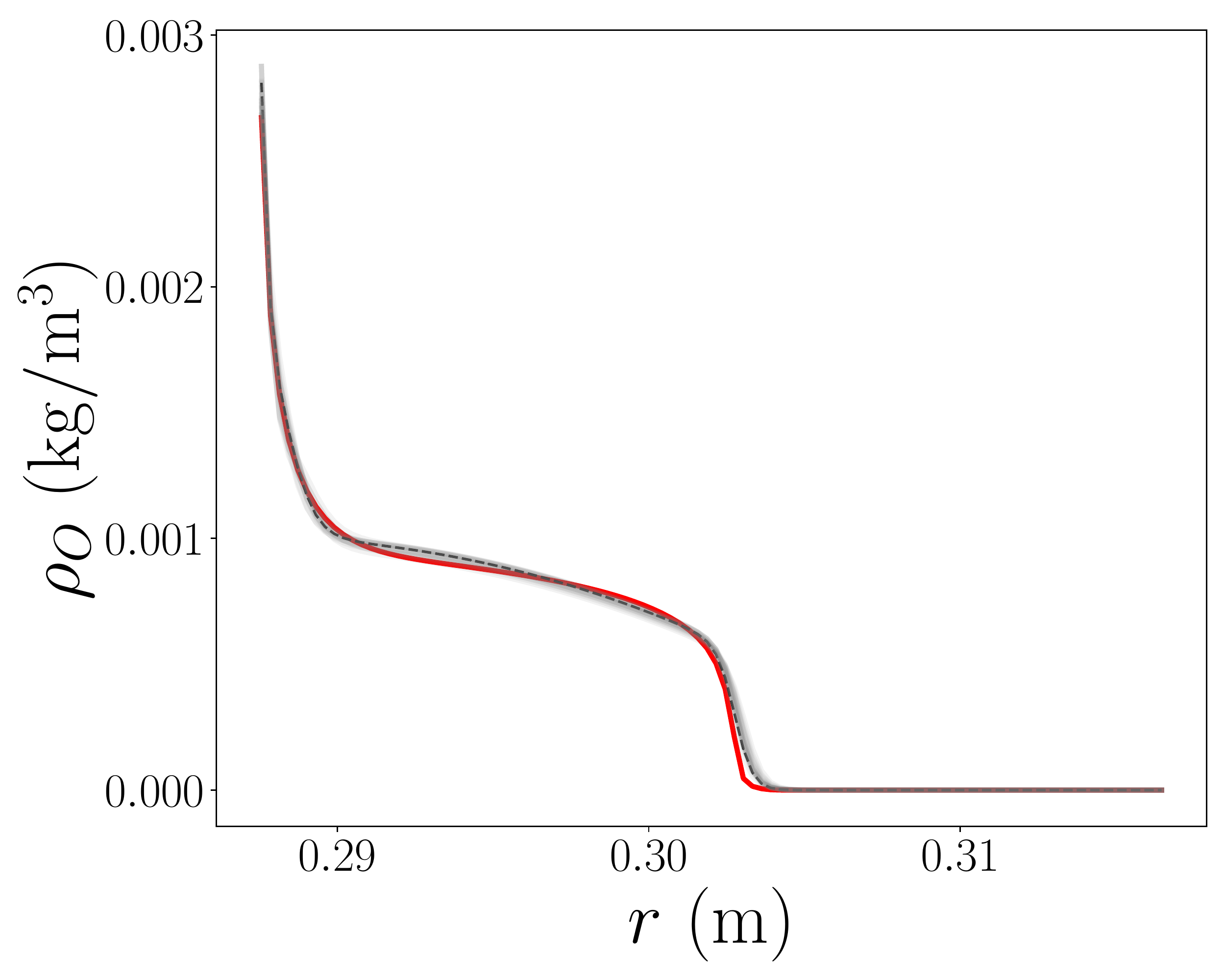}&
\includegraphics[width=58mm]{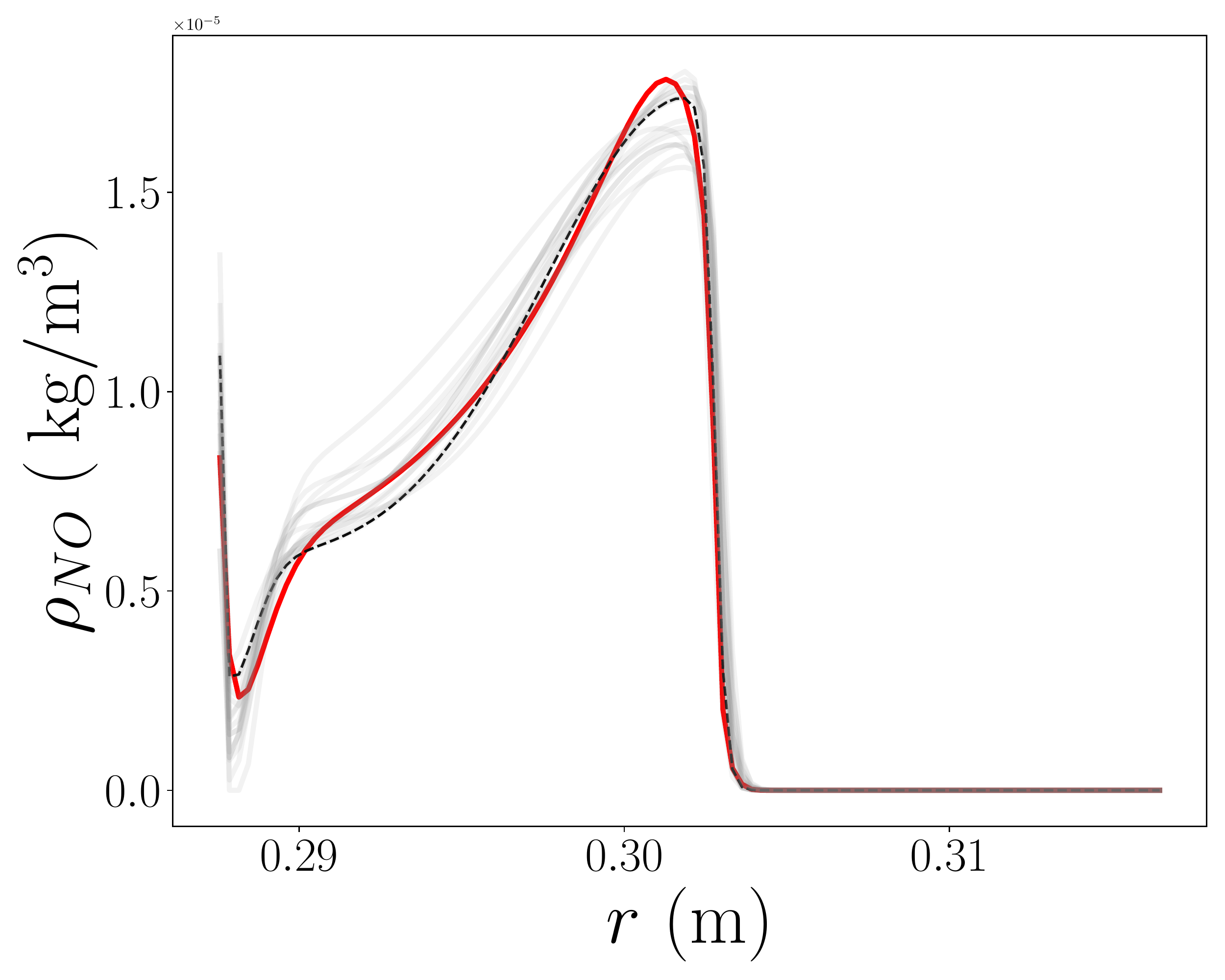}
\end{array}$
\end{center}

\begin{center}$
\begin{array}{rr}
\includegraphics[width=58mm]{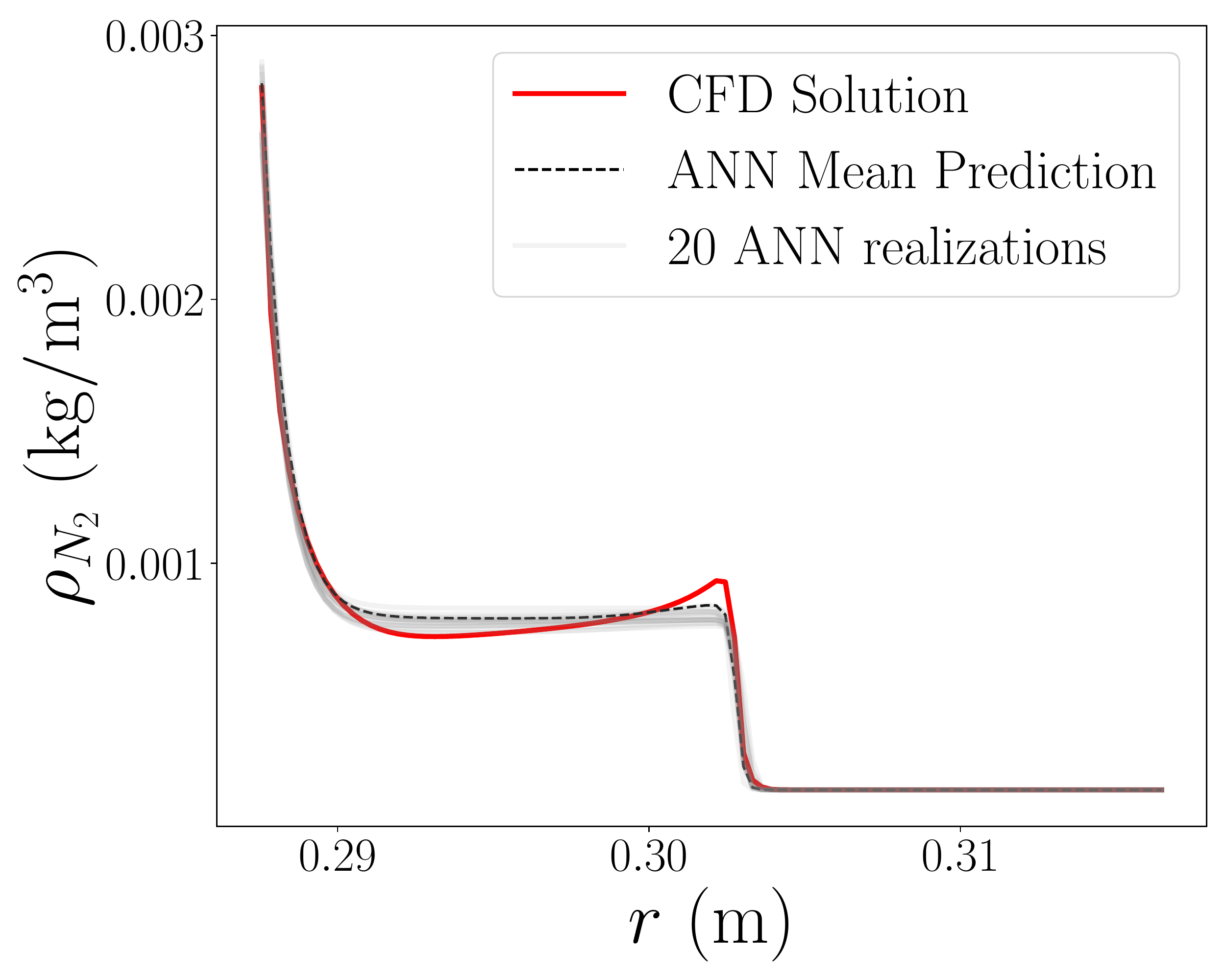}&
\includegraphics[width=58mm]{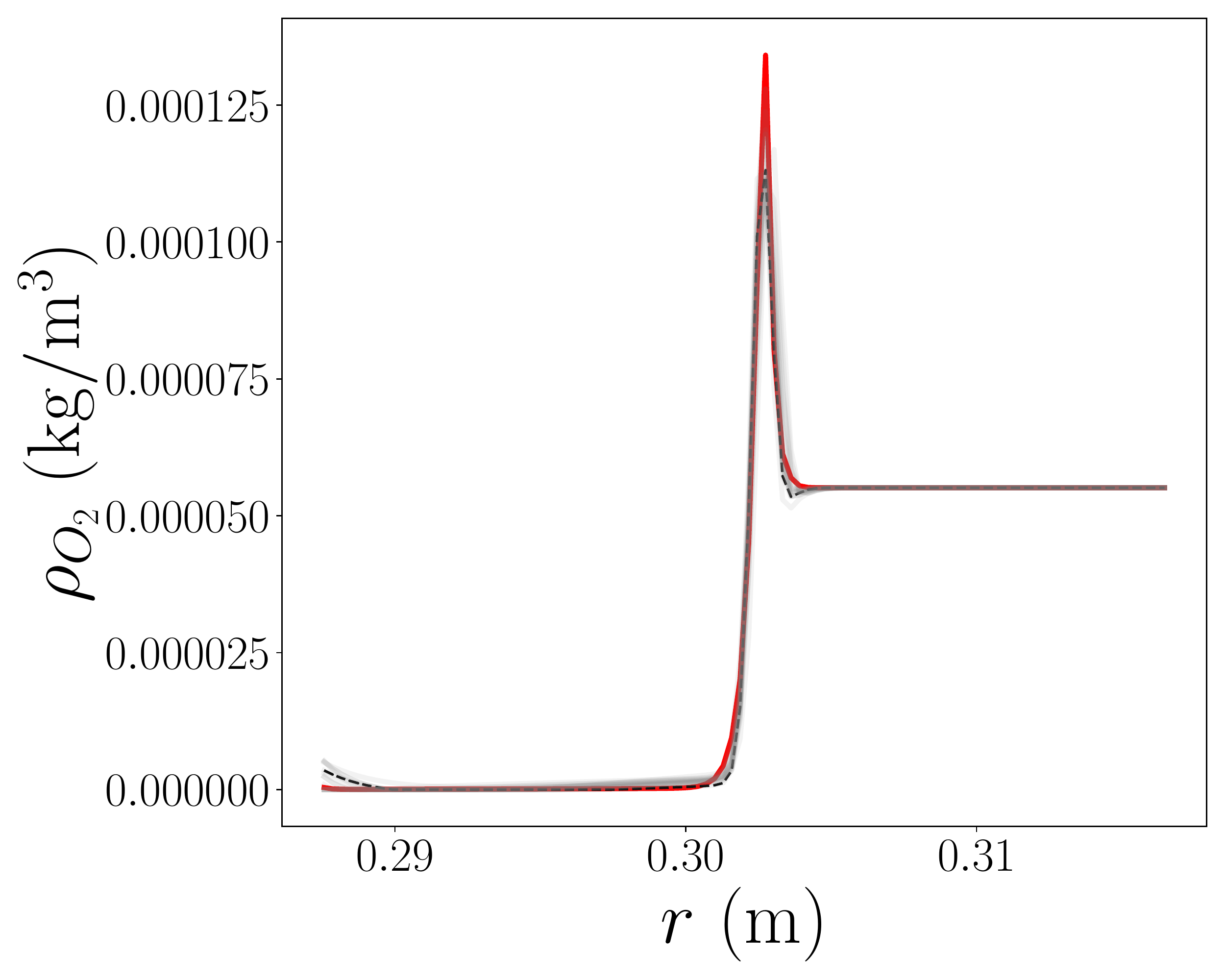}
\end{array}$
\end{center}

\caption{N2 network predictions for $\bm{min(J)}$ validation test case $\mathbf{u}=[63.6 \; \rm{km}, 7.90  \; \rm{km/s}, 0.288  \; \rm{m}, 2526  \; \rm{K}]^T$: Highly accurate reproduction of flow features for all 5 chemical species, over 4 orders of magnitude.}
\label{bestrhoi}
\end{figure}

We perform the same validation for network N2. The averaged $R^2$ errors for the five species concentrations are collected in \cref{res2} while \cref{histrhoi} shows the scatter matrix for the loss value distribution. It should be noted that the loss values of the N2 network, shown in the scatter matrix of \cref{histrhoi}, are not comparable with those of N1. This can be seen from \cref{loss}, which incorporates a different parameterization for N1 and N2 through $\mathbf{w}$ as well as the weights $g$ for each case.
However, \cref{res2} indicates a lower accuracy of this network compared to N1, deteriorating at low Mach and Eckert numbers. The main challenge in these conditions is the low concentrations of \ce{N, O, NO}, which alleviate through the nonlinear scaling of \cref{tabscale}. However, given the low \ce{N, O, NO} values, the lack of accuracy in these areas can be considered to be of secondary importance. This is depicted in \cref{worstrhoi}, which corresponds to $M=13.10$ and $Ec=7.98$. The network performs poorly in predicting the density of $N$ and $NO$, while the observed uncertainty grows substantially. On the other hand, $N$ and $NO$ are accurately predicted for the case with the lowest loss value in \cref{worstrhoi}, corresponding to $M=25.73$ and $Ec=27.13$. 

\begin{table}[!htb]
\renewcommand*{\arraystretch}{1.5}
\centering
\footnotesize
\caption{Average $\mb{R^2}$ values for species partial densities worst and best N2 network realization: Significant difference on \ce{NO} accuracy.}
\label{res2}
\begin{center}
\begin{tabular}{lcccccc}
\toprule \toprule
%row1
& $J$ 
& $R^2_{\hat{\rho}_{\ce{N}}}$ 
&  $R^2_{\hat{\rho}_{\ce{O}}}$ &  $R^2_{\hat{\rho}_{\ce{NO}}}$  &  $R^2_{\hat{\rho}_{\ce{N_2}}}$ &  $R^2_{\hat{\rho}_{\ce{O_2}}}$  \\
\midrule
%next row
 $min(J)$ fold  & $1.79$ & $95.03$ & $98.09$ & $86.94$ & $89.08$  & $92.19$\\
% row 3
 $max(J)$ fold  & $3.07$ & $95.05$ & $97.93$ & $81.57$ & $87.65$ & $92.07$\\
% row 3
\bottomrule
\bottomrule
\end{tabular}
\end{center}
\end{table}

\subsection{Heat and mass wall fluxes for non-reactive and carbon-ablation surfaces (ANNs N3, N4)}

\subsubsection{Conductive heat on nonreactive surfaces (ANN N3)}
\label{Conv_nonab}

\begin{figure}[!htb]
\begin{center}$
\begin{array}{ll}
\includegraphics[width=70mm,height=6cm]{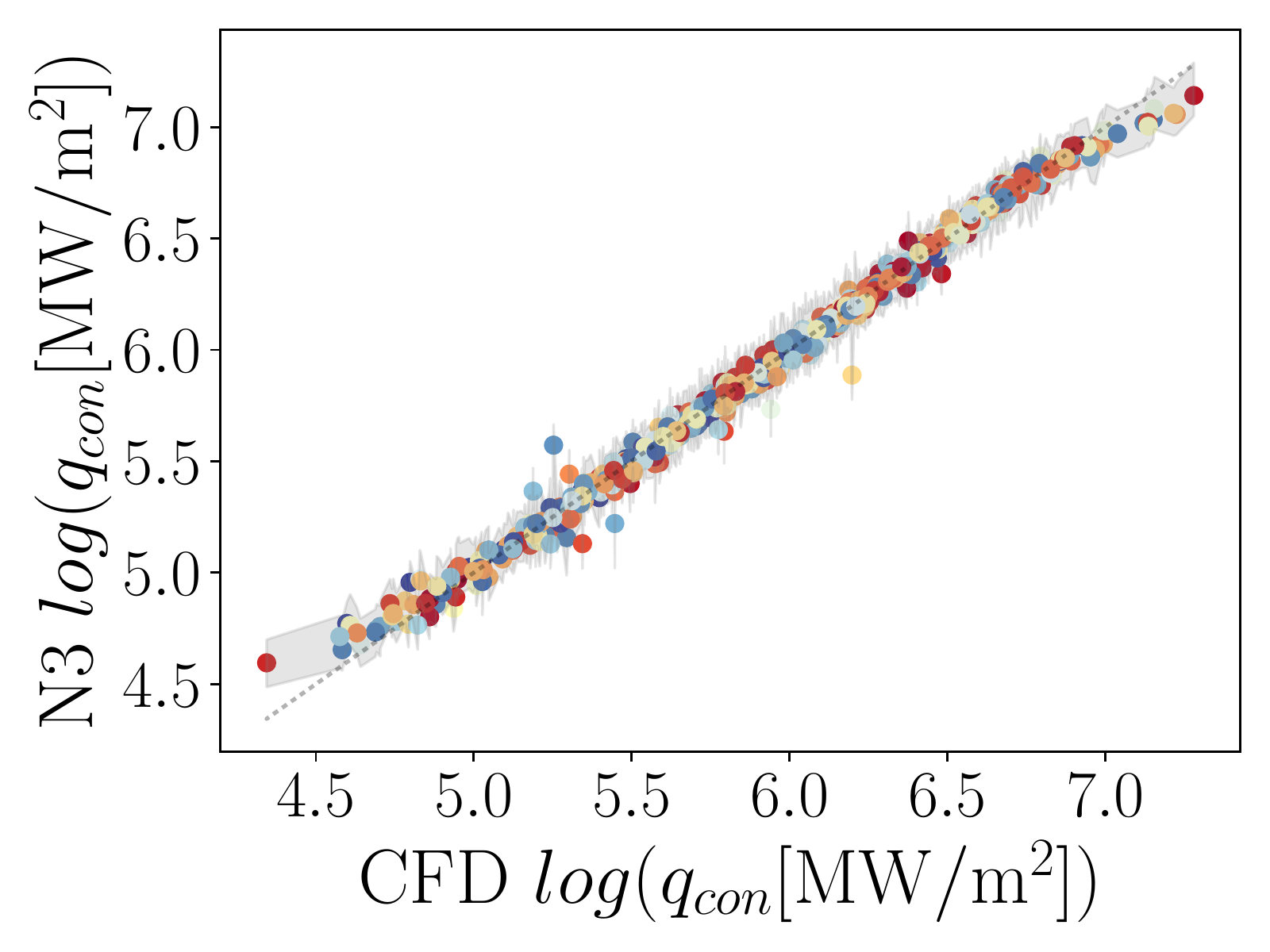}&
\includegraphics[width=10mm,height=6cm]{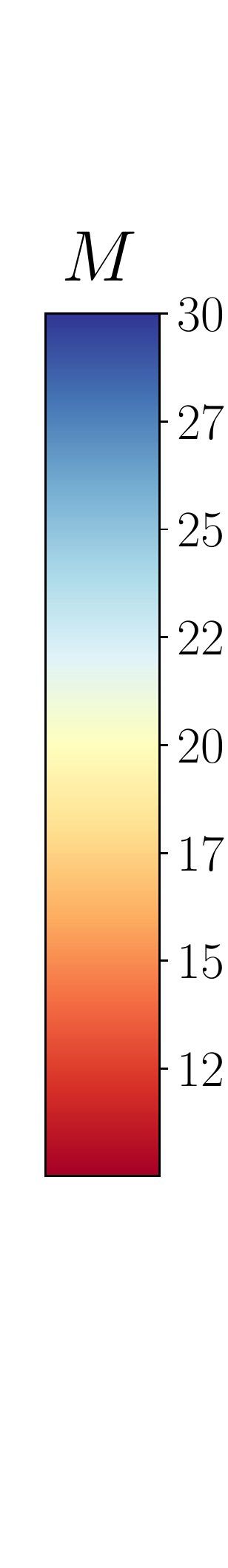}
\end{array}$
\end{center}
\caption{$45^\degree$ line plot for nonablative conduction heat flux at the wall: N3 predictions with $95 \%$ confidence interval (shaded area) follow closely the CFD values for the validation testcases.}
\label{45deg_na}
\end{figure}

The validation study for the ANN N3, predicting the conductive heat flux for the nonreactive surfaces (Section \ref{wallflux}), is presented here.
The fully connected ANN model N3  \cref{anncoef2} is trained $k=20$ times with randomly selected training data. The $R^2$ accuracy on the validation set for the different realizations is $ 85.37 \pm 3.64 \%$ for a $95\%$ confidence interval, assuming a Gaussian distribution of the $R^2$ score. 
To visualize the accuracy of prediction for the best N3 realization, the CFD results are plotted against the N3 predictions for the validation test-cases in \cref{45deg_na}. The predictions are colored by the value of the free-stream Mach number, and the $95 \%$ confidence interval is shown in grey. It can be observed that all predictions fall close to the identity line over the full range of investigated conditions, and no source of systematic error can be identified with respect to the Mach number. A larger deviation from the $45^\degree$ is observed at the lowest and the highest conductive heat fluxes since they rarely occur under the examined free-stream conditions. The potential of the N3 results is further highlighted through a comparison with employed semianalytical correlations for given reentry trajectories in Section \ref{S:5}.

\subsubsection{Heat and mass wall fluxes on carbon-based ablative surfaces  (ANN N4)}
\label{wall_ab}

As presented in Section \ref{wallflux}, in the regime of carbon ablation we are interested in predicting the three heat-flux components of gas conduction, diffusion and blowing, as well as the mass blowing flux at the ablative wall. Network N4 \cref{anncoef2} is trained and the average testing error over the 20 realizations for the output quantities of interest is given in \mref{abl_tab}. 
Note that the blowing heat flux spans almost 4 orders of magnitude in the original training data. 
This has an impact on the overall accuracy of the associated data-driven model, which shows the highest average error. 

\begin{table}[tb]
\footnotesize
\centering
\caption{Average N4 heat and mass fluxes error ($\mathbf{100 \%}$) over 20 folds, for validation cases.}
\label{abl_tab}

\begin{center}
\begin{tabular}{lcccc}
\toprule
\toprule
& $q_{\mr{con}}$ 
&  $q_{\mr{dif}}$ &  $q_{\mr{bl}}$  &  $ \dot m $ \\
\midrule
%next row
& $8.34$ &$9.54$ & $16.02$ & $7.24$\\
\bottomrule
\bottomrule
\end{tabular}
\end{center}
\end{table}

\begin{figure}
\centering
\begin{minipage}[6cm]{0.45\linewidth}
\subfloat[Conductive heat flux]{\includegraphics[width=6cm]{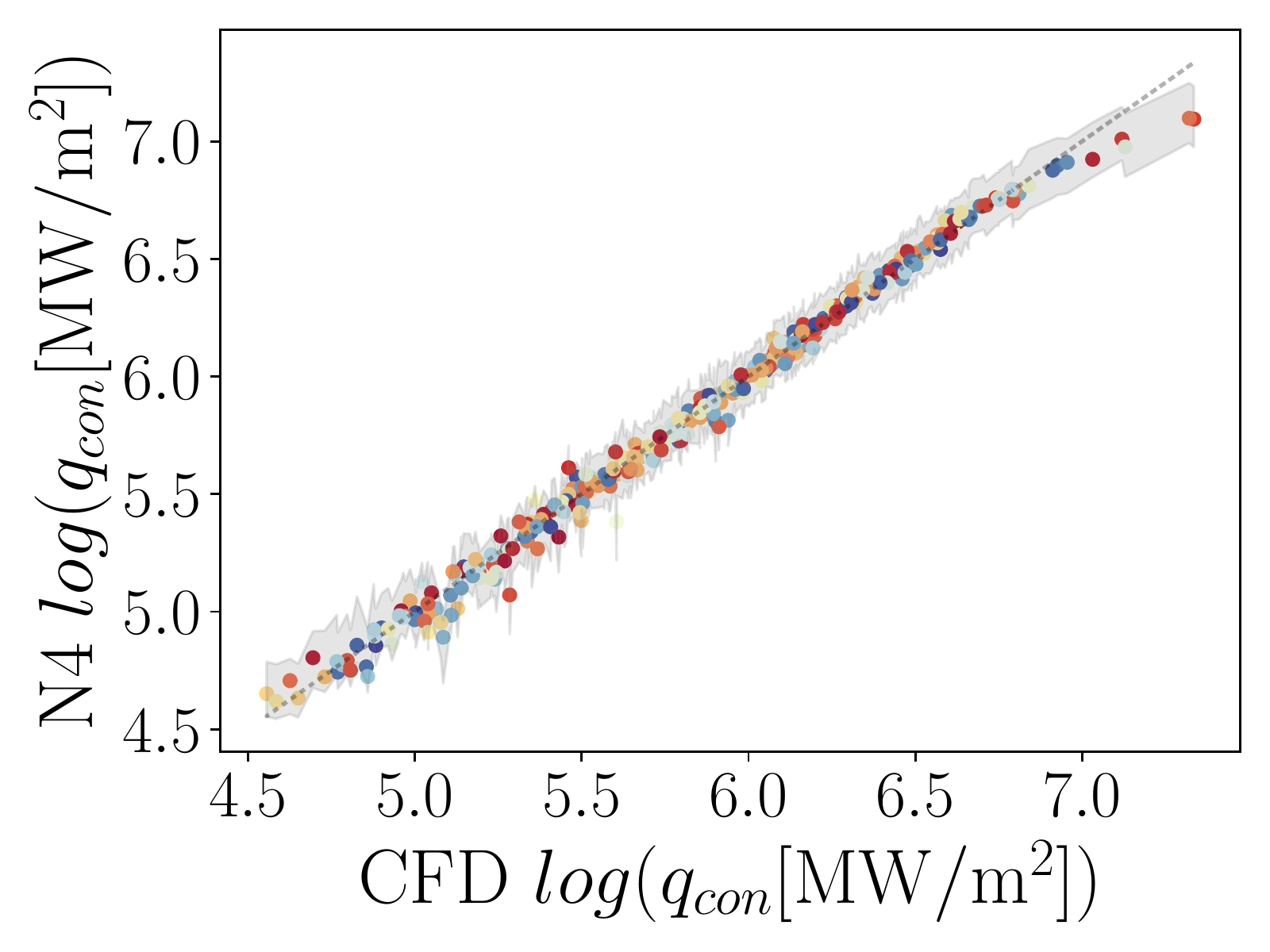}}

\vfill
\subfloat[Diffusive heat flux]{\includegraphics[width=6cm]{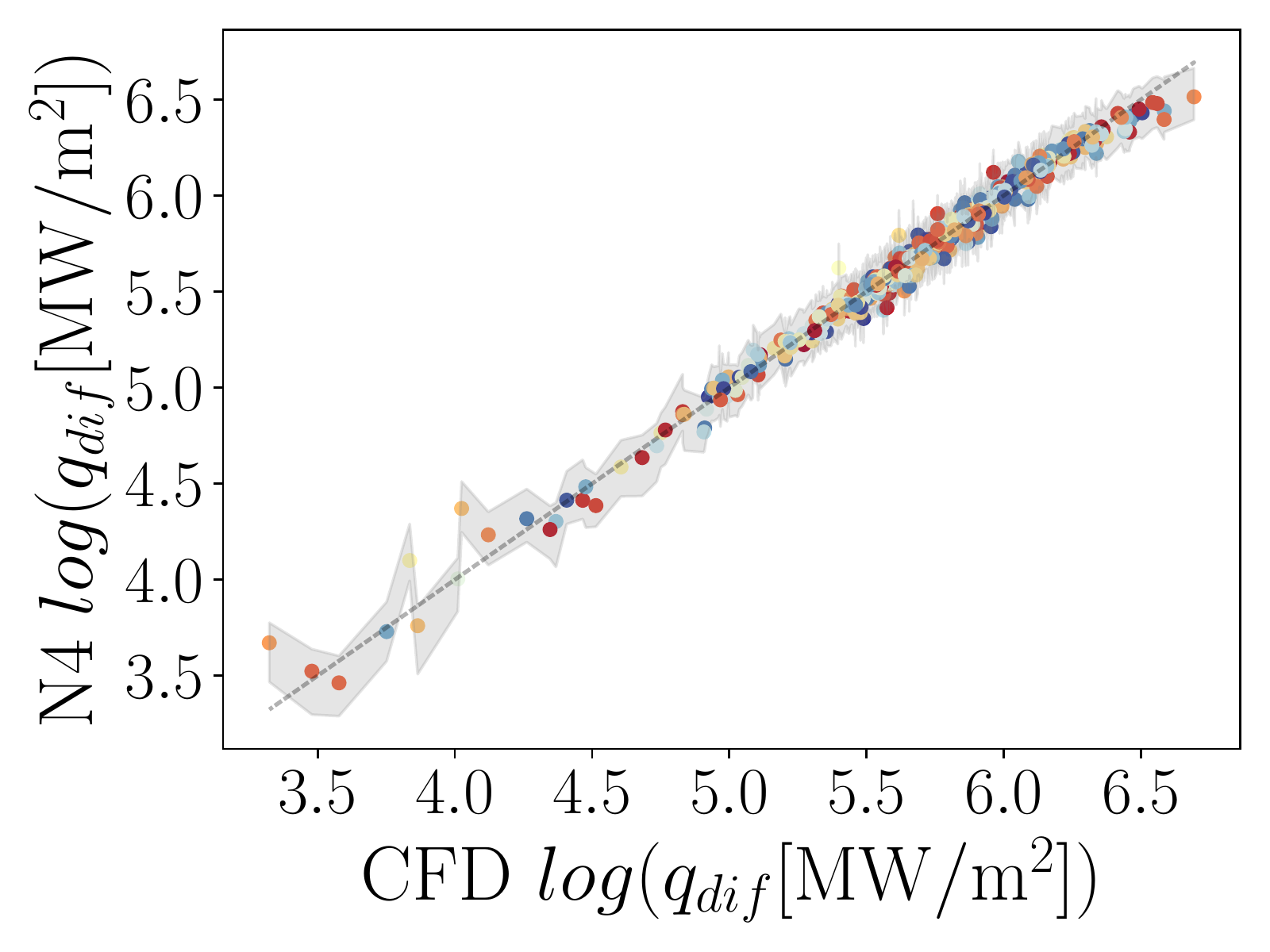}}
\end{minipage}%
\begin{minipage}[6cm]{0.45\linewidth}
\subfloat[Blowing heat flux]{\includegraphics[width=6cm]{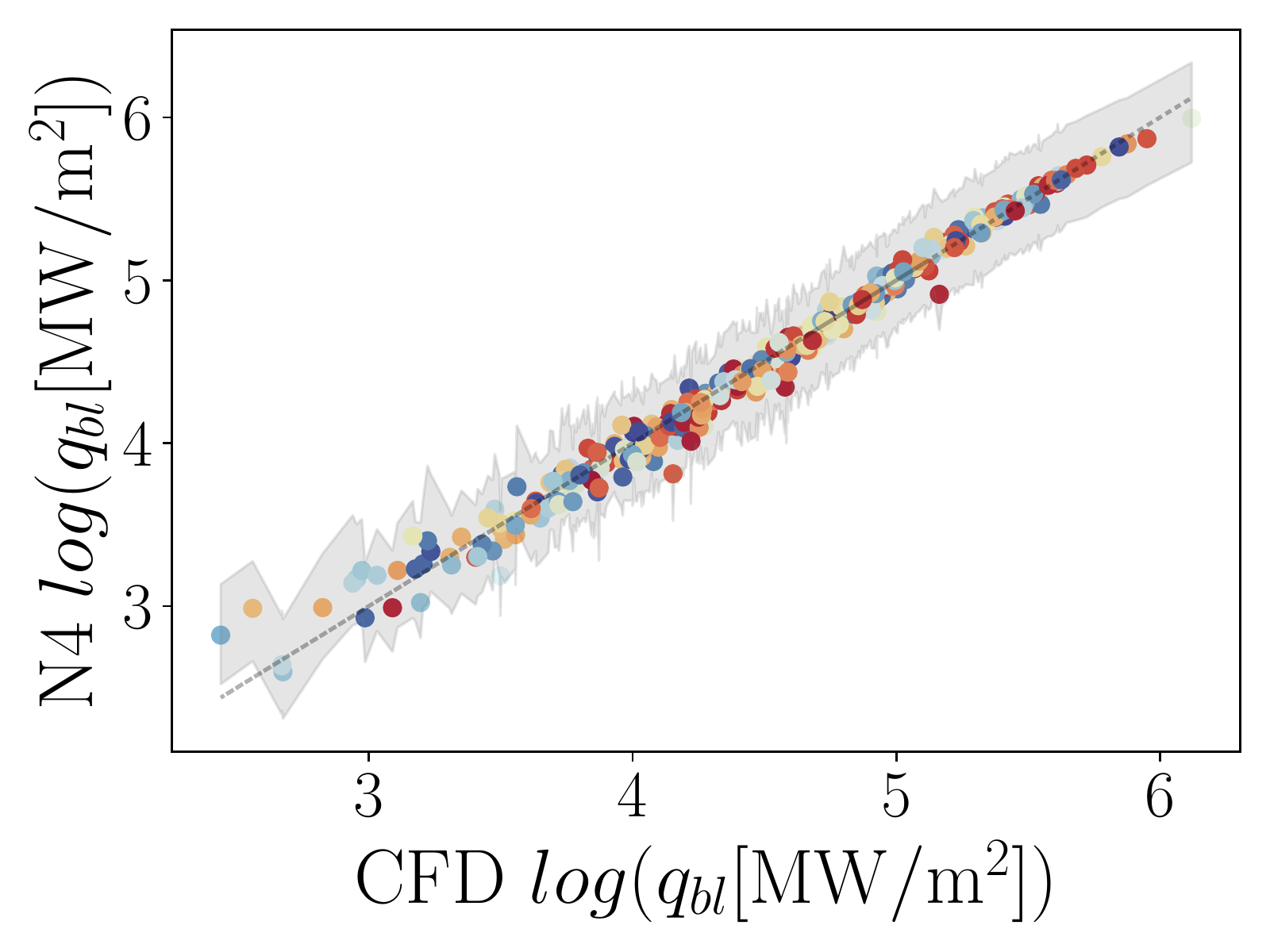}}

\vfill
\subfloat[Mass blowing]{\includegraphics[width=6cm]{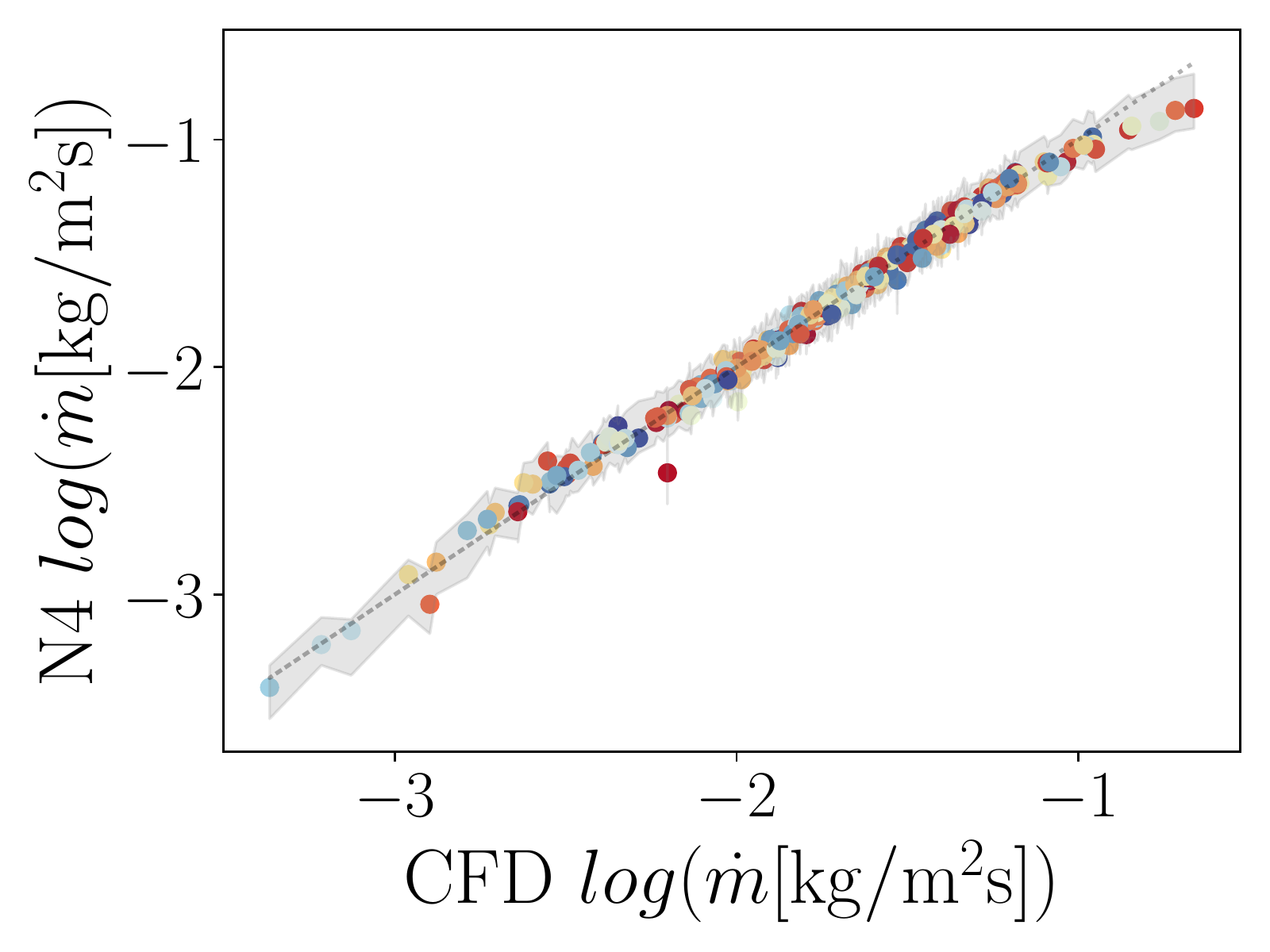}}
\end{minipage}%
\begin{minipage}[1.2cm]{0.08\linewidth}
\subfloat{\includegraphics[width=1.2cm]{FIGURES/ch4/Ma_qcond_abl.pdf}}
\end{minipage}
\caption{$45^\degree$ line plot for carbon ablation heat flux components and mass flux at the wall: N4 predictions with $95 \%$ confidence interval (shaded area) follow closely the CFD results for validation testcases.}\label{45_abl}
\end{figure}

As in the nonablative case, we illustrate the N4 results for the four outputs of interest (for the best N4 realization) against the CFD predictions in \cref{45_abl}. An increased uncertainty is recorded for the blowing heat flux predictions, also depicted in the average error over the 20 folds (\cref{abl_tab}). Assigning a Mach number color bar to the predictions indicates that the N4 error is randomly distributed over the free stream conditions for all heat flux components and the mass flux at the wall. However, in combination with \cref{abl_tab}, the N4 network predictions are considered sufficiently accurate. A detailed analysis of the results over reentry trajectories as well as an investigation of the carbon ablation mechanism is given in the following section.

\FloatBarrier
\section{Application to space debris trajectories analysis}
\label{S:5}

\addAT{We here present an application for data-driven models N1 to N4, for the computation of selected quantities of interest along precomputed trajectories. The trajectory data, input to the data-driven models, were obtained through the ESA's Debris Risk Assessment and Mitigation Analysis (DRAMA) software \protect\cite{DRAMA2019,Martin2004} and correspond to the reentry of the following debris-like spherical tanks:}

\begin{figure}
     \centering
     \begin{subfigure}[b]{0.48\textwidth}
         \centering
         \includegraphics[width=\textwidth]{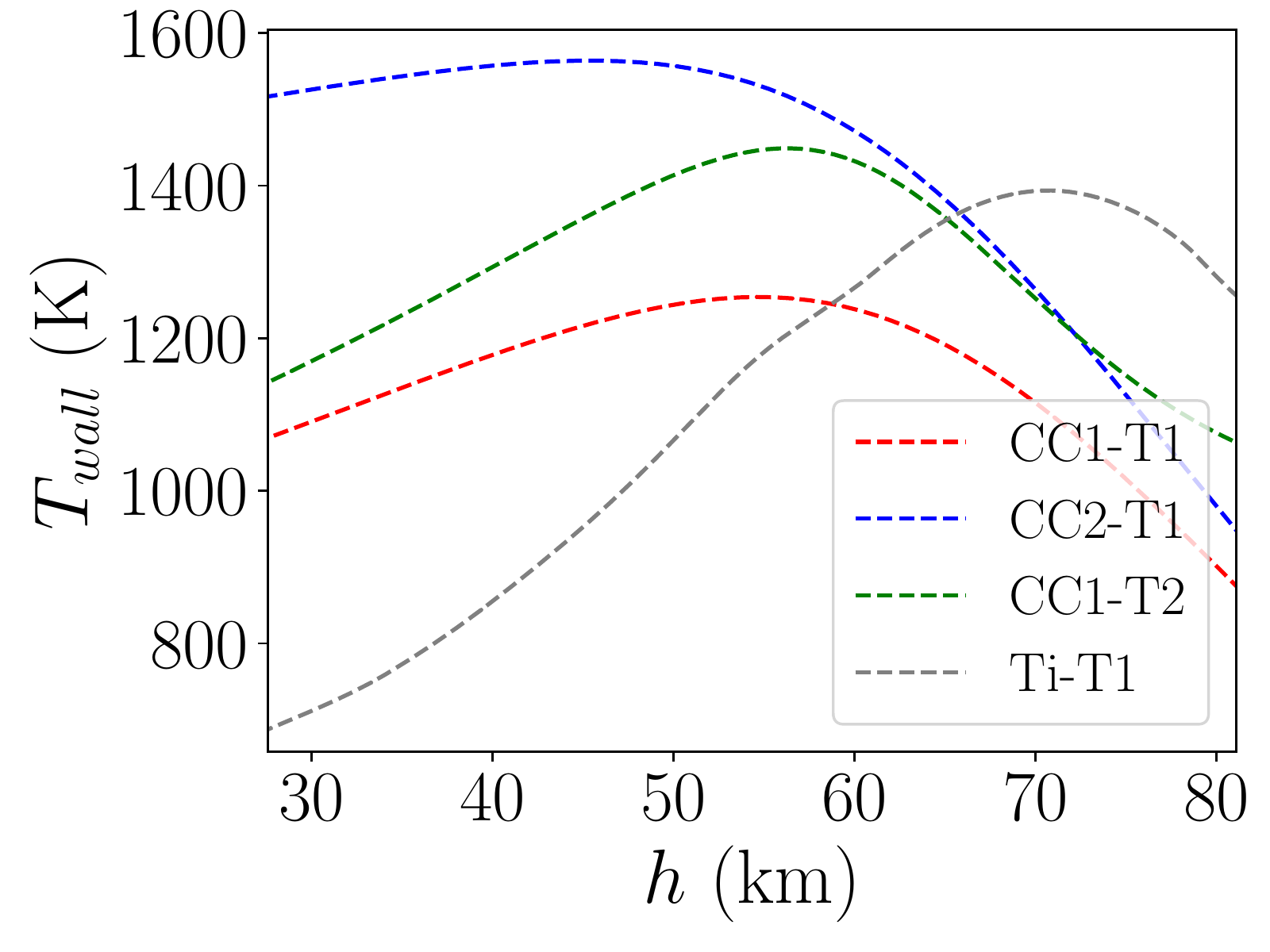}
         \caption{Temperature}
         \label{fig:drama_temp}
     \end{subfigure}
     \hfill
     \begin{subfigure}[b]{0.48\textwidth}
         \centering
         \includegraphics[width=\textwidth]{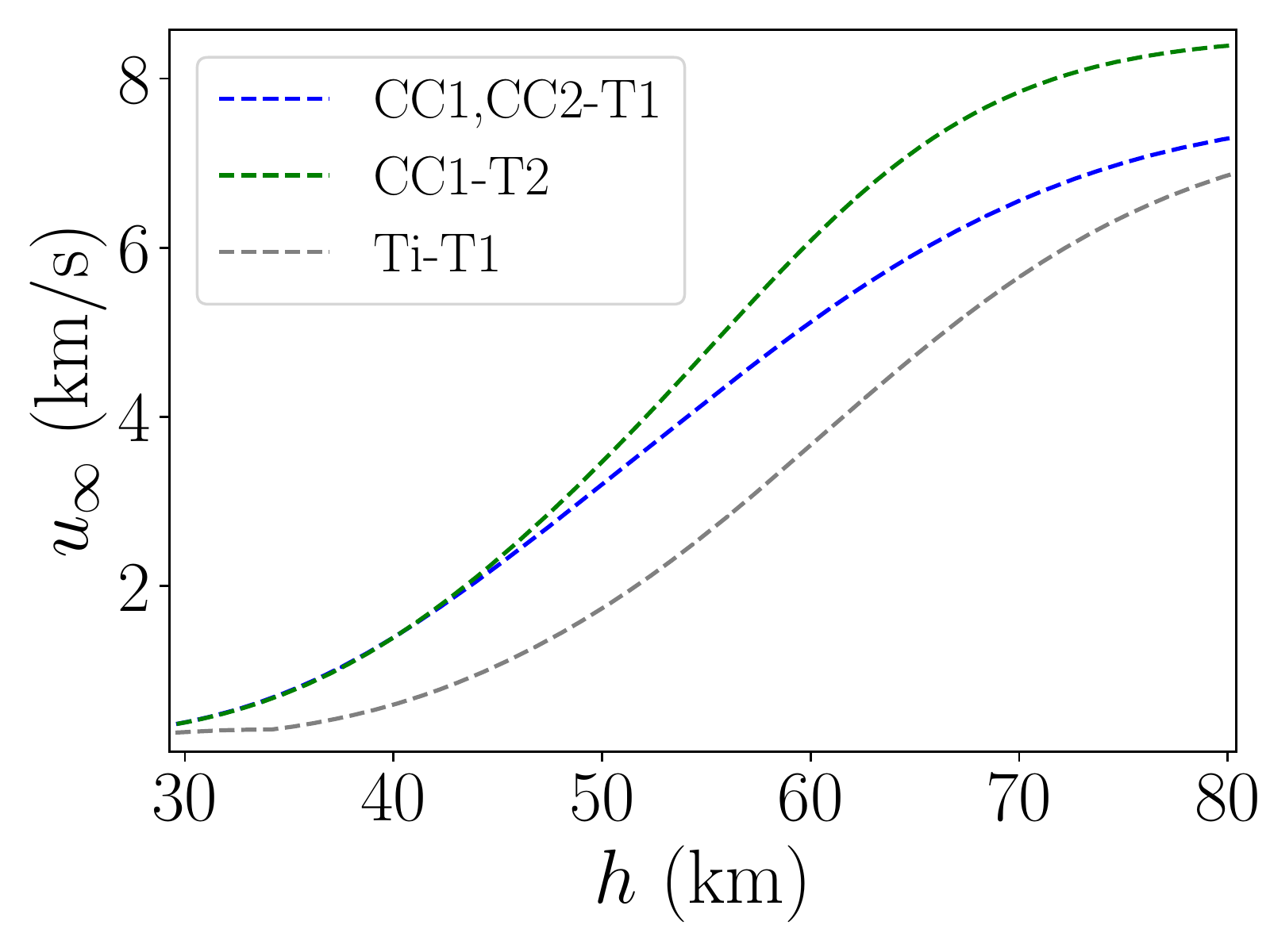}
         \caption{Velocity}
         \label{fig:drama_vel}
     \end{subfigure}
      \caption{DRAMA's trajectories wall temperature and velocity versus altitude.}
      \label{fig:cas}
\end{figure}

\begin{itemize}
    \item {CC1: carbon-carbon spherical tank with 0.5-m radius and 1-cm wall thickness: material density \SI{1770.0}{kg/m^3}, tank mass 54.501 kg.}
    \item CC2: carbon-carbon tank identical to CC1 but with an artificially-reduced surface emissivity with respect to the value provided in the DRAMA database.
    \item Ti: ``titanium'' (TiAl6v4) spherical tank with 0.28-m radius and 1.5-mm wall thickness: material density \SI{4417.0}{kg/m^3}, tank mass 6.493 kg.
\end{itemize}
\addAT{A first reentry condition was used to compute the trajectories of the above three objects using DRAMA \textemdash obviously, CC1 and CC2 have identical trajectories, but it will be clarified in the following why two separate simulations were performed. This reentry condition, labeled as T1, consisted in an reentry velocity of \num{7.7} \si{km/s} from \num{125} \si{km} altitude with a flight path angle of approximately \num{-0.1}\si{\degree}.
A second reentry condition, T2, characteristic of a steeper reentry, was used to compute a second trajectory for the CC1 object. This second condition consisted in: reentry velocity of \num{8.5} \si{km/s}, flight path angle of \num{-3}\si{\degree}, and initial altitude of \num{102} \si{km}. 
Besides demonstrating the ability to compute critical flow field properties, the computation of surface quantities using models N3 and N4 (Sections \ref{Conv_nonab} and \ref{wall_ab}) were also of interest. 
To compare the computed stagnation-point heat flux with DRAMA's correlation-based estimates, the surface temperatures computed by DRAMA \textemdash, i.e., through the lumped mass model using the incoming surface average heat flux, the available material properties and the defined object geometry \textemdash were also used as input to N3 and N4. 
This was considered appropriate to mimic a possible coupling between DRAMA and the data-driven models proposed here, where the latter provides surface heat-flux estimates as input for the computation of the object temperature performed by DRAMA.
Following this, the interest in studying the CC1 and CC2 objects under the T1 conditions was to assess the influence of different surface temperatures on the stagnation-point heat flux evaluated through the data-driven model.}

\subsection{Stagnation-line and stagnation-point quantities for nonreactive surfaces}
\label{nabps}

It should be recalled that, by construction, the validity of the data-driven models is limited to the range of flight and surface conditions given in \mref{tabin}. Therefore, only the trajectory intervals with quantities respecting these limits were evaluated. The main limitation appeared to be the lower bound of the flight velocity in \mref{tabin}, i.e., \num{3} \si{km/s}, which limited the application of the data-driven models to trajectory points above approximately \num{50} \si{km} and  \num{60} \si{km} for the carbon-carbon and titanium tanks, respectively.

\Mref{nonab_DRAMA} shows the predictions of N1 and N2 for three selected quantities of interest --i.e., temperature, pressure and atomic oxygen density-- at $\hat{x}=0.02$ along the computable part of each DRAMA's trajectory, along with corresponding uncertainties (Eqs. \ref{term_1}, \ref{term_2}). Note that this non-dimensional distance (see \ref{subsPS}) corresponds to different dimensional distances for objects of different radii, i.e., CCx vs Ti. However, these points always fall within the chemical equilibrium zone past the shock for all objects and analyzed trajectories.

For the T1 reentry conditions, objects CC1 and CC2, going through identical entry trajectories, show practically the same values for the three analyzed quantities. A detectable yet minor difference is visible only in the temperature values. This difference is due to the different surface temperatures imposed on the two objects in the data-driven model. Nevertheless, this deviation is practically negligible and falls well within the given confidence interval of the data-driven reconstruction of the flow-field temperature.

When reentering at condition T2, CC1 presents the highest post-shock temperature, pressure, and atomic-oxygen density, due to the more severe entry conditions. On the other hand, when Ti reenters at condition T1, it presents the least severe post-shock quantities, given that it is slowed down quite early along its trajectory compared to C1 and C2.

\begin{figure}[!htbp]
\center
\includegraphics[width=0.5\columnwidth]{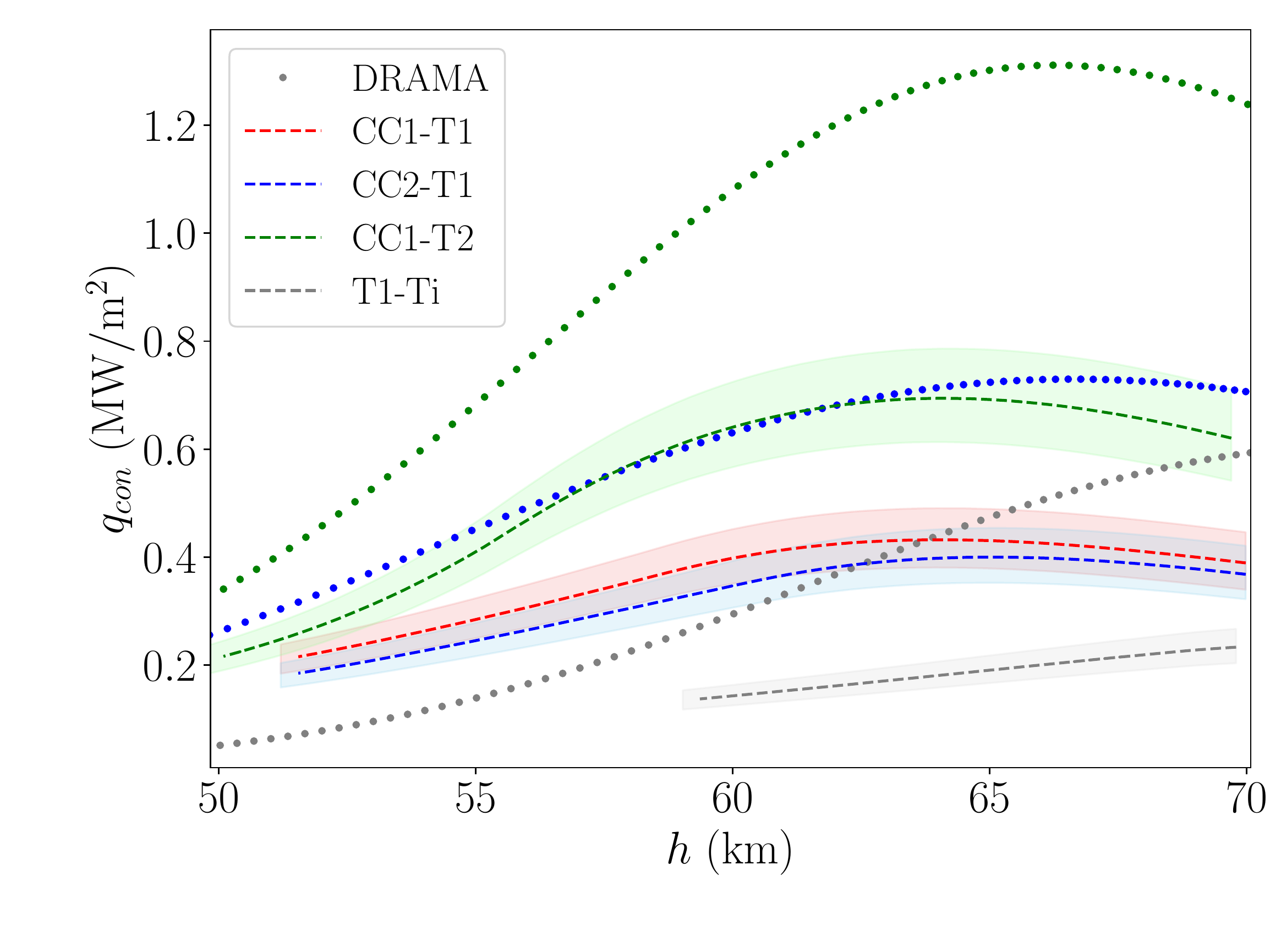}
\caption{Comparison between semiempirical correlation and data-driven predictions: The N3 model captures the wall temperature effect (90 $\%$ uncertainty bounds).}
\label{fig:qc_NA}
\end{figure}

Predictions of the incoming conductive heat flux at a non-ablative wall were made by means of model N3, described in Section \ref{Conv_nonab}. \Mref{fig:qc_NA} shows the comparison of this quantity as given by DRAMA (i.e., based on the Detra-Kemp-Riddell (DKR) correlation) and as computed by N3, with prediction uncertainties. One can quickly notice the significant difference between the estimates for any of the calculated objects and conditions. Besides this nonnegligible quantitative disagreement, which suggests errors exceeding 100\% in the heat-flux estimate of DRAMA, a slight qualitative displacement of the peak heating to lower altitudes for CC1 (both trajectories) and CC2 can also be detected. However, the same qualitative consideration does not hold for Ti due to peak heating outside the domain of applicability of N3.
Last, we would note that the difference between the results of CC1 and CC2 for the entry condition T1 is due exclusively to the different surface temperatures of these objects. The surface temperature is not considered in the DKR correlation. Hence DRAMA returned the same heat flux for these two objects.

\newcommand{\figOne}{FIGURES/ch4/T_DRAMA_traj.pdf}
\newcommand{\captionOne}{Temperature}
\newcommand{\labelOne}{}
\newcommand{\figTwo}{FIGURES/ch4/p_DRAMA_traj.pdf}
\newcommand{\captionTwo}{Pressure}
\newcommand{\labelTwo}{}
\newcommand{\figThree}{FIGURES/ch4/rhoO_DRAMA_traj.pdf}
\newcommand{\captionThree}{\ce{O} density}
\newcommand{\labelThree}{}
\newcommand{\captionGlobal}{N1, N2 predictions for temperature, pressure and atomic oxygen density across four trajectories computed with the ESA's DRAMA code, at $\hat{x}=0.02$. The selected position corresponds to the post-shock chemical equilibrium region of the flow (90 $\%$ uncertainty bounds).}
\newcommand{\labelGlobal}{nonab_DRAMA}
\begin{figure}
     \centering
     \begin{subfigure}[b]{0.48\textwidth}
         \centering
         \includegraphics[width=\textwidth]{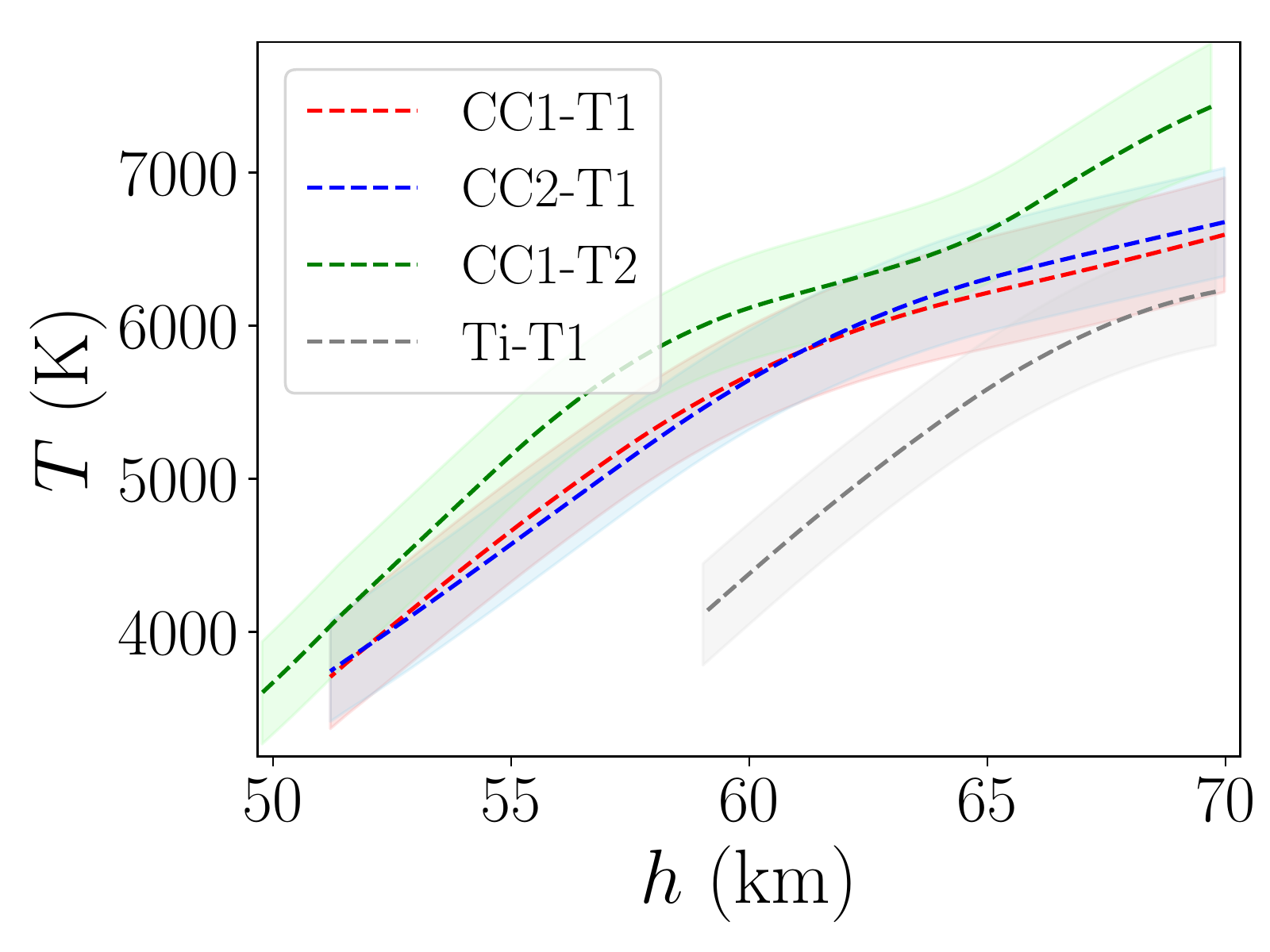}
         \caption{\captionOne}
         \label{\labelOne}
     \end{subfigure}
     \hfill
     \begin{subfigure}[b]{0.48\textwidth}
         \centering
         \includegraphics[width=\textwidth]{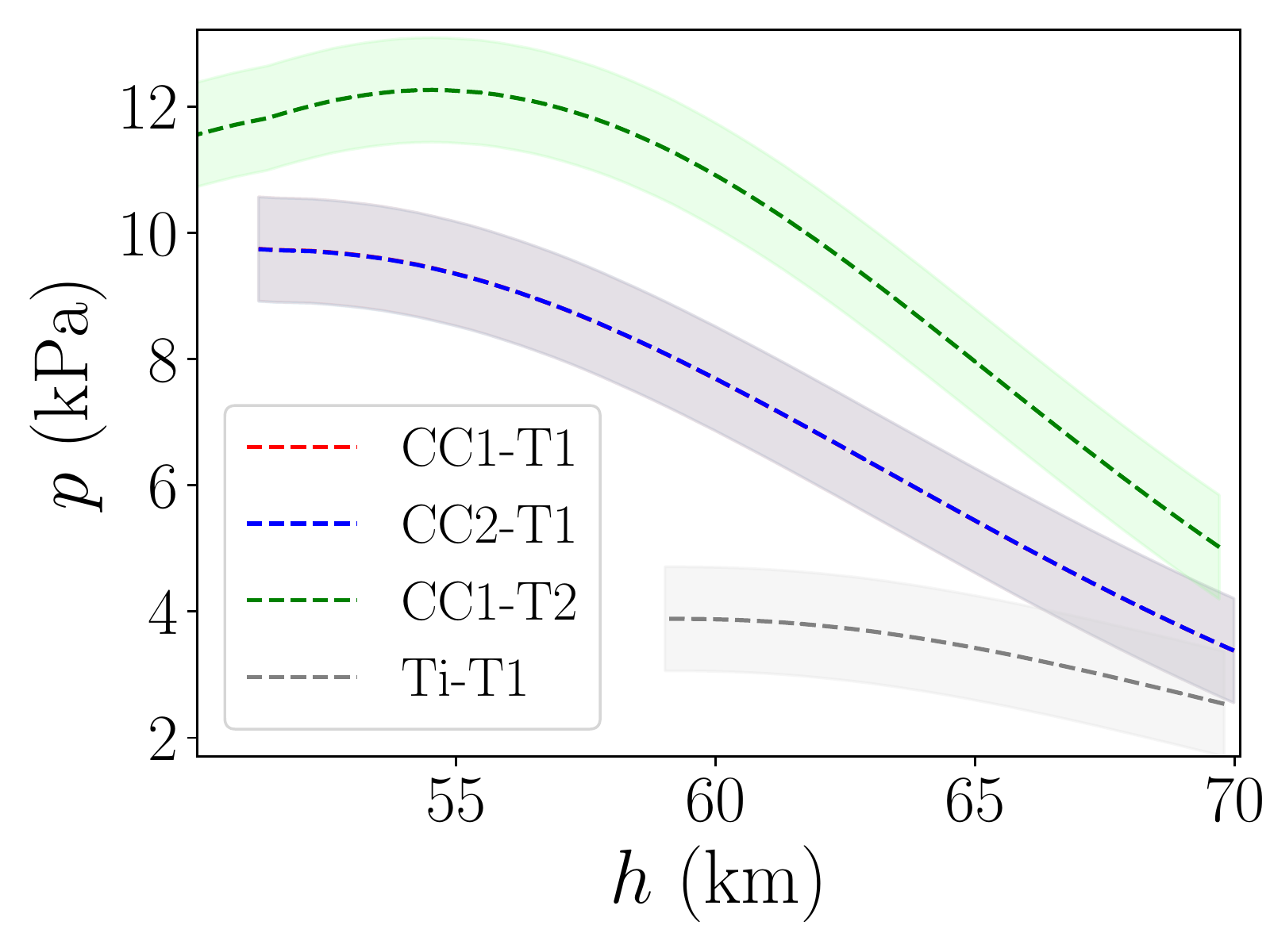}
         \caption{\captionTwo}
         \label{\labelTwo}
     \end{subfigure}
     \hfill \\ \vspace{5mm}
     \begin{subfigure}[b]{0.48\textwidth}
         \centering
         \includegraphics[width=\textwidth]{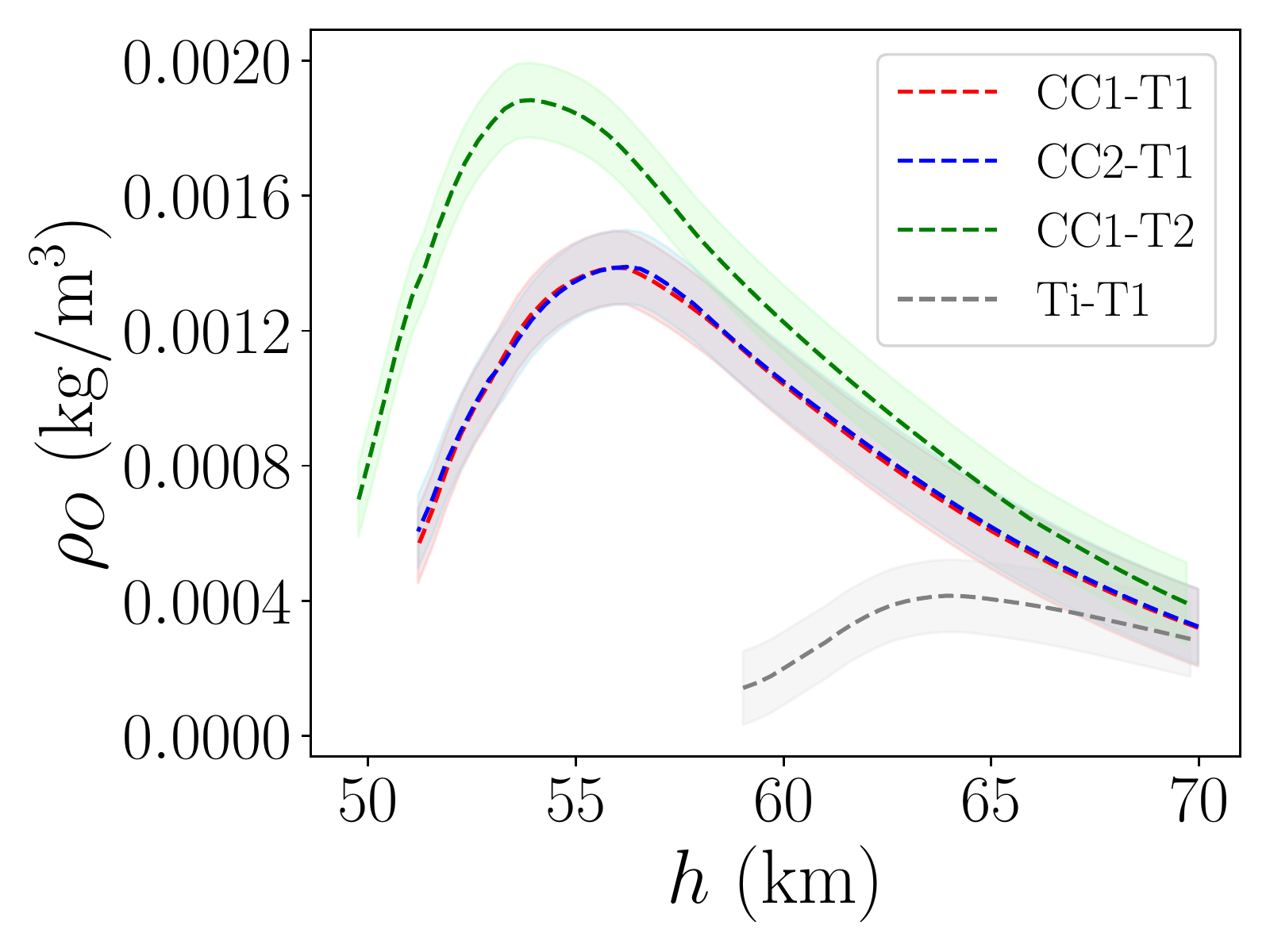}
         \caption{\captionThree}
         \label{\labelThree}
     \end{subfigure}
      \caption{\captionGlobal}
      \label{\labelGlobal}
\end{figure}

\subsection{Stagnation-point quantities for ablative carbon surfaces}
\label{abps}

\Mref{drama_abl} presents the output of the carbon ablation data-driven model (N4) for object CC1 (trajectories T1 and T2) and object CC2 (trajectory T2).
Like the nonreacting surface analysis, the object CC1 going through trajectory T2 encounters the most severe conditions in terms of incoming aerothermal heat flux, composed here by conduction and diffusion. The outgoing blowing heat flux ($q_{\mr{bl}}$) remains significantly lower than the incoming aerothermal one for all the cases.
The surface mass blowing in the case of aerothermal ablation is mainly driven by the incoming conductive heat flux, while the diffusive heat flux is due to the injection of the ablative species in the boundary layer and the contribution from the considered catalytic recombination of atomic nitrogen.
These dependencies appear evident in \mref{drama_abl}, where the mass blowing flux (\mref{fig:abla_mass}) and the diffusive flux (\mref{fig:abla_diff}) show very similar trends, both in qualitative agreement with the conductive heat flux (\mref{fig:abla_cond}).
This latter, similarly to the non-reacting surface (\mref{fig:qc_NA}), appears well correlated with the velocity profiles shown in \mref{fig:drama_vel}.
The blowing heat flux (\mref{fig:abla_blow}), directly influenced by both the mass blowing flux (\mref{fig:abla_mass}) and the surface temperature (\mref{fig:drama_temp}), shows a stronger decay for CC1-T2, resembling its surface temperature variation and approaching CC1-T1 towards the lower limit of the analyzed trajectory patch. 

\renewcommand{\figOne}{FIGURES/ch4/qcond_DRAMA.pdf}
\renewcommand{\captionOne}{Conductive heat flux}
\renewcommand{\labelOne}{fig:abla_cond}
\renewcommand{\figTwo}{FIGURES/ch4/qdif_DRAMA.pdf}
\renewcommand{\captionTwo}{Diffusive heat flux}
\renewcommand{\labelTwo}{fig:abla_diff}
\renewcommand{\figThree}{FIGURES/ch4/qabl_DRAMA.pdf}
\renewcommand{\captionThree}{Blowing heat flux}
\renewcommand{\labelThree}{fig:abla_blow}
\newcommand{\figFour}{FIGURES/ch4/mabl_DRAMA.pdf}
\newcommand{\captionFour}{Mass blowing flux}
\newcommand{\labelFour}{fig:abla_mass}
\renewcommand{\captionGlobal}{N4 predictions for mass blowing rate, diffusive and blowing heat flux at the wall (90 $\%$ uncertainty bounds).}
\renewcommand{\labelGlobal}{drama_abl}
\begin{figure}[tb]
     \centering
     \begin{subfigure}[b]{0.48\textwidth}
         \centering
         \includegraphics[width=\textwidth]{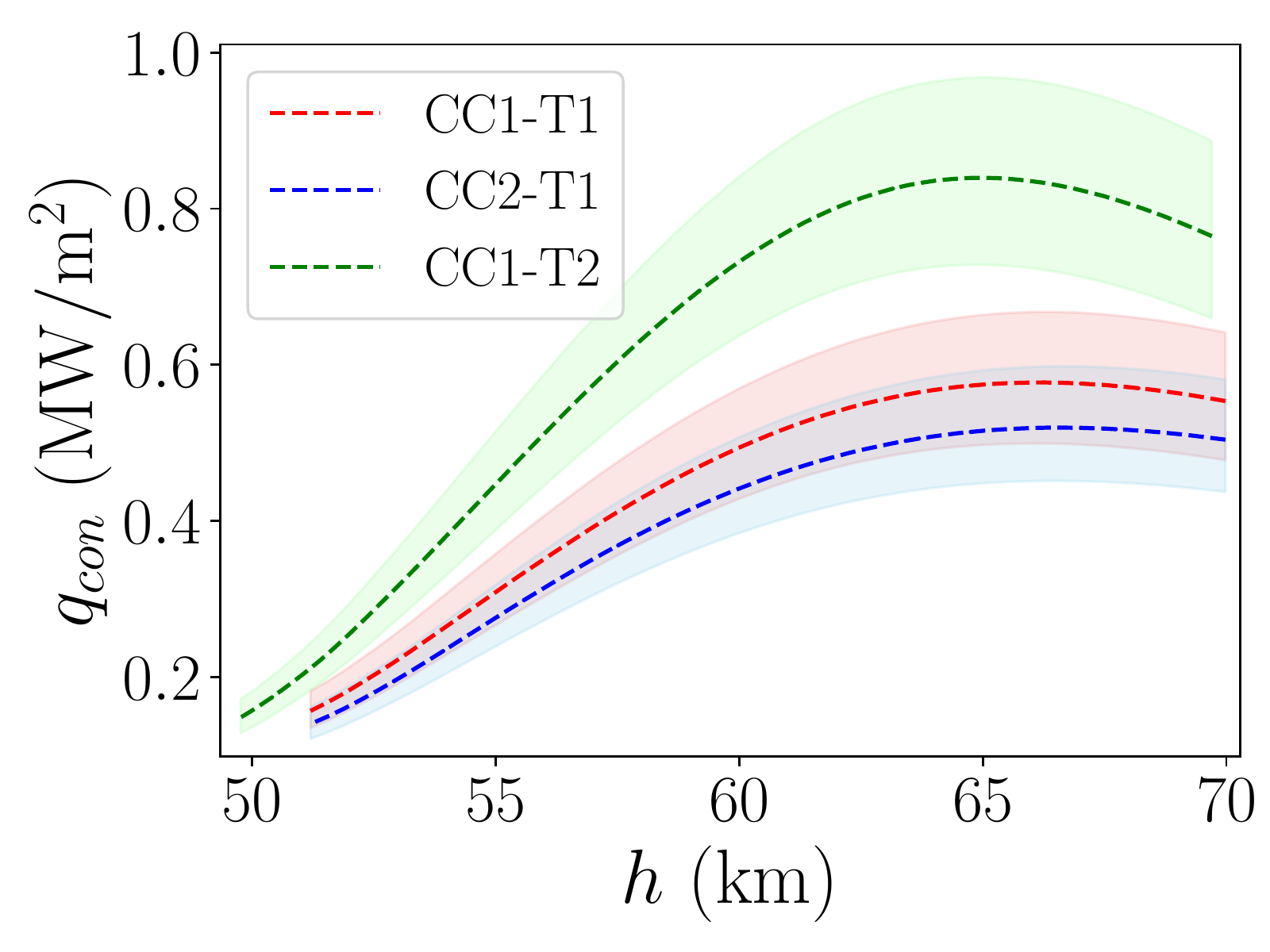}
         \caption{\captionOne}
         \label{\labelOne}
     \end{subfigure}
     \hfill
     \begin{subfigure}[b]{0.48\textwidth}
         \centering
         \includegraphics[width=\textwidth]{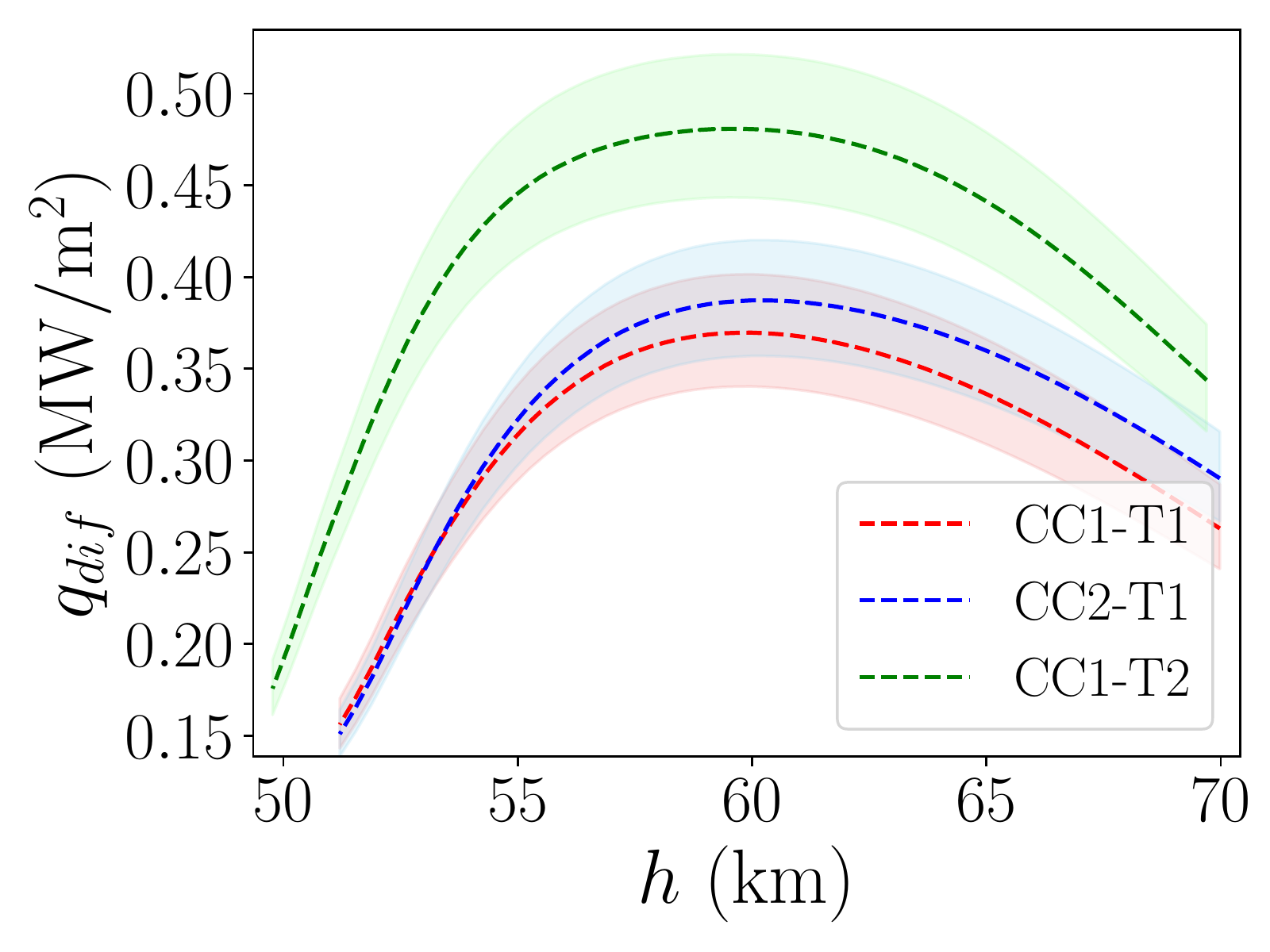}
         \caption{\captionTwo}
         \label{\labelTwo}
     \end{subfigure}
     \hfill \\ \vspace{5mm}
     \begin{subfigure}[b]{0.48\textwidth}
         \centering
         \includegraphics[width=\textwidth]{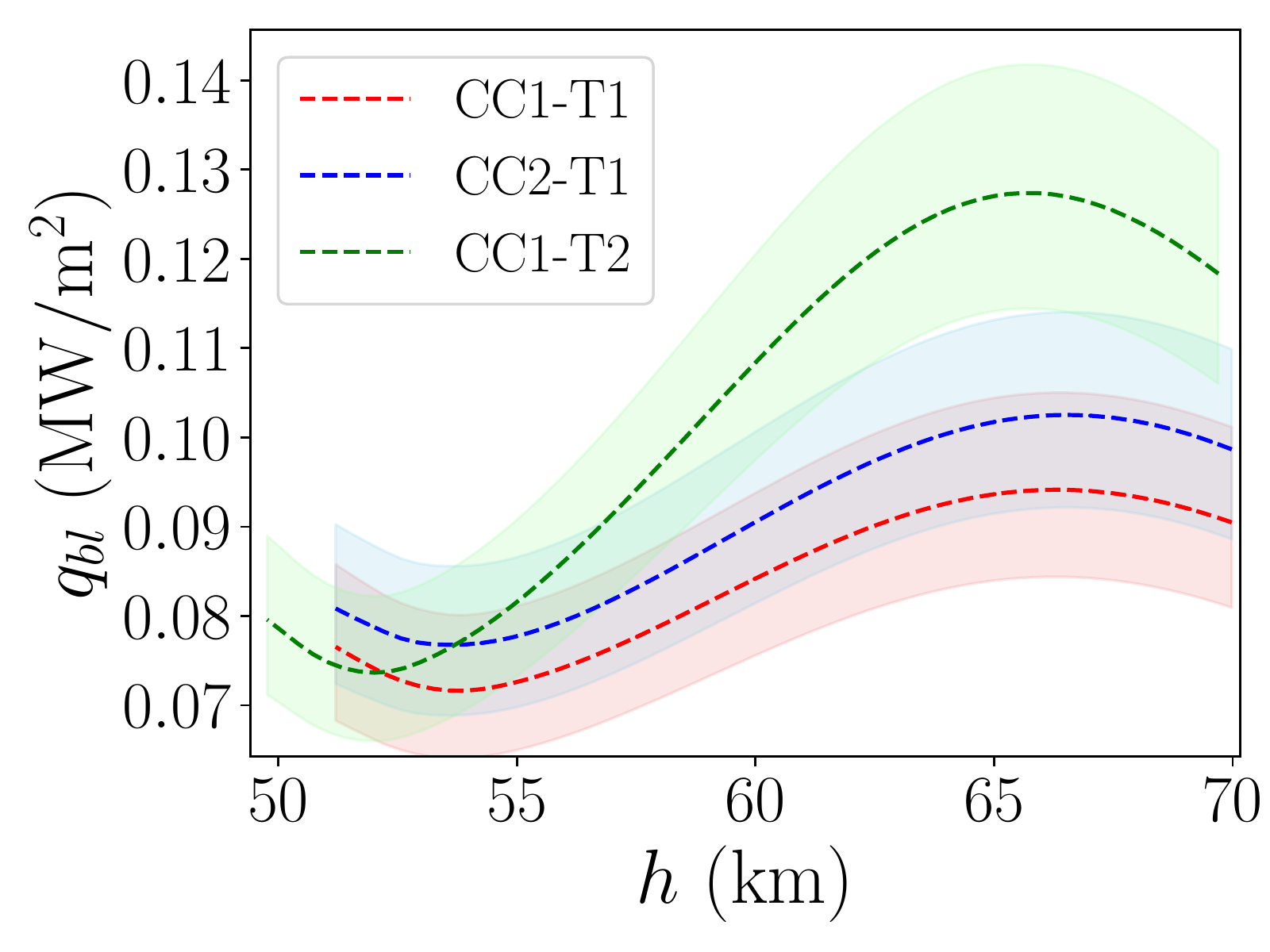}
         \caption{\captionThree}
         \label{\labelThree}
     \end{subfigure}
     \hfill
     \begin{subfigure}[b]{0.48\textwidth}
         \centering
         \includegraphics[width=\textwidth]{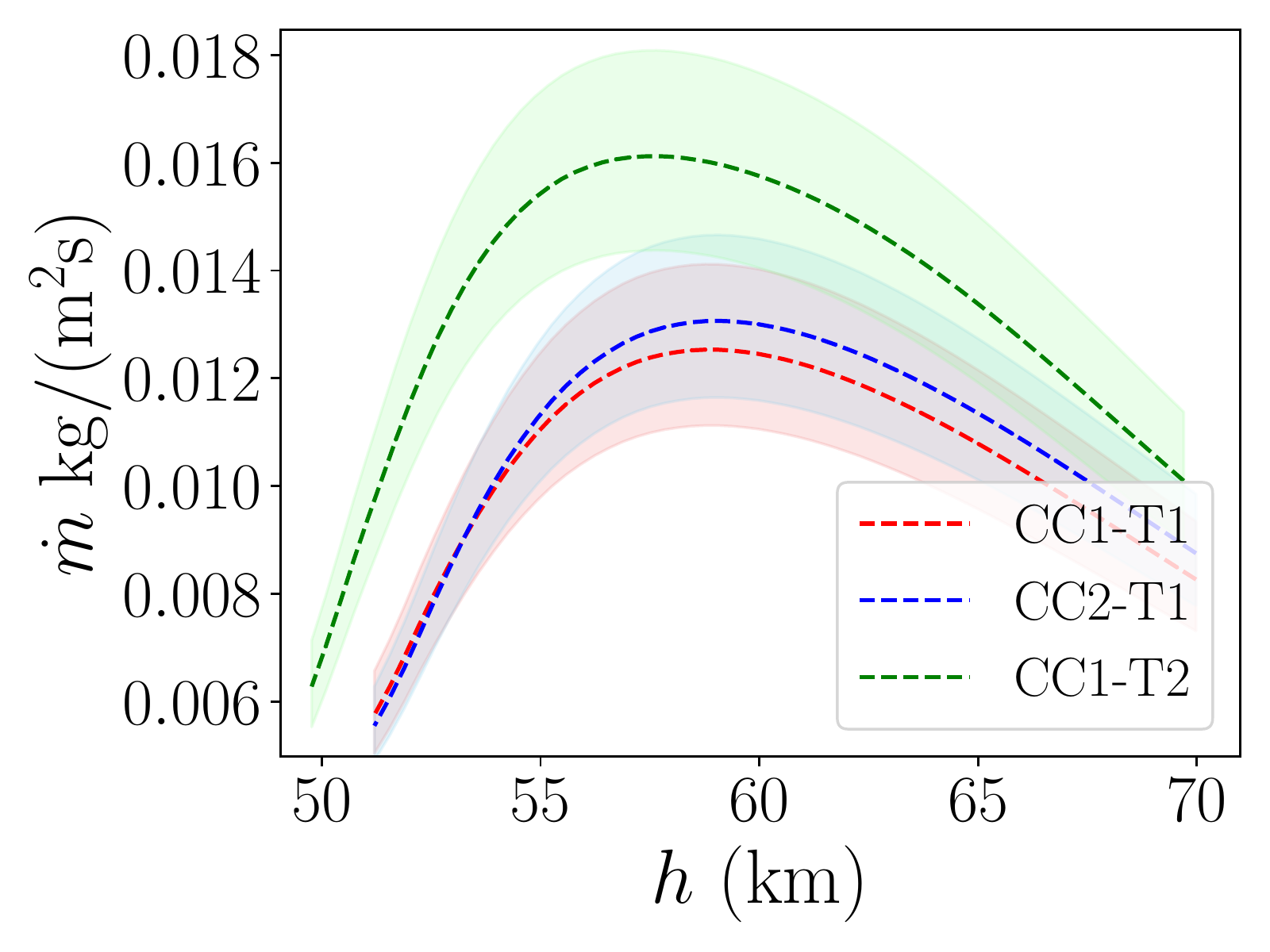}
         \caption{\captionFour}
         \label{\labelFour}
     \end{subfigure}
      \caption{\captionGlobal}
      \label{\labelGlobal}
\end{figure}

Finally, \mref{fig:cas} presents a comparison of the conductive and total heat fluxes between the nonreactive and the ablative surfaces.
Here we observe that higher conductive heat flux is established when ablation is included.
Generally speaking, this appears counterintuitive since it is well known that the injection of ablative species in the boundary layer generates the so-called ``blockage effect'', which should reduce the incoming conduction \cite{Turchi2015}.
However, further verification studies that compared the results obtained for the two surfaces on a representative trajectory point with the CFD code used to generate the training data confirmed the predictions by the data-driven models N3 and N4.
The analysis of the CFD ablative results suggests that one specific gas reaction is responsible for this unusual behaviour. In particular, the third-body dissociation of \ce{CN} (i.e., \ce{CN + M <=> C + N + M}) seems to be extremely active in the vicinity of the surface, ``neutralizing'' the possible beneficial effect of the blowing with a strong release of energy that overall increases the conduction to the surface.
The expected trend, with the conductive heat flux for the ablative surface falling below the one for the case of nonreactive surface, is retrieved only towards the lower limit of the analyzed trajectory patch, where the increase of pressure and the decrease of surface temperature probably prevent the \ce{CN} dissociation. Interestingly, in this case, the blockage seems efficient despite the low magnitude of the mass blowing flux (\mref{fig:abla_mass}).
The picture is similar if we analyze the total heat flux (\mref{fig:ablaNOabla_tot}), which corresponds to the sole conduction for the nonreactive surface.
As evident from \mref{drama_abl}, the diffusive heat flux generated by the mass injection outweighs the cooling effect of the blowing heat flux. Therefore, the total heat flux in the case of ablation is significantly higher than the one obtained for the nonreactive surface.

The analysis of the results for the presented application suggests that the developed models provide a robust method for the predictions of relevant quantities on both nonreactive and ablative carbon surfaces.
This approach could also offer a valuable way to predict wall heat and mass fluxes along reentry trajectories based on the above results. It is particularly interesting to highlight that the developed model helped identify an unusual behaviour along with reentry (\mref{fig:cas}). The achieved accuracy and predictions speed-up allowed for capturing this trend, the cause of which was then more closely examined through CFD.
Eventually, the method could be used as an alternative to semiempirical correlations, directly coupled to reentry analysis tools for space debris, e.g., DRAMA. 

\renewcommand{\figOne}{FIGURES/ch4/comparative_abl_nonabl.pdf}
\renewcommand{\captionOne}{Conductive heat flux}
\renewcommand{\labelOne}{fig:ablaNOabla_cond}
\renewcommand{\figTwo}{FIGURES/ch4/tot_comparative_abl_nonabl.pdf}
\renewcommand{\captionTwo}{Total heat flux}
\renewcommand{\labelTwo}{fig:ablaNOabla_tot}
\renewcommand{\captionGlobal}{Comparison between ablative (N4) and nonablative (N3) regime predictions for wall conductive and total heat flux (90 $\%$ uncertainty bounds).}
\renewcommand{\labelGlobal}{fig:cas}
\begin{figure}
     \centering
     \begin{subfigure}[b]{0.48\textwidth}
         \centering
         \includegraphics[width=\textwidth]{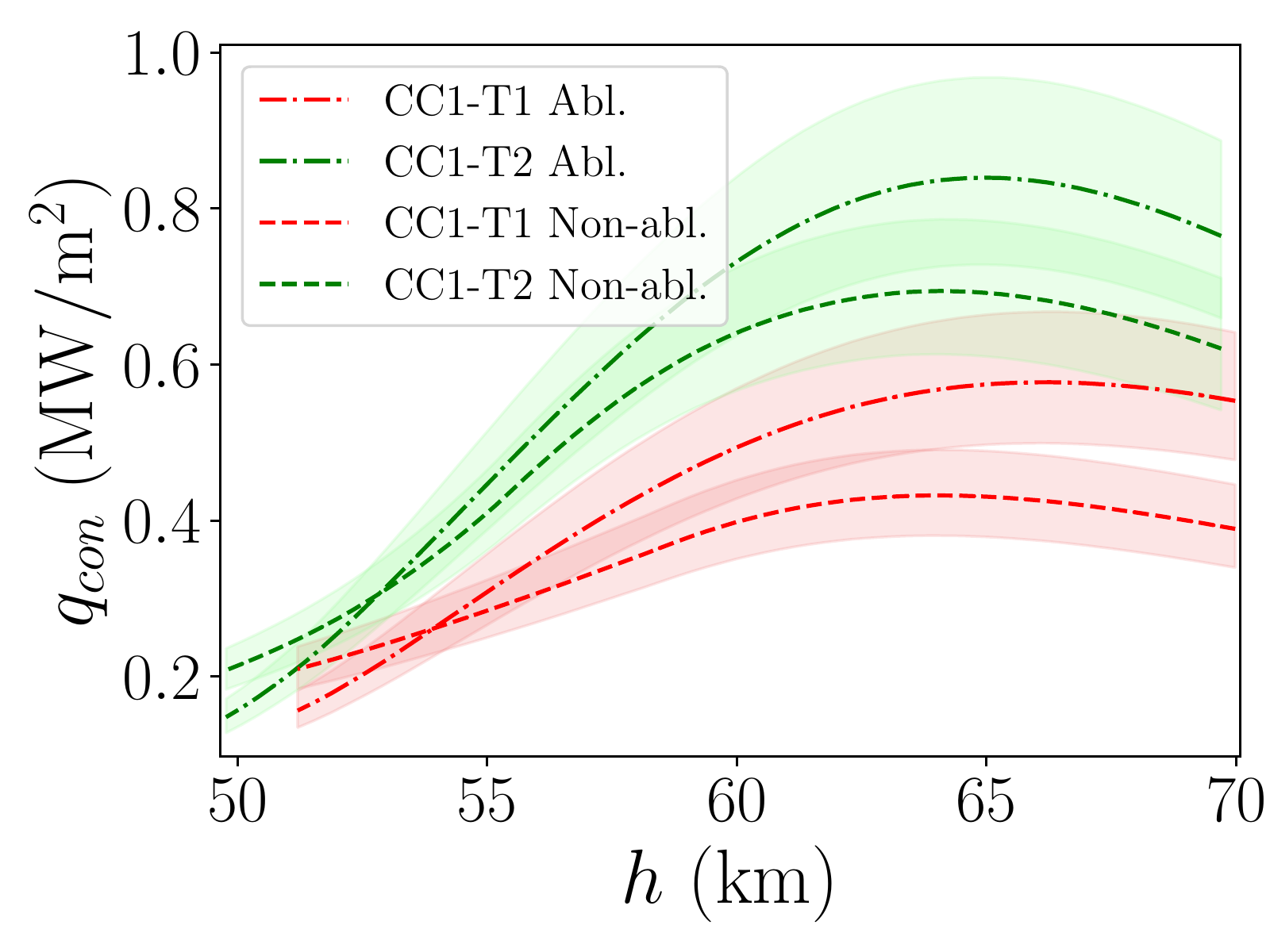}
         \caption{\captionOne}
         \label{\labelOne}
     \end{subfigure}
     \hfill
     \begin{subfigure}[b]{0.48\textwidth}
         \centering
         \includegraphics[width=\textwidth]{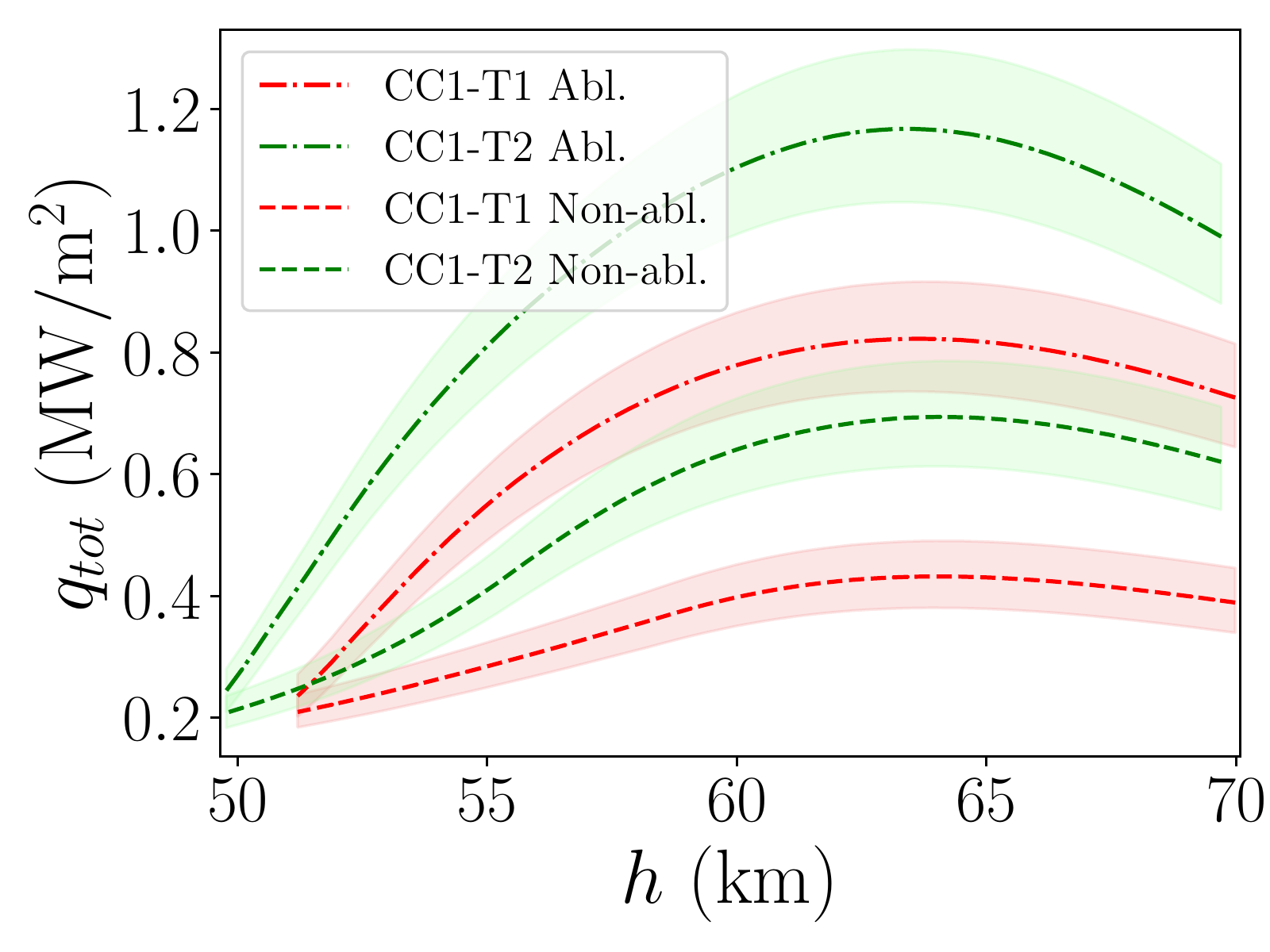}
         \caption{\captionTwo}
         \label{\labelTwo}
     \end{subfigure}
      \caption{\captionGlobal}
      \label{\labelGlobal}
\end{figure}

\section{Conclusion}
\label{S:6}

We presented data-driven models for the wall fluxes and the primary flow quantities along the stagnation streamline of a hypersonic blunt body atmospheric reentry.

Nondimensional analysis of the governing equations was carried out to set the problem in dimensionless form. A data collection campaign was performed with a one-dimensional CFD solver over the range of inputs relevant to atmospheric reentry, providing the foundation of the data-driven models. We also extended the methodology from nonablative cases to the prediction of wall fluxes in the regime of carbon ablation. Ad-hoc parametric representations of the flow quantities along the stagnation line and species partial densities were then proposed. 

Curve-fitting the functions to the corresponding CFD solutions yields an average $R^2$ coefficient higher than $96 \%$ for all examined quantities. Subsequently, four neural networks (N1, N2, N3 and N4) were proposed to construct data-driven functions between the flight parameters and the outputs of interest. These were designed to predict the flow field representation parameters for mixture quantities, species partial densities, nonablative heat flux at the wall, and carbon ablation heat and mass fluxes at the wall. 

All networks were trained in an ensemble learning approach to estimate the epistemic uncertainties. Promising results were obtained during the validation phase of the ANNs, with average accuracy above $95 \%$ for the best N1 realization, $86 \%$ for the best N2 realization and over $80 \%$ for wall fluxes predictions (N3 and N4). Furthermore, the potential of the developed models was finally showcased along reentry trajectories computed with the ESA DRAMA code. From the speed-up in reentry predictions, valuable insights on reentry mechanisms were obtained, along with a mismatch with currently employed correlations for heat fluxes was reported.

The developed methodology could comprise an independent tool for fast and accurate predictions of the stagnation streamline flow field and wall heat and mass fluxes. Furthermore, by reducing computational overhead through closed-form input-to-output functions, a future coupling with multiphysics reentry analysis tools of thermal analysis or trajectory dynamics is made feasible. Other reentry regimes of ablation and evaporation of materials with high practical interest could also be examined following the proposed methodology, with minor modifications considering the analyzed parametrization. 

Finally, an extension of the current one-dimensional formulation could be of interest to accommodate the more realistic three-dimensional flows over reentering bodies of different shapes. Overall, the presented data-driven model provides an efficient alternative to CFD or semianalytical correlations, enhancing multidisciplinary analyses, optimization studies or real-time applications of a blunt body hypersonic reentry.

\appendix
\section*{Appendices}
\label{appex}

\subsection{Derivation of nondimensional system of equations}
\label{appex1}

The derivation of the non-dimensional form of the system of equations \ref{syst} analyzed in Refs. \cite{Klomfass} and \cite{Munafo2014} is hereby presented. Since the final solution of the time-marching strategy is steady-state, time dependencies can be neglected. All used dimensionless quantities used are given in Table~\ref{refq}.

\begin{table}[!htb]
\begin{center}
\caption{Nondimensional formulation for system \ref{syst}:   ${D_i}$ is the diffusion coefficient, ${k_i}$ is a reference reaction rate for each species.}
\begin{tabular}{lllll}
 &  Quantity & Reference value \\
\hline \hline
 & $r$ & $R$   \\
 & $u$ &  $u_{\infty}$  \\
 & $v$ & $u_{\infty}$  \\
 & $T$ & $(T_{\mr{wall}} - T_{\infty})$   \\
 & $p$ & ${\rho_{\infty} u_{\infty}^{2}}$   \\
 & $\rho_i$ & $\rho_\infty$   \\
 & $j_i$ & $D_i/R$  \\
 & $\omega_i$ & $k_i$  \\
 \hline \hline
\end{tabular}
\label{refq}
\end{center}
\end{table}

\noindent Considering the mass conservation equation for species $i$, density variations are manipulated as follows:

\begin{equation}
\frac{\delta \rho}{\rho} \approx \frac{\delta p}{a^{2} \rho}=M^{2}\,.
\end{equation}

\noindent Based on the above, the non-dimensional form of the mass conservation equation for species $i$, becomes:

\begin{equation}
\label{cont1}
{M}^{2} \frac{d \hat{\rho}_{i}}{d \hat{r}} \hat{u}+\frac{d \hat{u}}{d \hat{r}} \hat{\rho_{i}}+2 \hat{\rho_{i}} \frac{(\hat{u}+\hat{v})}{\hat{r}}-\frac{1}{Bo_i}\left(\frac{d \hat{j}_{i}}{d \hat{r}}-\frac{2}{\hat{r}} \hat{j}_{i}\right)=D a_{i} \hat{\omega_{i}}\,,
\end{equation} having introduced the Damköhler numbers $Da_i={k_i R}/{\rho_\infty u_\infty}$ and the Bodenstein numbers $Bo_i={D_i}/{\rho_\infty u_\infty R}$ for each species, accounting for their chemical and transport properties accordingly.

\noindent Substituting the stress tensor components $\tau_{\phi\phi}$, $\tau_{r\phi}$, $\tau_{rr}$ by their expression through Newton's constitutional law, as extensively presented in \cite{Klomfass}, the first momentum conservation equation is given by

\begin{equation}
\label{mom1}
\frac{d \left(\hat{\rho} \hat{u}^{2}\right)}{d \hat{r}}+\frac{\hat{\rho} \hat{u}(\hat{u}+\hat{v})}{\hat{r}}+\frac{d \hat{p}}{d \hat{r}}-\frac{1}{R e} \hat{A}=0 \,,
\end{equation} where $Re={u_\infty R}/{\nu}$ is the Reynolds number. The term $\hat{A}$ originates from the stress tensor components and is given here for completeness:

\begin{equation}
\label{mom2}
\hat{A}=\frac{4}{3}\left(\frac{d^{2} \hat{u}}{d \hat{r}^{2}}-\frac{d\left(\frac{\hat{u}+\hat{v}}{\hat{r}}\right)}{d \hat{r}}\right)+\left(4 \frac{d \hat{u}}{d \hat{r}} \frac{1}{\hat{r}}-6 \frac{\hat{u}+\hat{v}}{\hat{r}^{2}}+2 \frac{d \hat{v}}{d \hat{r}} \frac{1}{\hat{r}}\right)\,.
\end{equation}

 Similar steps are taken for the second momentum conservation equation, obtaining:

\begin{equation}
\frac{d (\hat{\rho} \hat{u} \hat{v})}{d \hat{r}}+3 \frac{\hat{\rho} \hat{v}(\hat{u}+\hat{v})}{\hat{r}}-2 \frac{\hat{p}}{\hat{r}}-\frac{1}{R e} \hat{B}=0 \,.
\end{equation}

The term $\hat{B}$ is here given by:  
\begin{equation}
\hat{B}=\left(\frac{d^{2} \hat{v}}{d \hat{r}^{2}}-\frac{d \left(\frac{\hat{u}+\hat{v}}{\hat{r}}\right)}{d \hat{r}}\right)+\left(9 \frac{d \hat{v}}{d \hat{r}} \frac{1}{\hat{r}}-11 \frac{\hat{u}+\hat{v}}{\hat{r}^{2}}+2 \frac{d \hat{u}}{d \hat{r}} \frac{1}{\hat{r}}\right)\,.
\end{equation}

 Finally, the energy conservation equation is considered. After the necessary manipulation, the non-dimensional version of the energy equation will be:

\begin{equation}
\label{ener}
\frac{d (\hat{\rho} \hat{u} \hat{\theta})}{d \hat{r}}+\frac{2 \hat{\rho} \hat{\theta}(\hat{u}+\hat{v})}{\hat{r}}-\frac{1}{P e}\left(2 \frac{d^{2} \hat{\theta}}{d \hat{r}^{2}}+\frac{1}{\hat{r}} \frac{d \hat{\theta}}{d \hat{r}}\right)-\sum_{i} \frac{C_{p_{i}}}{C_{p}}\left(\frac{d \hat{\jmath}_{i}}{d \hat{r}}+\hat{\jmath}_{i} \frac{1}{\hat{r}}\right)-\frac{E c}{R e}\left(\frac{4}{3} \frac{d \hat{C}}{d \hat{r}}+\frac{\hat{D}}{\hat{r}}\right)=0 \,,
\end{equation} where $Pe={{u_\infty R}/a}$ with $a$ being the thermal diffusivity and $Ec={{u_\infty}^{2}}/{(C_p (T_{wall}- T_\infty))}$. Terms $\hat{C}$ and $\hat{D}$ are similarly originating from the expression of the stress tensor components: 

\begin{equation}
\hat{C}=\frac{d \hat{u}}{d \hat{r}} \hat{u}-\frac{\hat{u}^{2}}{\hat{r}}-\frac{\hat{u} \hat{v}}{\hat{r}} 
\end{equation}

\begin{equation}
\hat{D}=\frac{1}{3}\left(4 \frac{d \hat{u}}{d \hat{r}} \hat{u}-7 \frac{\hat{u}^{2}}{\hat{r}}-5 \frac{\hat{u} \hat{v}}{\hat{r}}+3 \frac{d \hat{v}}{d \hat{r}} \hat{u}-2 \frac{d \hat{u}}{d \hat{r}} \hat{v}+2 \frac{\hat{v}^{2}}{\hat{r}}\right)
\end{equation}

\noindent Reviewing equations \ref{cont1}, \ref{mom1}, \ref{mom2} and \ref{ener}, the non-dimensional numbers that govern the problem are identified ($M,Bo_i,Da_i,Re,Ec,Pe,Cp_i/Cp$).

\subsection{Parameters physical interpretation}
\label{appex2}

The discussed approach provides parametric regression functions for flow field quantities along the stagnation streamline. It is of interest to investigate the relation of three curve-fitted parameters from Section \ref{hfm} with respect to the simulation inputs. Such relations produce valuable insight on the physical interpretation of the proposed parametrization.

\begin{figure}[!htb]
\begin{center}$
\begin{array}{lll}
\includegraphics[width=54mm]{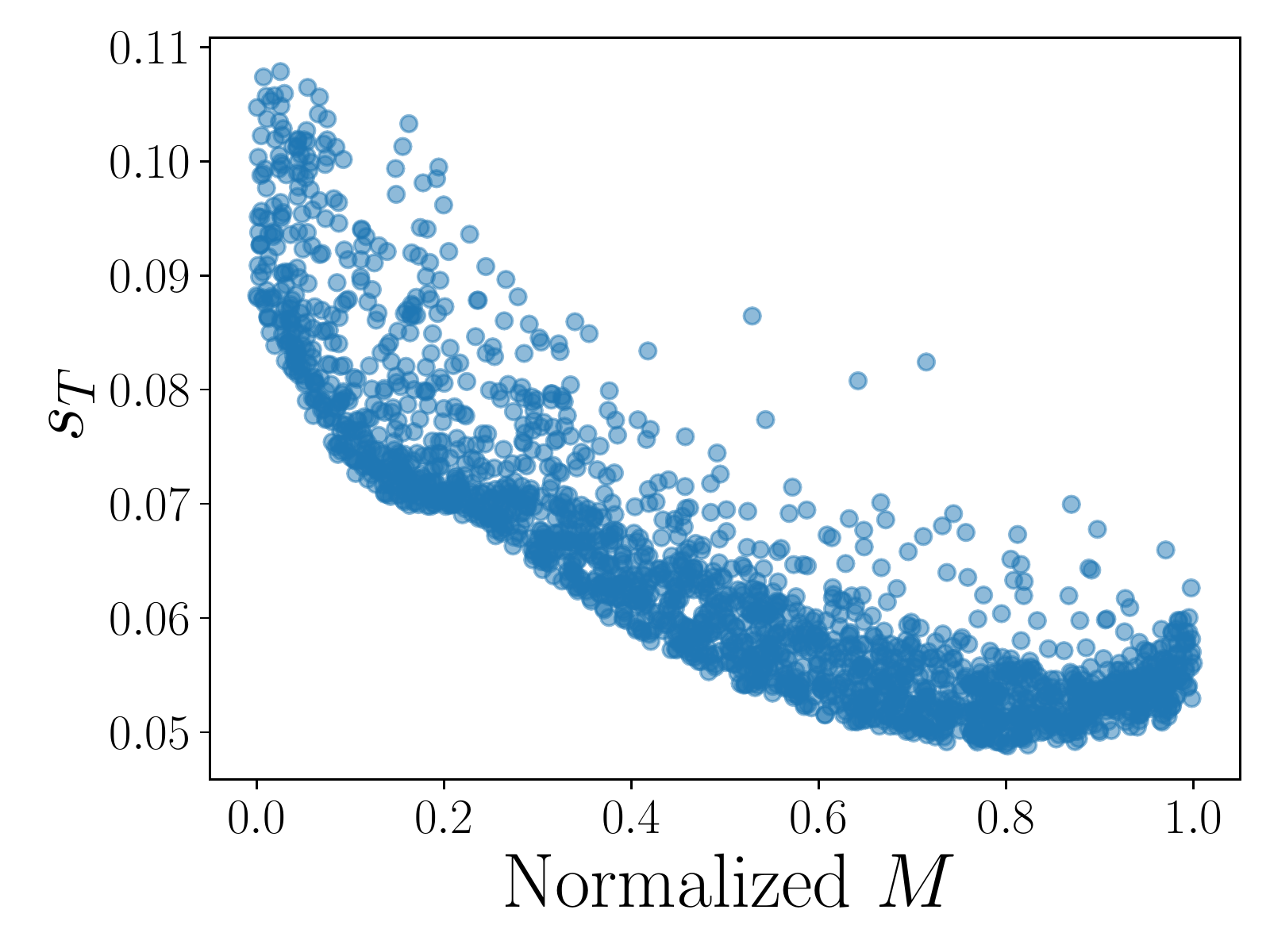}&
\includegraphics[width=54mm]{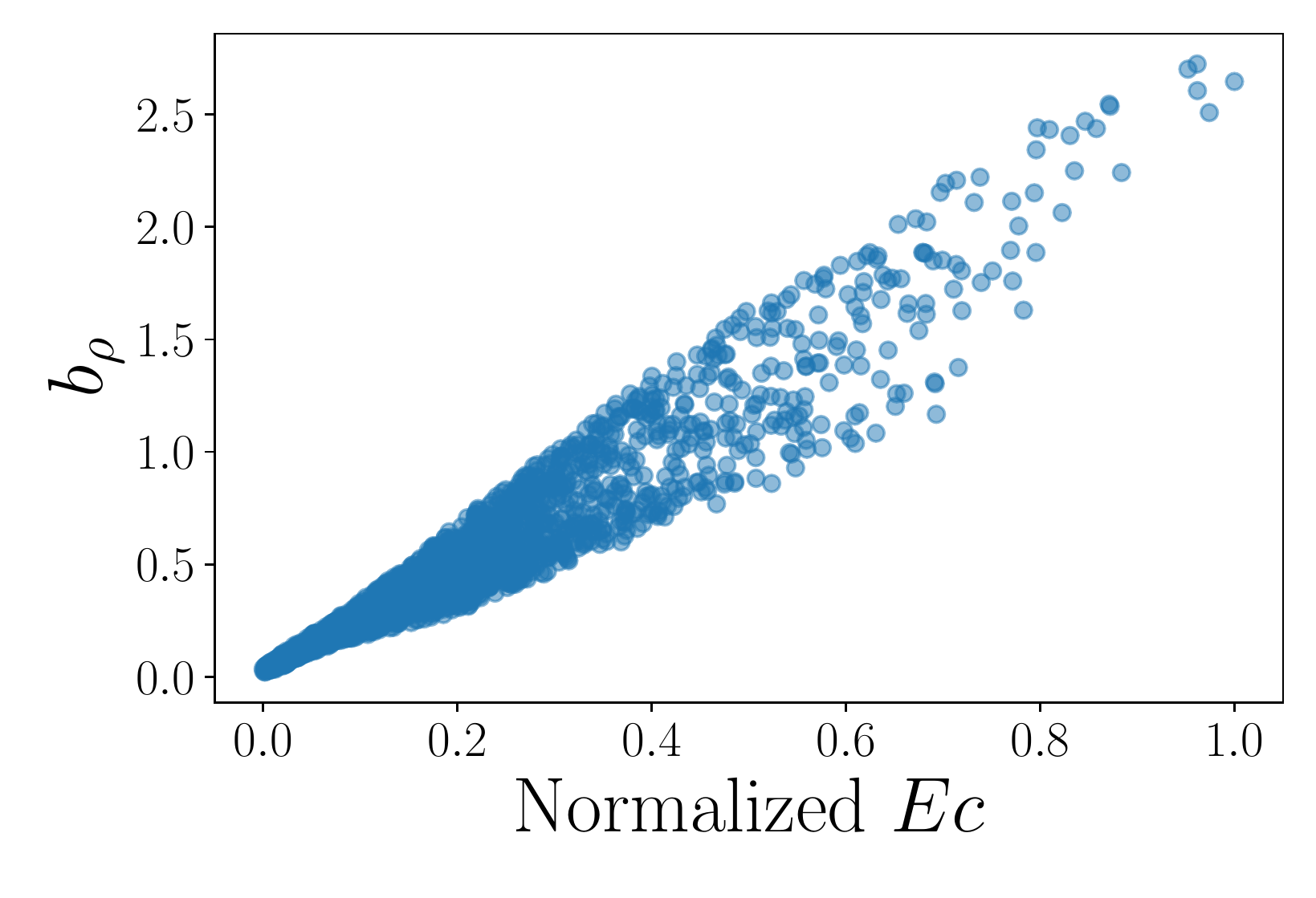}&
\includegraphics[width=54mm]{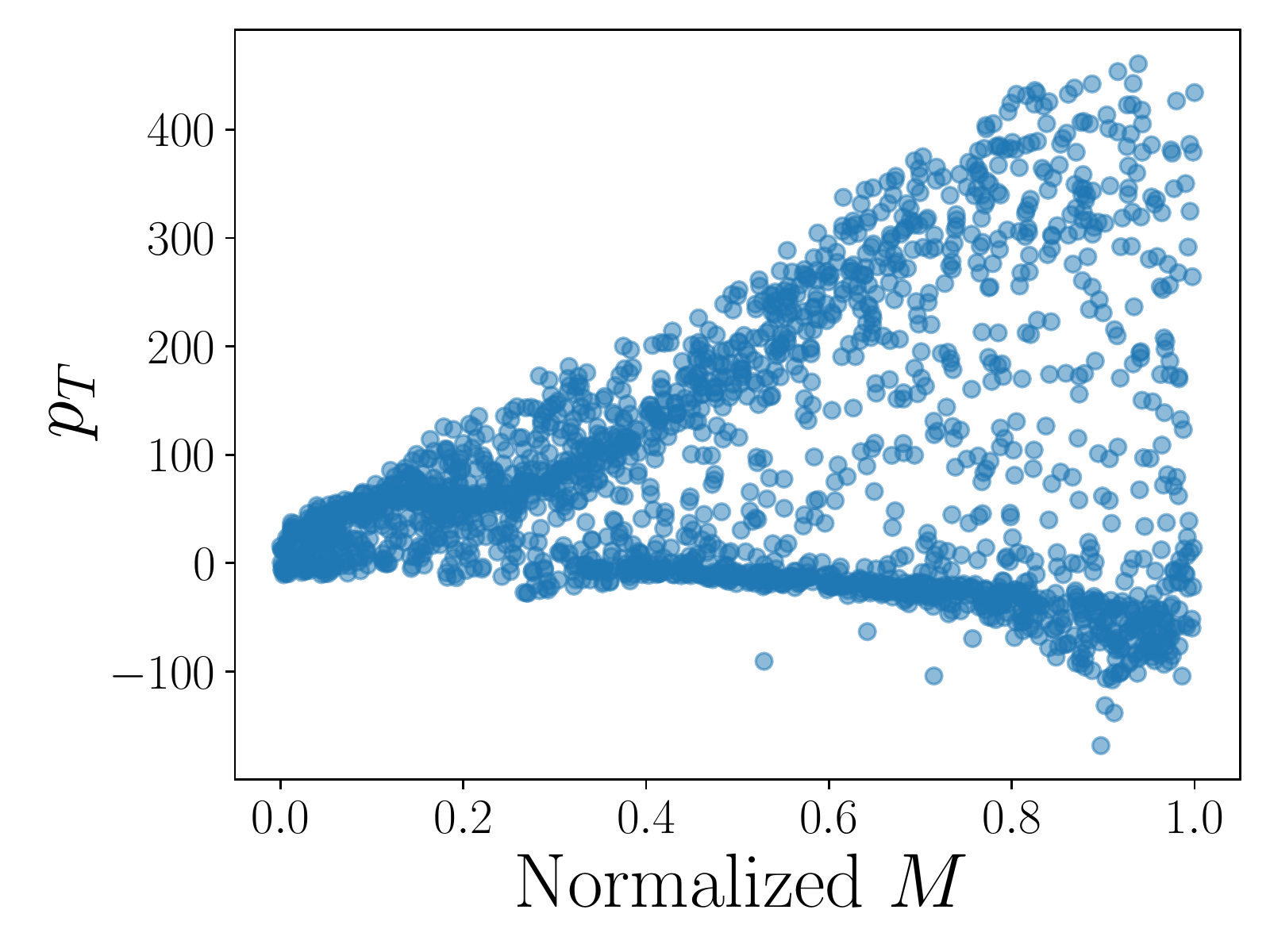}
\end{array}$
\caption{Examples of output parameters projection to problem inputs: shock position $s$, density decrease along thermal boundary layer $b_\rho$, post-shock temperature slope $p_T$.} \label{fig:proj}
\end{center}
\end{figure}

Firstly, parameter $s$ corresponding to the shock position, is highly correlated to the Mach number (Figure \ref{fig:proj}), as expected \cite{Billig1967}. It is also observed that for very high Mach numbers, the shock standoff distance shows an increasing trend, possibly due to the changes in chemistry after the shock. Second, an increased correlation is observed for coefficient $b_\rho$ (Eq.~\ref{rhohat}) to the Eckert number. This parameter is linked to the decrease in density from the wall to the post-shock region. This correlation is expected by the physical interpretation of the Eckert number being inversely proportional to the heat dissipation potential. Finally, a pitchfork-like shape of coefficient $p_T$ (\cref{temhat}) with respect to the Mach number is observed. This shape is also identified in the parameters $a_p$, $u_T$, $u_\rho$, all related to the post-shock behaviour of the flow. At the examined altitude range of 30 to 70 km, the Earth's atmosphere is such that two altitude values have the same temperature $T_\infty$ \cite{Layers}. As a result, for a given free-stream velocity, the same Mach number can lead to two different solutions.

\bibliography{sample}

\end{document}

%% file: review.tex
\usepackage[dvipsnames,svgnames,x11names]{xcolor}
 \usepackage[final]{changes}

\definechangesauthor[color=BrickRed]{AT}
\definechangesauthor[color=NavyBlue]{LD}
\definechangesauthor[color=orange]{BD}
\definechangesauthor[color=green]{MM}

\newcommand{\addAT}[2]{\added[id=AT]{#1}}

\usepackage{todonotes}
\setcommentmarkup{\todo[color={authorcolor!20},size=\scriptsize]{#3: #1}}

\usepackage{xargs}                      % Use more than one optional parameter in a new commands
\newcommandx{\AT}[2][1=]{\todo[linecolor=BrickRed,backgroundcolor=BrickRed!25,bordercolor=BrickRed,#1]{#2}}
\newcommandx{\LD}[2][1=]{\todo[linecolor=NavyBlue,backgroundcolor=NavyBlue!25,bordercolor=NavyBlue,#1]{#2}}
\newcommandx{\BD}[2][1=]{\todo[linecolor=orange,backgroundcolor=orange!25,bordercolor=orange,#1]{#2}}
\newcommandx{\MM}[2][1=]{\todo[linecolor=green,backgroundcolor=green!25,bordercolor=green,#1]{#2}}